\documentclass[a4paper,12pt]{article}
\usepackage{amssymb,amsmath}
\usepackage[dvips]{lscape,graphicx}

\newcommand{\bc}{\begin{center}}
\newcommand{\ec}{\end{center}}
\newcommand{\bd}{\begin{displaymath}}
\newcommand{\ed}{\end{displaymath}}
\newcommand{\be}{\begin{equation}}
\newcommand{\ee}{\end{equation}}
\newcommand{\ba}{\begin{array}}
\newcommand{\ea}{\end{array}}
\newcommand{\bt}{\begin{tabular}}
\newcommand{\et}{\end{tabular}}

\newcommand{\ds}{\displaystyle}

\begin{document}

\bibliographystyle{OurBibTeX}

\begin{titlepage}

\vspace*{-15mm}
\begin{flushright}
UH-511-1193-12
\end{flushright}
\vspace*{5mm}

\begin{center}
{\sffamily
\LARGE
$E_6$ inspired SUSY models with exact custodial symmetry
}
\\[8mm]
Roman Nevzorov$^{a,\,b,\,}$\footnote{E-mail: \texttt{nevzorov@itep.ru}}\quad
\\[3mm]
{\small\it
$^a$ Department of Physics and Astronomy, University of Hawaii, \\
Honolulu, HI 96822, USA\\[2mm]
$^b$ Institute for Theoretical and Experimental Physics,\\ 
Moscow, 117218, Russia 
}\\[1mm]
\end{center}
\vspace*{0.75cm}



\begin{abstract}{

\noindent
The breakdown of $E_6$ gauge symmetry at high energies may lead to supersymmetric
(SUSY) models based on the Standard Model (SM) gauge group together with extra
$U(1)_{\psi}$ and $U(1)_{\chi}$ gauge symmetries. To ensure anomaly cancellation
the particle content of these $E_6$ inspired models involves extra exotic states
that generically give rise to non-diagonal flavour transitions and rapid proton decay.
We argue that a single discrete $\tilde{Z}^{H}_2$ symmetry can be used to forbid
tree-level flavor-changing transitions, as well as the most dangerous baryon and lepton
number violating operators. We present $5D$ and $6D$ orbifold GUT constructions that 
lead to the $E_6$ inspired SUSY models of this type. The breakdown of $U(1)_{\psi}$
and $U(1)_{\chi}$ gauge symmetries that preserves $E_6$ matter parity assignment
guarantees that ordinary quarks and leptons and their superpartners, as well as
the exotic states which originate from $27$ representations of $E_6$ survive to low 
energies. These $E_6$ inspired models contain two dark-matter candidates
and must also include additional TeV scale vectorlike lepton or vectorlike down type
quark states to render the lightest exotic quark unstable. We examine gauge
coupling unification in these models and discuss their implications for collider
phenomenology and cosmology.}
\end{abstract}

\end{titlepage}

\newpage
\section{Introduction}

$E_6$ inspired models are well motivated extensions of the Standard
Model (SM).  Indeed, supersymmetric (SUSY) models based on the $E_6$
gauge symmetry or its subgroup can originate from the ten--dimensional
heterotic superstring theory \cite{1}.  Within this framework gauge and
gravitational anomaly cancellation was found to occur for the gauge
groups $SO(32)$ or $E_8\times E'_8$. However only $E_8\times E'_8$ can
contain the SM since it allows for chiral fermions while $SO(32)$ does
not. Compactification of the extra dimensions results in the breakdown
of $E_8$ up to $E_6$ or one of its subgroups in the observable sector
\cite{2}. The remaining $E'_8$ couples to the usual matter
representations of the $E_6$ only by virtue of gravitational
interactions and comprises a hidden sector that is thought to be
responsible for the spontaneous breakdown of local SUSY
(supergravity). At low energies the hidden sector decouples from the
observable sector of quarks and leptons, the gauge and Higgs bosons and
their superpartners. Its only manifest effect is a set of soft SUSY
breaking terms which spoil the degeneracy between bosons and fermions
within one supermultiplet \cite{3}.  The scale of soft SUSY breaking
terms  is set by the gravitino mass, $m_{3/2}$.  In the simplest SUSY
extensions of the SM these terms also determine the electroweak (EW)
scale. A large mass hierarchy between $m_{3/2}$ and Planck scale can be
caused by the non--perturbative effects in the hidden sector that may
trigger the breakdown of supergravity (SUGRA) \cite{4}.

Since $E_6$ is a rank - 6 group the breakdown of $E_6$ symmetry may
result in low energy models based on rank - 5 or rank - 6 gauge groups,
with one or two additional $U(1)$ gauge group factors in comparison to
the SM. Indeed, $E_6$ contains the maximal subgroup $SO(10)\times
U(1)_{\psi}$ while $SO(10)$ can be decomposed in terms of the
$SU(5)\times U(1)_{\chi}$ subgroup \cite{5}--\cite{Langacker:2008yv}. By
means of the Hosotani mechanism \cite{6} $E_6$ can be broken directly to
$$
E_6\to SU(3)_C\times SU(2)_W\times U(1)_Y\times U(1)_{\psi}\times U(1)_{\chi}
$$
which has rank--6. This rank--6 model may be reduced further to an effective
rank--5 model with only one extra gauge symmetry $U(1)'$ which is a linear
combination of $U(1)_{\chi}$ and $U(1)_{\psi}$:
\be
U(1)'=U(1)_{\chi}\cos\theta+U(1)_{\psi}\sin\theta\,.
\label{0}
\ee

In the models based on rank - 6 or rank - 5 subgroups of $E_6$ the
anomalies are automatically cancelled if the low energy particle
spectrum consists of a complete representations of $E_6$.
Consequently, in $E_6$-inspired SUSY models one is forced
to augment the minimal particle spectrum by a number of exotics which,
together with ordinary quarks and leptons, form complete fundamental
$27$ representations of $E_6$. Thus we will assume that the
particle content of these models includes at least three fundamental
representations of $E_6$ at low energies. These multiplets decompose
under the $SU(5)\times U(1)_{\psi}\times U(1)_{\chi}$ subgroup of $E_6$
as follows: \be \ba{rcl} 27_i &\to &
\ds\left(10,\,\dfrac{1}{\sqrt{24}},\,-\dfrac{1}{\sqrt{40}}\right)_i
+\left(5^{*},\,\dfrac{1}{\sqrt{24}},\,\dfrac{3}{\sqrt{40}}\right)_i
+\left(5^{*},\,-\dfrac{2}{\sqrt{24}},\,-\dfrac{2}{\sqrt{40}}\right)_i
\\[3mm] & + &
\ds\left(5,\,-\dfrac{2}{\sqrt{24}},\,\dfrac{2}{\sqrt{40}}\right)_i
+\left(1,\,\dfrac{4}{\sqrt{24}},\,0\right)_i
+\left(1,\,\dfrac{1}{\sqrt{24}},\,-\dfrac{5}{\sqrt{40}}\right)_i\,.  \ea
\label{1}
\ee The first, second and third quantities in brackets are the $SU(5)$
representation and extra $U(1)_{\psi}$ and $U(1)_{\chi}$ charges
respectively, while $i$ is a family index that runs from 1 to 3. An
ordinary SM family, which contains the doublets of left--handed quarks
$Q_i$ and leptons $L_i$, right-handed up-- and down--quarks ($u^c_i$ and
$d^c_i$) as well as right--handed charged leptons $(e^c_i)$, is assigned
to $\left(10,\,\dfrac{1}{\sqrt{24}},\,-\dfrac{1}{\sqrt{40}}\right)_i$ +
$\left(5^{*},\,\dfrac{1}{\sqrt{24}},\,\dfrac{3}{\sqrt{40}}\right)_i$.
Right-handed neutrinos $N^c_i$ are associated with the last term in
Eq.~(\ref{1}),
$\left(1,\,\dfrac{1}{\sqrt{24}},\,-\dfrac{5}{\sqrt{40}}\right)_i$.  The
next-to-last term, $\left(1,\,\dfrac{4}{\sqrt{24}},\,0\right)_i$,
represents new SM-singlet fields $S_i$, with non-zero $U(1)_{\psi}$
charges that therefore survive down to the EW scale. The pair of
$SU(2)_W$--doublets ($H^d_{i}$ and $H^u_{i}$) that are contained in
$\left(5^{*},\,-\dfrac{2}{\sqrt{24}},\,-\dfrac{2}{\sqrt{40}}\right)_i$
and $\ds\left(5,\,-\dfrac{2}{\sqrt{24}},\,\dfrac{2}{\sqrt{40}}\right)_i$
have the quantum numbers of Higgs doublets. They form either Higgs or
Inert Higgs $SU(2)_W$ multiplets.~\footnote{We use the terminology
``Inert Higgs'' to denote Higgs--like doublets that do not develop
VEVs.} Other components of these $SU(5)$ multiplets form colour
triplets of exotic quarks $\overline{D}_i$ and $D_i$ with electric
charges $+ 1/3$ and $-1/3$ respectively. These exotic quark states carry
a $B-L$ charge $\left(\pm\dfrac{2}{3}\right)$ twice larger than that of
ordinary ones.  In phenomenologically viable $E_6$ inspired models they
can be either diquarks or leptoquarks.

The presence of the $Z'$ bosons associated with extra $U(1)$ gauge
symmetries and exotic matter in the low-energy spectrum stimulated the
extensive studies of the $E_6$ inspired SUSY models over the years
\cite{5},~\cite{7}. Recently, the latest Tevatron and early
LHC $Z'$ mass limits in these models have been discussed in
\cite{Accomando:2010fz} while different aspects of phenomenology of
exotic quarks and squarks have been considered in \cite{Kang:2007ib}.
Also the implications of the $E_6$ inspired SUSY models have been
studied for EW symmetry breaking (EWSB)
\cite{Langacker:1998tc}--\cite{Daikoku:2000ep}, neutrino physics
\cite{Kang:2004ix}--\cite{Ma:1995xk}, leptogenesis
\cite{Hambye:2000bn}--\cite{King:2008qb}, EW baryogenesis
\cite{baryogen}, muon anomalous magnetic moment \cite{g-2}, electric
dipole moment of electron \cite{Suematsu:1997tv} and tau lepton
\cite{GutierrezRodriguez:2006hb}, lepton flavour violating processes
like $\mu\to e\gamma$ \cite{Suematsu:1997qt} and CP-violation in the
Higgs sector \cite{Ham:2008fx}. The neutralino sector in $E_6$ inspired
SUSY models was analysed previously in \cite{Keith:1997zb},
\cite{Suematsu:1997tv}--\cite{Suematsu:1997qt},
\cite{Suematsu:1997au}--\cite{E6neutralino-higgs}.  Such models have
also been proposed as the solution to the tachyon problems of anomaly
mediated SUSY breaking, via $U(1)^\prime$ D-term contributions
\cite{Asano:2008ju}, and used in combination with a generation symmetry
to construct a model explaining fermion mass hierarchy and mixing
\cite{Stech:2008wd}.  An important feature of $E_6$ inspired SUSY models
is that the mass of the lightest Higgs particle can be substantially
larger in these models than in the minimal supersymmetric standard model
(MSSM) and next-to-minimal supersymmetric standard model (NMSSM)
\cite{Daikoku:2000ep}, \cite{King:2005jy}--\cite{Accomando:2006ga}.  The
Higgs sector in these models was examined recently in
\cite{E6neutralino-higgs}, \cite{King:2005jy}, \cite{E6-higgs}.

Within the class of rank - 5 $E_6$ inspired SUSY models, there is a
unique choice of Abelian $U(1)_{N}$ gauge symmetry that allows zero
charges for right-handed neutrinos and thus a high scale see-saw mechanism.
This corresponds to $\theta=\arctan\sqrt{15}$.
Only in this Exceptional Supersymmetric Standard Model (E$_6$SSM)
\cite{King:2005jy}--\cite{King:2005my} right--handed neutrinos may be
superheavy, shedding light on the origin of the mass hierarchy in the lepton sector
and providing a mechanism for the generation of the baryon asymmetry in the Universe
via leptogenesis \cite{Hambye:2000bn}--\cite{King:2008qb}. Indeed, the heavy Majorana
right-handed neutrinos may decay into final states with lepton number
$L=\pm 1$, thereby creating a lepton asymmetry in the early universe.
Since in the E$_6$SSM the Yukawa couplings of the new exotic particles
are not constrained by neutrino oscillation data, substantial values of
the CP--asymmetries can be induced even for a relatively small mass of
the lightest right--handed neutrino ($M_1 \sim 10^6\,\mbox{GeV}$) so
that successful thermal leptogenesis may be achieved without
encountering a gravitino problem \cite{King:2008qb}.

Supersymmetric models with an additional $U(1)_{N}$ gauge symmetry have
been studied in \cite{Ma:1995xk} in the context of non--standard
neutrino models with extra singlets, in \cite{Suematsu:1997au} from the
point of view of $Z-Z'$ mixing, in \cite{Keith:1997zb} and
\cite{Suematsu:1997au}--\cite{Keith:1996fv} where the neutralino sector
was explored, in \cite{Keith:1997zb}, \cite{King:2007uj} where the
renormalisation group (RG) flow of couplings was examined and in
\cite{Suematsu:1994qm}--\cite{Daikoku:2000ep} where EWSB was
studied. The presence of a $Z'$ boson and of exotic quarks predicted by
the Exceptional SUSY model provides spectacular new physics signals at
the LHC which were analysed in
\cite{King:2005jy}--\cite{Accomando:2006ga}, \cite{Howl:2007zi}. The
presence of light exotic particles in the E$_6$SSM spectrum also lead to
the nonstandard decays of the SM--like Higgs boson that were discussed
in details in \cite{Hall:2010ix}.  Recently the particle spectrum and
collider signatures associated with it were studied within the
constrained version of the E$_6$SSM \cite{8}.

Although the presence of TeV scale exotic matter in $E_6$ inspired SUSY
models gives rise to specatucular collider signatures, it also causes
some serious problems. In particular, light exotic states generically
lead to non--diagonal flavour transitions and rapid proton decay. To
suppress flavour changing processes as well as baryon and lepton number
violating operators one can impose a set of discrete symmetries. For
example, one can impose an approximate $Z^{H}_2$ symmetry, under which
all superfields except one pair of $H^{d}_{i}$ and $H^{u}_{i}$ (say
$H_d\equiv H^{d}_{3}$ and $H_u\equiv H^{u}_{3}$) and one SM-type singlet
field ($S\equiv S_3$) are odd
\cite{King:2005jy}--\cite{King:2005my}. When all $Z^{H}_2$ symmetry
violating couplings are small this discrete symmetry allows to suppress
flavour changing processes. If the Lagrangian of the $E_6$
inspired SUSY models is invariant with respect to either a $Z_2^L$
symmetry, under which all superfields except leptons are even (Model I),
or a $Z_2^B$ discrete symmetry that implies
that exotic quark and lepton superfields are odd whereas the others
remain even (Model II), then the most dangerous baryon and lepton
number violating operators get forbidden and proton is sufficiently
longlived \cite{King:2005jy}--\cite{King:2005my}. The symmetries
$Z^{H}_2$, $Z_2^L$ and $Z_2^B$ obviously do not commute with $E_6$
because different components of fundamental representations of $E_6$
transform differently under these symmetries.

The necessity of introducing multiple discrete symmetries to ameliorate
phenomenological problems that generically arise due to the presence of
low mass exotics is an undesirable feature of these models.
In this paper we consider rank - 6 $E_6$ inspired SUSY models in which a
{\em single} discrete $\tilde{Z}^{H}_2$ symmetry serves to
simultaneously forbid tree--level
flavor--changing transitions and the most dangerous baryon and lepton
number violating operators. We consider models where the $U(1)_{\psi}$
and $U(1)_{\chi}$ gauge symmetries are spontaneously broken at some
intermediate scale so that the matter parity,
\be Z_{2}^{M}=(-1)^{3(B-L)}\;,
\label{2}
\ee
is preserved. As a consequence the low-energy spectrum of the models
will include {\em two} stable weakly interacting particles that
potentially contribute to the dark matter density of our Universe.
The invariance of the Lagrangian with respect to $Z_{2}^{M}$ and
$\tilde{Z}^{H}_2$ symmetries leads to unusual collider signatures
associated with exotic states that originate from $27$--plets.
These signatures have not been studied in details before. In
addition to the exotic matter multiplets that stem from the fundamental
$27$ representations of $E_6$ the considered models predict the
existence of a set of vector-like supermultiplets. In particular the
low-energy spectrum of the models involves either a doublet of
vector-like leptons or a triplet of vector-like down type quarks. If
these extra states are relatively light, they will manifest themselves
at the LHC in the near future.

The layout of this paper is as follows. In Section 2 we specify the 
rank--6 $E_6$ inspired SUSY models with exact custodial symmetry.
In Section 3 we present five--dimensional ($5D$) and six--dimensional 
($6D$) orbifold Grand Unified theories (GUTs) that lead to the rank--6 
$E_6$ inspired SUSY models that we propose. In Sections 4 and 5 the 
RG flow of gauge couplings and implications for collider phenomenology 
and cosmology are discussed. Our results are summarized in Section 6.

\section{$E_6$ inspired SUSY models with exact custodial
$\tilde{Z}^{H}_2$ symmetry}

In our analysis we concentrate on the rank--6 $E_6$ inspired SUSY models with two
extra $U(1)$ gauge symmetries --- $U(1)_{\chi}$ and $U(1)_{\psi}$. In other words
we assume that near the GUT or string scale $E_6$ or its subgroup is broken
down to $SU(3)_C\times SU(2)_W\times U(1)_Y\times U(1)_{\psi}\times U(1)_{\chi}$. In the
next section we argue that this breakdown can be achieved within orbifold GUT models.
We also allow three copies of 27-plets to survive to low energies so that anomalies get
cancelled generation by generation within each complete $27_i$ representation of $E_6$.
In $E_6$ models the renormalisable part of the superpotential comes from the
$27\times 27\times 27$ decomposition of the $E_6$ fundamental representation and
can be written as
\be
\ba{rcl}
W_{E_6}&=&W_0+W_1+W_2,\\[3mm]
W_0&=&\lambda_{ijk}S_i(H^d_{j}H^u_{k})+\kappa_{ijk}S_i(D_j\overline{D}_k)+
h^N_{ijk} N_i^c (H^u_{j} L_k)+ h^U_{ijk} u^c_{i} (H^u_{j} Q_k)+\\[2mm]
&&+h^D_{ijk} d^c_i (H^d_{j} Q_k) + h^E_{ijk} e^c_{i} (H^d_{j} L_k)\,,\\[3mm]
W_1&=& g^Q_{ijk} D_{i} (Q_j Q_k)+g^{q}_{ijk}\overline{D}_i d^c_j u^c_k\,,\\[3mm]
W_2&=& g^N_{ijk}N_i^c D_j d^c_k+g^E_{ijk} e^c_i
D_j u^c_k+g^D_{ijk} (Q_i L_j) \overline{D}_k\,.
\ea
\label{3}
\ee
Here the summation over repeated family indexes ($i,j,k=1,2,3$) is implied.
In the considered models $B-L$ number is conserved automatically since the
corresponding global symmetry $U(1)_{B-L}$ is a linear superposition of
$U(1)_Y$ and $U(1)_{\chi}$. At the same time if terms in $W_1$ and $W_2$
are simultaneously present in the superpotential then baryon and lepton numbers
are violated. In other words one cannot define the baryon and lepton numbers of the
exotic quarks $D_i$ and $\overline{D}_i$ so that the complete Lagrangian is
invariant separately under $U(1)_{B}$ and $U(1)_{L}$ global
symmetries. In this case the Yukawa interactions in $W_1$ and $W_2$ give
rise to rapid proton decay.

Another problem is associated with the presence of three families of $H^u_{i}$
and $H^d_{i}$. All these Higgs--like doublets can couple to ordinary quarks
and charged leptons of different generations resulting in the phenomenologically
unwanted flavor changing transitions. For example, non--diagonal flavor
interactions contribute to the amplitude of $K^0-\overline{K}^0$ oscillations
and give rise to new channels of muon decay like $\mu\to e^{-}e^{+}e^{-}$.
In order to avoid the appearance of flavor changing neutral currents (FCNCs)
at the tree level and forbid the most dangerous baryon and lepton number violating
operators one can try to impose a single $\tilde{Z}^{H}_2$ discrete symmetry.
One should note that the imposition of additional discrete symmetry to stabilize
the proton is a generic feature of many phenomenologically viable SUSY models.

In our model building strategy we use $SU(5)$ SUSY GUT as a guideline.
Indeed, the low--energy spectrum of the MSSM, in addition to the complete $SU(5)$
multiplets, contains an extra pair of doublets from $5$ and $\overline{5}$ fundamental
representations, that play a role of the Higgs fields which break EW symmetry.
In the MSSM the potentially dangerous operators, that lead to the rapid proton decay,
are forbidden by the matter parity $Z_{2}^{M}$ under which Higgs doublets are even
while all matter superfields, that fill in complete $SU(5)$ representations,
are odd. Following this inspirational example we augment three
27-plets of $E_6$ by a number of components $M_{l}$ and $\overline{M}_l$ from
extra $27'_l$ and $\overline{27'}_l$ below the GUT scale. Because additional pairs
of multiplets $M_{l}$ and $\overline{M}_l$ have opposite $U(1)_{Y}$, $U(1)_{\psi}$
and $U(1)_{\chi}$ charges their contributions to the anomalies get cancelled
identically. As in the case of the MSSM we allow the set of multiplets $M_{l}$
to be used for the breakdown of gauge symmetry. If the corresponding set includes
$H^u\equiv H_u$, $H^d\equiv H_d$, $S$ and $N^c\equiv N^c_H$ then
the $SU(2)_W\times U(1)_Y\times U(1)_{\psi}\times U(1)_{\chi}$ symmetry can be
broken down to $U(1)_{em}$ associated with electromagnetism. The VEVs of $S$
and $N^c$ break $U(1)_{\psi}$ and $U(1)_{\chi}$ entirely while the
$SU(2)_W\times U(1)_Y$ symmetry remains intact. When the neutral components of
$H_u$ and $H_d$ acquire non--zero VEVs then $SU(2)_W\times U(1)_Y$ symmetry
gets broken to $U(1)_{em}$ and the masses of all quarks and charged leptons
are generated.

As in the case of the MSSM we assume that all multiplets $M_{l}$ are even
under $\tilde{Z}^{H}_2$ symmetry while three copies of the complete fundamental
representations of $E_6$ are odd. This forbids couplings in the superpotential
that come from $27_i \times 27_j \times 27_k$. On the other hand the $\tilde{Z}^{H}_2$
symmetry allows the Yukawa interactions that stem from $27'_l \times 27'_m \times 27'_n$,
and $27'_l \times 27_i \times 27_k$
The multiplets
$M_{l}$ have to be even under $\tilde{Z}^{H}_2$ symmetry because some of them are
expected to get VEVs. Otherwise the VEVs of the corresponding fields lead to the
breakdown of the discrete $\tilde{Z}^{H}_2$ symmetry giving rise to the baryon and
lepton number violating operators in general. If the set of multiplets $M_{l}$
includes only one pair of doublets $H_d$ and $H_u$ the $\tilde{Z}^{H}_2$ symmetry
defined above permits to suppress unwanted FCNC processes at the tree level since
down-type quarks and charged leptons couple to just one Higgs doublet $H_d$,
whereas the up-type quarks couple to $H_u$ only.

The superfields $\overline{M}_l$ can be either odd or even under this $\tilde{Z}^{H}_2$
symmetry. Depending on whether these fields are even or odd under $\tilde{Z}^{H}_2$
a subset of terms in the most general renormalizable superpotential can be written as
\be
\ba{c}
W_{\rm{total}}=Y'_{lmn} 27'_l 27'_m 27'_n + Y_{lij} 27'_l 27_i 27_j +
\tilde{Y}_{lmn} \overline{27'}_l \overline{27'}_m \overline{27'}_n +\\[2mm]
+ \mu'_{il} 27_i \overline{27'}_l + \tilde{\mu}'_{ml} 27'_m \overline{27'}_l...\,,
\ea
\label{4}
\ee
where $Y'_{lmn}$ and $Y_{lij}$ are Yukawa couplings and $\mu'_{il}$ and $\tilde{\mu}'_{ml}$
are mass parameters. Also one should keep in mind that only $M_{l}$ and $\overline{M}_l$
components of $27'_l$ and $\overline{27'}_l$ appear below the GUT scale.
If $\overline{M}_l$ is odd under $\tilde{Z}^{H}_2$ symmetry then the term
$\tilde{\mu}'_{ml} 27'_m \overline{27'}_l$ and
$\tilde{Y}_{lmn}\overline{27'}_l \overline{27'}_m \overline{27'}_n $ are forbidden
while $\mu'_{il}$ can have non-zero values. When $\overline{M}_l$ is even $\mu'_{il}$
vanish whereas $\tilde{\mu}'_{ml} 27'_m \overline{27'}_l$ and
$\tilde{Y}_{lmn}\overline{27'}_l \overline{27'}_m \overline{27'}_n $ are allowed by
$\tilde{Z}^{H}_2$ symmetry. In general mass parameters $\mu'_{il}$ and $\tilde{\mu}'_{ml}$ are
expected to be of the order of GUT scale. In order to allow some of the $\overline{M}_l$ multiplets
to survive to low energies we assume that the corresponding mass terms are forbidden
at high energies and get induced at some intermediate scale which is much lower than $M_X$.

The VEVs of the superfields $N^c_H$ and $\overline{N}_H^c$ (that originate from
$27'_N$ and $\overline{27'}_N$) can be used not only for the breakdown of $U(1)_{\psi}$
and $U(1)_{\chi}$ gauge symmetries, but also to generate Majorana masses for the
right--handed neutrinos that can be induced through interactions
\be
\Delta W_{N}=\ds\frac{\varkappa_{ij}}{M_{Pl}}(27_i\,\overline{27'}_{N})(27_j\,\overline{27'}_{N})\,.
\label{8}
\ee
The non--renormalizable operators (\ref{8}) give rise to the right--handed neutrino masses
which are substantially lower than the VEVs of $N^c_H$ and $\overline{N}_H^c$. Because
the observed pattern of the left--handed neutrino masses and mixings can be naturally
reproduced by means of seesaw mechanism if the right--handed neutrinos are superheavy,
the $N^c_H$ and $\overline{N}_H^c$ are expected to acquire VEVs
$<N_H^c>\simeq<\overline{N}_H^c>\lesssim M_X$. This implies that $U(1)_{\psi}\times U(1)_{\chi}$
symmetry is broken down to $U(1)_{N}$ near the GUT scale, where $U(1)_{N}$ symmetry
is a linear superposition of $U(1)_{\psi}$ and $U(1)_{\chi}$, i.e.
\be
U(1)_N=\dfrac{1}{4} U(1)_{\chi}+\dfrac{\sqrt{15}}{4} U(1)_{\psi}\,,
\label{7}
\ee
under which right-handed neutrinos have zero charges. Since $N^c_H$ and $\overline{N}_H^c$
acquire VEVs both supermultiplets must be even under $\tilde{Z}^{H}_2$ symmetry.


At the same time the VEVs of $N^c_H$ and $\overline{N}_H^c$ may break $U(1)_{B-L}$ symmetry.
In particular, as follows from Eq.~(\ref{3}) the VEV of $N^c_H$ can induce the bilinear
terms $M^{L}_{ij} (H^u_{i} L_{j})$ and $M^{B}_{ij} (D_i d^c_j)$ in the superpotential.
Although such breakdown of gauge symmetry might be possible
the extra particles tend to be rather heavy in the considered case and thus irrelevant for
collider phenomenology. Therefore we shall assume further that the couplings of $N^c_H$
to $27_i$ are forbidden. This, for example, can be achieved by imposing an extra discrete
symmetry $Z_n$. Although this symmetry can forbid the interactions of $N^c_H$ with
three complete $27_i$ representations of $E_6$ it should allow non--renormalizable
interactions (\ref{8}) that induce the large Majorana masses for right-handed neutrinos.
These requirements are fulfilled if Lagrangian is invariant under $Z_2$ symmetry transformations
$N^c_H\to -N^c_H$ and $\overline{N}_H^c\to -\overline{N}_H^c$. Alternatively, one can impose
$Z_n$ symmetry ($n>2$) under which only $N^c_H$ transforms. The invariance of the Lagrangian
with respect to $Z_n$ symmetry ($n>2$) under which only $N^c_H$ transforms implies that the
mass term $\mu_H N^c_H \overline{N}_H^c$ in the superpotential (\ref{4}) is forbidden.
On the other hand this symmetry allows non--renormalizable term in the superpotential
\be
\Delta W_{N^c_{H}} = \varkappa\, \ds\frac{(N^c_H \overline{N}_H^c)^n}{M^{2n-3}_{Pl}}\,,.
\label{5}
\ee
In this case $N^c_H$ and $\overline{N}_H^c$ can develop VEVs along the $D$--flat direction so that
\be
<N_H^c>\simeq<\overline{N}_H^c>\sim M_{Pl} \cdot
\biggl[ \ds\frac{1}{\varkappa}\frac{M_S}{M_{Pl}}\biggr]^{\ds\frac{1}{2n-2}}\,,
\label{6}
\ee
where $M_S$ is a low--energy supersymmetry breaking scale. This mechanism permits to
generate $<N_H^c>\, \gtrsim 10^{14}\,\mbox{GeV}$ resulting in right-handed neutrino masses
of order of
$$
\varkappa_{ij} M_{Pl} \cdot \biggl[ \ds\frac{1}{\varkappa}\frac{M_S}{M_{Pl}}\biggr]^{\ds\frac{1}{n-1}}
\gtrsim 10^{11}\,\mbox{GeV}\,.
$$

\begin{table}[ht]
\centering
\begin{tabular}{|c|c|c|c|c|c|c|c|c|c|}
\hline
                   &  $27_i$          &   $27_i$              &$27'_{H_u}$&$27'_{S}$&
$\overline{27'}_{H_u}$&$\overline{27'}_{S}$&$27'_N$&$27'_{L}$&$27'_{d}$\\
& & &$(27'_{H_d})$& &$(\overline{27'}_{H_d})$& &$(\overline{27'}_N)$&$(\overline{27'}_L)$&$(\overline{27'}_{d})$\\
\hline
                   &$Q_i,u^c_i,d^c_i,$&$\overline{D}_i,D_i,$  & $H_u$     & $S$     &
$\overline{H}_u$&$\overline{S}$&$N^c_H$&$L_4$&$d^c_4$\\
                   &$L_i,e^c_i,N^c_i$ &  $H^d_{i},H^u_{i},S_i$& $(H_d)$   &         &
$(\overline{H}_d)$&&$(\overline{N}_H^c)$&$(\overline{L}_4)$&$(\overline{d^c}_4)$\\
\hline
$\tilde{Z}^{H}_2$  & $-$              & $-$                   & $+$       & $+$     &
$-$&$\pm$&$+$&$+$&$+$\\
\hline
$Z_{2}^{M}$        & $-$              & $+$                   & $+$       & $+$     &
$+$&$+$&$-$&$-$&$-$\\
\hline
$Z_{2}^{E}$        & $+$              & $-$                   & $+$       & $+$     &
$-$&$\pm$&$-$&$-$&$-$\\
\hline
\end{tabular}
\caption{Transformation properties of different components of $E_6$ multiplets
under $\tilde{Z}^H_2$, $Z_{2}^{M}$ and $Z_{2}^{E}$ discrete symmetries.}
\label{tab1}
\end{table}

The mechanism of the gauge symmetry breaking discussed above ensures that
the low--energy effective Lagrangian is automatically invariant under the
matter parity $Z_{2}^{M}$. Such spontaneous breakdown of the
$U(1)_{\psi}\times U(1)_{\chi}$ gauge symmetry can occur because $Z_{2}^{M}$
is a discrete subgroup of $U(1)_{\psi}$ and $U(1)_{\chi}$. This follows
from the $U(1)_{\psi}$ and $U(1)_{\chi}$ charge assignments presented in Eq.~(\ref{1}).
Thus in the considered case the VEVs of $N^c_H$ and $\overline{N}_H^c$ break
$U(1)_{\psi}\times U(1)_{\chi}$ gauge symmetry down to $U(1)_{N}\times Z_{2}^{M}$.
As a consequence the low--energy effective Lagrangian is invariant under both
$Z_{2}^{M}$ and $\tilde{Z}^{H}_2$ discrete symmetries. Moreover the $\tilde{Z}^{H}_2$
symmetry is a product of
\be
\tilde{Z}^{H}_2 = Z_{2}^{M}\times Z_{2}^{E}\,,
\label{9}
\ee
where $Z_{2}^{E}$ is associated with most of the exotic states. In other words
all exotic quarks and squarks, Inert Higgs and Higgsino multiplets
as well as SM singlet and singlino states that do not get VEV are odd
under $Z_{2}^{E}$ symmetry. The transformation properties of different
components of $27_i$, $27'_l$ and $\overline{27'}_l$ multiplets under
the $\tilde{Z}^{H}_2$, $Z_{2}^{M}$ and $Z_{2}^{E}$ symmetries are summarized
in Table~\ref{tab1}. Since the Lagrangian of the considered $E_6$ inspired
models is invariant under $Z_{2}^{M}$ and $\tilde{Z}^{H}_2$ symmetries it is
also invariant under the transformations of $Z_{2}^{E}$ symmetry.
Because $Z_{2}^{E}$ is conserved the lightest exotic state, which is odd
under this symmetry, is absolutely stable and contributes to the relic
density of dark matter.

It is also well known that in SUSY models the lightest supersymmetric
particle (LSP), i.e. the lightest $R$--parity odd particle
($Z_{2}^{R}=(-1)^{3(B-L)+2s}$), must be stable. If in the considered
models the lightest exotic state (i.e. state with $Z_{2}^{E}=-1$) has
even $R$--parity then the lightest $R$--parity odd state cannot decay
as usual. When the lightest exotic state is $R$--parity odd particle
either the lightest $R$--parity even exotic state or the next-to-lightest
$R$--parity odd state with $Z_{2}^{E}=+1$ must be absolutely stable.
Thus the considered $E_6$ inspired SUSY models contain at least two
dark-matter candidates.

The residual extra $U(1)_{N}$ gauge symmetry gets broken by the VEV
of the SM--singlet superfield $S$ (and possibly $\overline{S}$).
The VEV of the field $S$ induces the mass of the $Z'$ associated with
$U(1)_{N}$ symmetry as well as the masses of all exotic quarks and
inert Higgsinos. If $S$ acquires VEV of order $10-100\,\mbox{TeV}$
(or even lower) the lightest exotic particles can be produced at the LHC.
This is the most interesting scenario that we are going to focus on here.
In some cases the superfield $\overline{S}$ may also acquire non--zero
VEV breaking $U(1)_{N}$ symmetry as we will discuss later. If this is
a case then $\overline{S}$ should be even under the $\tilde{Z}^{H}_2$
symmetry. Otherwise the superfield $\overline{S}$ can be $\tilde{Z}^{H}_2$
odd.

The above consideration indicate that the set of multiplets $M_{l}$ has
to contain at least $H_u$, $H_d$, $S$ and $N^c_H$ in order to guarantee
the appropriate breakdown of the gauge symmetry in the rank--6 $E_6$
inspired SUSY models. However if the set of $\tilde{Z}^{H}_2$ even
supermultiplets $M_{l}$ involve only $H_u$, $H_d$, $S$ and $N^c_H$ then
the lightest exotic quarks are extremely long--lived particles. Indeed,
in the considered case the $\tilde{Z}^{H}_2$ symmetry forbids all
Yukawa interactions in $W_1$ and $W_2$ that allow the lightest exotic
quarks to decay. Moreover the Lagrangian of such model is invariant not
only with respect to $U(1)_L$ and $U(1)_B$ but also under $U(1)_D$
symmetry transformations
\be
D\to e^{i\alpha} D\,,\qquad\qquad
\overline{D}\to e^{-i\alpha}\overline{D}\,.
\label{10}
\ee
The $U(1)_D$ invariance ensures that the lightest exotic quark is very
long--lived. The $U(1)_L$, $U(1)_B$ and $U(1)_D$ global symmetries are expected
to be broken by a set of non--renormalizable operators which are suppressed
by inverse power of the GUT scale $M_X$ or $M_{Pl}$. These operators give rise
to the decays of the exotic quarks but do not lead to the rapid proton decay.
Since the extended gauge symmetry in the considered rank--6 $E_6$ inspired
SUSY models forbids any dimension five operators that break $U(1)_D$ global
symmetry the lifetime of the lightest exotic quarks is expected to be of
order of
\be
\tau_D\gtrsim M_X^4/\mu_D^5\,,
\label{11}
\ee
where $\mu_D$ is the mass of the lightest exotic quark. When
$\mu_D\simeq \mbox{TeV}$ the lifetime of the lightest exotic quarks
$\tau_D\gtrsim 10^{49}\,\mbox{GeV}^{-1}\sim 10^{17}\,\mbox{years}$, i.e.
considerably larger than the age of the Universe.

The long--lived exotic quarks would have been copiously produced
during the very early epochs of the Big Bang. Those lightest exotic
quarks which survive annihilation would subsequently have been confined
in heavy hadrons which would annihilate further. The remaining heavy
hadrons originating from the Big Bang should be present in terrestrial
matter. There are very strong upper limits on the abundances of
nuclear isotopes which contain such stable relics in the mass range
from $1\,\mbox{GeV}$ to $10\,\mbox{TeV}$. Different experiments set
limits on their relative concentrations from $10^{-15}$ to $10^{-30}$
per nucleon \cite{42}. At the same time various theoretical
estimations \cite{43} show that if remnant particles would exist in
nature today their concentration is expected to be at the level of
$10^{-10}$ per nucleon. Therefore $E_6$ inspired models with
very long--lived exotic quarks are ruled out.

To ensure that the lightest exotic quarks decay within a reasonable
time the set of $\tilde{Z}^{H}_2$ even supermultiplets $M_{l}$
needs to be supplemented by some components of $27$-plet that carry
$SU(3)_C$ color or lepton number. In this context we
consider two scenarios that lead to different collider signatures
associated with the exotic quarks. In the simplest case ({\bf scenario A})
the set of $\tilde{Z}^{H}_2$ even supermultiplets $M_{l}$ involves lepton
superfields $L_4$ and/or $e^c_4$ that survive to low energies. This implies
that $\overline{D}_i$ and $D_i$ can interact with leptons and quarks
only while the couplings of these exotic quarks to a pair of quarks are
forbidden by the postulated $\tilde{Z}^{H}_2$ symmetry. Then baryon
number is conserved and exotic quarks are leptoquarks.

In this paper we restrict our consideration to the $E_6$ inspired
SUSY models that lead to the approximate unification of the $SU(3)_C$,
$SU(2)_W$ and $U(1)_Y$ gauge couplings at some high energy scale $M_X$.
This requirement implies that in the one--loop approximation the
gauge coupling unification is expected to be almost exact. On the
other hand it is well known that the one--loop gauge coupling
unification in SUSY models remains intact if the MSSM particle content
is supplemented by the complete representations of $SU(5)$ (see for
example \cite{Hempfling:1995rb}). Thus we require that the extra matter
beyond the MSSM fill in complete $SU(5)$ representations.
In the {\bf scenario A} this requirement can be fulfilled if
$\overline{H}_u$ and $\overline{H}_d$ are odd under the $\tilde{Z}^{H}_2$
symmetry while $\overline{L}_4$ is $\tilde{Z}^{H}_2$ even
supermultiplet. Then $\overline{H}_u$ and $\overline{H}_d$ from the
$\overline{27'}_l$ can get combined with the superposition of the
corresponding components from $27_i$ so that the resulting
vectorlike states gain masses of order of $M_X$. The supermultiplets
$L_4$ and $\overline{L}_4$ are also expected to form vectorlike states.
However these states are required to be light enough to ensure that
the lightest exotic quarks decay sufficiently fast\footnote{Note that
the superfields $e^c_4$ and $\overline{e^c}_4$ are not allowed to
survive to low energies because they spoil the one--loop gauge coupling
unification.}. The appropriate mass term $\mu_L L_4\overline{L}_4$ in
the superpotential can be induced within SUGRA models just after the
breakdown of local SUSY if the K\"ahler potential contains an extra
term $(Z_L (L_4\overline{L}_4)+h.c)$\cite{45}.


The presence of the bosonic and fermionic components of $\overline{S}$
at low energies is not constrained by the unification of the $SU(3)_C$,
$SU(2)_W$ and $U(1)_Y$ gauge couplings since $\overline{S}$ is the SM
singlet superfield. If $\overline{S}$ is odd under the $\tilde{Z}^{H}_2$
symmetry then it can get combined with the superposition of the
appropriate components of $27_i$. The corresponding vectorlike states
may be either superheavy ($\sim M_X$) or gain TeV scale masses.
When $\overline{S}$ is $\tilde{Z}^{H}_2$ even superfield then its scalar
component is expected to acquire a non-zero VEV breaking $U(1)_N$
gauge symmetry.

Thus {\bf scenario A} implies that in the simplest case the low energy
matter content of the considered $E_6$ inspired SUSY models involves:
\be
\ba{c}
3\left[(Q_i,\,u^c_i,\,d^c_i,\,L_i,\,e^c_i,\,N_i^c)\right]
+3(D_i,\,\bar{D}_i)+2(S_{\alpha})+2(H^u_{\alpha})+2(H^d_{\alpha})\\[2mm]
+L_4+\overline{L}_4+N_H^c+\overline{N}_H^c+S+H_u+H_d\,,
\ea
\label{12}
\ee
where the right--handed neutrinos $N^c_i$ are expected to gain masses
at some intermediate scale, while the remaining matter survives
down to the EW scale. In Eq.~(\ref{12}) $\alpha=1,2$ and $i=1,2,3$.
Integrating out $N^c_i$, $N^c_H$ and $\overline{N}_H^c$ as well as
neglecting all suppressed non-renormalisable interactions one gets
an explicit expression for the superpotential in the considered case
\be
\ba{c}
W_{A} = \lambda S (H_u H_d) + \lambda_{\alpha\beta} S (H^d_{\alpha} H^u_{\beta})
+ \kappa_{ij} S (D_{i} \overline{D}_{j}) + \tilde{f}_{\alpha\beta} S_{\alpha} (H^d_{\beta} H_u)
+ f_{\alpha\beta} S_{\alpha} (H_d H^u_{\beta}) \\[2mm]
+ g^D_{ij} (Q_i L_4) \overline{D}_j
+ h^E_{i\alpha} e^c_{i} (H^d_{\alpha} L_4) + \mu_L L_4\overline{L}_4
+ W_{MSSM}(\mu=0)\,.
\ea
\label{13}
\ee

A second scenario, that allows the lightest exotic quarks to decay
within a reasonable time and prevents rapid proton decay, is realized
when the set of multiplets $M_{l}$ together with $H_u$, $H_d$, $S$ and
$N^c_H$ contains an extra $d^c_4$ superfield (instead of $L_4$) from $27'_{d}$.
If the $\tilde{Z}^{H}_2$ even supermultiplet $d^c_4$ survives to low
energies then exotic quarks are allowed to have non-zero Yukawa
couplings with pair of quarks which permit their decays. They can also
interact with $d^c_4$ and right-handed neutrinos. However if Majorana
right-handed neutrinos are very heavy ($\sim M_X$) then the interactions
of exotic quarks with leptons are extremely suppressed. As a consequence
in this {\bf scenario B} $\overline{D}_i$ and $D_i$ manifest themselves
in the Yukawa interactions as superfields with baryon number
$\left(\pm\dfrac{2}{3}\right)$.

Although in the {\bf scenario B} the baryon and lepton number violating
operators are expected to be suppressed by inverse powers of the masses
of the right--handed neutrinos they can still lead to the rapid proton
decay. The Yukawa interactions of the $\tilde{Z}^{H}_2$ even superfield
$d^c_4$ with other supermultiplets of ordinary and exotic matter
can be written in the following form
\be
\Delta W_{d^c_4} = h^D_{i k} d^c_4 (H^d_{i} Q_k) +
g^{q}_{ij}\overline{D}_i d^c_4 u^c_j+ g^N_{ij} N_i^c D_j d^c_4\,.
\label{14}
\ee
Integrating out Majorana right-handed neutrinos
one obtains in the leading approximation
\be
\Delta W_{d^c_4} \to h^D_{i k} d^c_4 (H^d_{i} Q_k) +
g^{q}_{ij}\overline{D}_i d^c_4 u^c_j+
\frac{\tilde{\varkappa}_{ij}}{M_N} (L_i H_u) (D_j d^c_4)\,,
\label{15}
\ee
where $M_N$ is an effective seesaw scale which is determined
by the masses and couplings of $N^c_i$ and
$\tilde{\varkappa}_{ij}\sim g^N_{ij}$. In the considered case
the baryon and lepton number violation takes place only when
all three terms in Eqs.~(\ref{14})--(\ref{15}) are present
in the superpotential. If $g^N_{ij}=0$
($\tilde{\varkappa}_{ij}=0$) or $g^{q}_{ij}=0$ the baryon and
lepton number conservation requires exotic quarks to be either
diquarks or leptoquarks respectively. When $h^D_{i k}$
vanish the conservation of the baryon and lepton numbers
implies that the superfields $D_i$, $\overline{D}_i$ and $d^c_4$
have the following $U(1)_L$ and $U(1)_B$ charges
$B_D=-B_{\overline{D}}=-B_{d^c_4}=-1/6$ and
$L_D=-L_{\overline{D}}=L_{d^c_4}=-1/2$. This consideration
indicates that in the case when all three terms are present
in Eqs.~(\ref{14})--(\ref{15}) the $U(1)_L$ and $U(1)_B$
global symmetries can not be preserved. It means that
in the leading approximation the proton decay rate is caused
by all three types of the corresponding Yukawa couplings and
has to go to zero when the Yukawa couplings of at least one
type of Yukawa interactions vanish. In practice, the proton
lifetime is determined by the one--loop box diagram that
leads to the dimension seven operator
\be
\mathcal{L}_{p}\simeq \left(\frac{c_{ijkl}}{M_S^2}\right)
\left(\frac{\langle H_u \rangle}{M_N}\right)
\Biggl[\epsilon_{\alpha\beta\gamma}\overline{u^c}_{\alpha i} d_{\beta j}
\overline{\nu}_{k} d_{\gamma l}\Biggr]\,,
\label{16}
\ee
where $\langle H_u \rangle = v_2/\sqrt{2}$ and
$c_{ijkl}\propto \tilde{\varkappa}\, g^{q}\, (h^D)^2$.
In Eq.~(\ref{16}) Greek indices denote the color degrees of freedom while
$SU(2)$ indices are suppressed. Here we assume that all
particles propagating in the loop have masses of the order of $M_S$.
For $M_N\gtrsim 10^{11}\,\mbox{GeV}$ and
$h^D_{i k} \sim g^{q}_{ij} \sim g^N_{ij}$ the appropriate suppression
of the proton decay rate can be achieved if the corresponding Yukawa
couplings are less than $10^{-5}$.

Once again, the requirement of the approximate unification of the
$SU(3)_C$, $SU(2)_W$ and $U(1)_Y$ gauge couplings constrains the low
energy matter content in the {\bf scenario B}. The concept of gauge
coupling unification implies that the perturbation theory method
provides an adequate description of the RG flow of gauge couplings
up to the GUT scale $M_X$ at least. The requirement of the validity of
perturbation theory up to the scale $M_X$ sets stringent constraint
on the number of extra $SU(2)_W$ and $SU(3)_C$ supermultiplets that
can survive to low energies in addition to three complete fundamental
representations of $E_6$. For example, the applicability of perturbation
theory up to the high energies permits only one extra pair of $SU(3)_C$
triplet superfields to have mass of the order of TeV scale. The same
requirement limits the number of pairs of $SU(2)_W$ doublets to two.

Because in the {\bf scenario B} the $\tilde{Z}^{H}_2$ even supermultiplets
$d^c_4$ and $\overline{d^c}_4$ are expected to form vectorlike states
which have to have TeV scale masses the limit caused by the validity of
perturbation theory up to the scale $M_X$ is saturated. Then in order to
ensure that the extra matter beyond the MSSM fill in complete $SU(5)$
representations $\overline{H}_u$ and $\overline{H}_d$ should survive to
the TeV scale as well. As before we assume that these supermultiplets are
odd under the $\tilde{Z}^{H}_2$ symmetry so that they can get combined
with the superposition of the corresponding components from $27_i$ at
low energies forming vectorlike states. Again the superfield $\overline{S}$
may or may not survive to the TeV scale. It can be either even or
odd under the $\tilde{Z}^{H}_2$ symmetry. If $\overline{S}$ is
$\tilde{Z}^{H}_2$ even, it should survive to low energies
and its scalar component is expected to get a VEV.

Following the above discussion the low energy matter content in the
simplest case of the {\bf scenario B} may be summarized as:
\be
\ba{c}
3\left[(Q_i,\,u^c_i,\,d^c_i,\,L_i,\,e^c_i,\,N_i^c)\right]
+3(D_i,\,\bar{D}_i)+3(H^u_{i})+3(H^d_{i})+2(S_{\alpha})\\[2mm]
+d^c_4+\overline{d^c}_4+N_H^c+\overline{N}_H^c+H_u+\overline{H}_u
+H_d+\overline{H}_d+S\,.
\ea
\label{17}
\ee
All states in Eq.~(\ref{17}) are expected to be considerably lighter than
the GUT scale $M_X$. Assuming that $N^c_i$, $N^c_H$ and $\overline{N}_H^c$
gain intermediate scale masses the renormalizable part of the
TeV scale superpotential associated with the {\bf scenario B}
can be written as
\be
\ba{c}
W_{B} = \lambda S (H_u H_d) + \lambda_{ij} S (H^d_{i} H^u_{j})
+ \kappa_{ij} S (D_{i} \overline{D}_{j}) + \tilde{f}_{\alpha i} S_{\alpha} (H^d_{i} H_u)
+ f_{\alpha i} S_{\alpha} (H_d H^u_{i}) \\[2mm]
+ g^{q}_{ij}\overline{D}_i d^c_4 u^c_j + h^D_{ij} d^c_4 (H^d_{i} Q_j)
+ \mu_d d^c_4\overline{d^c}_4 + \mu^u_{i} H^u_{i} \overline{H}_u
+ \mu^d_{i} H^d_{i} \overline{H}_d + W_{MSSM}(\mu=0)\,.
\ea
\label{18}
\ee

The superpotential (\ref{18}) contains a set of the TeV scale mass
parameters, i.e. $\mu_d$, $\mu^u_{i}$, $\mu^d_{i}$. These are introduced
to avoid massless fermionic states associated with $d^c_4$, $\overline{d^c}_4$,
$\overline{H}_u$ and $\overline{H}_d$ supermultiplets and can be induced
after the breakdown of local SUSY as it has been discussed earlier.
On the other hand the superpotential (\ref{18}) also contains the Yukawa
couplings $g^{q}_{ij}$ and $h^D_{ij}$ which are expected to be small
in order to avoid rapid proton decay. The appropriate suppression of the
corresponding Yukawa couplings and mass parameters $\mu_d$, $\mu^u_{i}$
and $\mu^d_{i}$ can be achieved if the Lagrangian of the $E_6$ inspired
model is invariant under the discrete $Z_k$ symmetry which gets broken
spontaneously at the intermediate scale. As an example one can consider
the model with extra SM singlet superfield $\Phi$ which transforms
under the discrete $Z_k$ symmetry. For concreteness here we assume that
at high energies the Lagrangian of the model is invariant under the
$Z_6$ symmetry transformations
\be
\Phi\to \omega\,\Phi\,,\qquad d^c_4\to \omega^5\, d^c_4\,,\qquad
\overline{d^c}_4\to \omega^3\, \overline{d^c}_4\,,\qquad
\overline{H}_u\to \omega^2\,\overline{H}_u\,,\qquad
\overline{H}_d\to \omega^2\,\overline{H}_d\,,
\label{19}
\ee
where $\omega=e^{i\pi/3}$. Then the part of the superpotential
that depends on the $d^c_4$, $\overline{d^c}_4$, $\overline{H}_u$,
$\overline{H}_d$ and $\Phi$ takes the form
\be
\ba{rcl}
\Delta W_{Z_6} &=& \ds\frac{\Phi}{M_{Pl}}\biggl[\sigma_{ij} d^c_4 (H^d_{i} Q_j) +
\tilde{\sigma}_{ij} \overline{D}_i d^c_4 u^c_j+ \hat{\sigma}_{ij} N_i^c D_j d^c_4\biggr]\\[3mm]
&+&\ds\frac{\Phi^4}{M_{Pl}^{3}}\biggl[\eta_d d^c_4\overline{d^c}_4 +
\eta^u_{i} H^u_{i} \overline{H}_u + \eta^d_{i} H^d_{i} \overline{H}_d \biggr] +
\sigma\ds\frac{\Phi^6}{M_{Pl}^{3}} + ...\,.
\ea
\label{20}
\ee
At the intermediate scale the imposed $Z_6$ symmetry may be broken spontaneously
by the VEV of the superfield $\Phi$
\be
<\Phi>\sim \biggl[\ds\frac{M_S}{M_{Pl}}\biggr]^{1/4} M_{Pl}\simeq 10^{14}\,\mbox{GeV}
\label{21}
\ee
inducing bilinear mass terms in the superpotential and small Yukawa
couplings of the $d^c_4$ supermultiplet to other superfields. The corresponding
Yukawa couplings and mass parameters are given by \footnote{The same mechanism
can be used for the generation of the mass term $\mu_L L_4\overline{L}_4$
in the scenario A.}
\be
\mu_d \sim \mu^u_{i} \sim \mu^d_{i} \sim \dfrac{<\Phi^4>}{M_{Pl}^3}\simeq M_S\,,
\qquad
h^D_{i k} \sim g^{q}_{ij} \sim g^N_{ij} \lesssim \dfrac{<\Phi>}{M_{Pl}}\sim 10^{-4}\,.
\label{22}
\ee

Although {\bf scenarios A} and {\bf B} discussed in this section
allow us to suppress baryon and lepton number violating operators and
non-diagonal flavor transitions they have at least one drawback.
Both scenarios imply that a number of incomplete $E_6$ multiplets
survive below the scale $M_X$. In fact, the number of incomplete $E_6$
multiplets tends to be larger than the number of generations.
Therefore the origin and mechanism resulting in the incomplete
$E_6$ representations requires further justification. The splitting
of GUT multiplets can be naturally achieved in the framework of
orbifold GUTs. In the next section we present
$5D$ and $6D$ orbifold GUT models that can lead to the
{\bf scenarios A} and {\bf B} just below the GUT scale.

\section{5D and 6D orbifold GUT models}

The structure of the $E_6$ inspired SUSY models discussed in the previous
section, its gauge group and field content, points towards an underlying
GUT model based on the $E_6$ or its subgroup. The breaking of these GUT groups
down to the $SU(3)_C\times SU(2)_W\times U(1)_Y\times U(1)_{\psi}\times U(1)_{\chi}$
is in general rather involved and requires often large Higgs representations.
In particular, the splitting of GUT multiplets (like doublet-triplet splitting
within SU(5) GUT) requires either fine--tuning of parameters or additional,
sophisticated mechanisms \cite{Masiero:1982fe}--\cite{Altarelli:2000fu}.

Higher--dimensional theories offer new possibilities
to describe gauge symmetry breaking. A simple and elegant scheme is provided
by orbifold compactifications which have been considered for SUSY GUT models
in five dimensions \cite{Kawamura:2000ev}--\cite{Braam:2010sy} and six
dimensions \cite{5d+6d-susy-ogut}--\cite{Buchmuller:2004eg}. These models
apply ideas that first appeared in string--motivated work \cite{Candelas:1985en}:
the gauge symmetry is broken by identifications imposed on the gauge fields
under the spacetime symmetries of an orbifold. In these models many good
properties of GUT's like gauge coupling unification and charge quantization
are maintained while some unsatisfactory properties of the conventional breaking
mechanism, like doublet-triplet splitting, are avoided. Recently, orbifold
compactifications of the heterotic string have been constructed which can
account for the SM in four dimensions and which have five--dimensional or
six--dimensional GUT structures as intermediate step very similar to orbifold
GUT models \cite{Buchmuller:2005jr}. Hence, orbifold compactifications provide
an attractive starting point for attempts to embed the SM into higher dimensional
string theories.

\subsection{$SU(5)\times U(1)_{\chi}\times U(1)_{\psi}$ model in five dimensions}

The simplest GUT group which unifies the gauge interactions of the SM is $SU(5)$
\cite{Georgi:1974sy}. Therefore we first analyze the higher dimensional SUSY GUT model
based on the $SU(5)\times U(1)_{\chi}\times U(1)_{\psi}$ gauge group which is a rank--6
subgroup of $E_6$. For simplicity we consider a single compact extra dimension $S^1$,
$y (=x_5)$, and assume a fixed radius with size given by the GUT scale ($R\sim 1/M_X$).
The orbifold $S^1/Z_2$ is obtained by dividing the circle $S^1$ with a $Z_2$
transformation which acts on $S^1$ according to $y\to -y$. The components of the
$SU(5)$ supermultiplets that propagate in 5 dimensions transform under the specified
$Z_2$ action as $\Phi(x_{\mu}, -y) = P \Phi(x_{\mu}, y)$, where $P$ acts on each
component of the $SU(5)$ representation $\Phi$, making some components positive
and some components negative, i.e. $P=(+,+,...-, -,...)$. The Lagrangian should be
invariant under the $Z_2$ transformations\footnote{It is worth to point out that the
$Z_2$ invariance of the Lagrangian does not require that $P=\pm I$, where $I$
is the unit matrix. In general, matrix $P$ should satisfy the condition $P^2=I$.}.
The $Z_2$ transformation can be regarded as equivalence relation that allows to reduce
the circle $S^1$ to the interval $y\in [0, \pi R]$.

Here we consider a 5-dimensional space-time factorized into a product of the
ordinary $4D$ Minkowski space time $M^4$ and the orbifold $S^1/(Z_2\times Z'_2)$.
The orbifold $S^1/(Z_2\times Z'_2)$ is obtained by dividing $S^1/Z_2$ with
another $Z_2$ transformation, denoted by $Z'_2$, which acts as $y'\to -y'$,
with $y'\equiv y - \pi R/2$. Each reflection symmetry, $y\to -y$ and $y'\to -y'$,
has its own orbifold parity, $P$ and $P'$, which are defined by
\be
\ba{c}
\Phi(x,y)\to \Phi(x, -y) = P \Phi(x_{\mu}, y)\,,\\[2mm]
\Phi(x,y')\to \Phi(x, -y') = P' \Phi(x_{\mu}, y')
\ea
\label{23}
\ee
where $\Phi(x,y)$ is an $SU(5)$ multiplet field living in the $5D$ bulk,
while $P$ and $P'$ are matrix representations of the two $Z_2$ operator
actions which have eigenvalues $\pm 1$. All interactions must be invariant
under $Z_2\times Z'_2$ symmetry.

Each reflection also introduces special points, $O$ and $O'$, located at
$y=0$ and $y=\pi R/2\equiv \ell$ which are fixed points of the transformations.
The equivalences associated with the two reflection symmetries allow to
work with the theory obtained by truncating to the physically irreducible
interval $y\in [0, \ell]$ with the two $4D$ walls (branes) placed at the
fixed points $y=0$ and $y=\ell$. These are only two inequivalent branes
(the branes at $y=\pi R$ and $y=-\pi R/2$ are identified with those at $y=0$
and $y=\pi R/2$, respectively). Thus physical space reduces to the
interval $[0, \ell]$ with a length of $\pi R/2$.

Denoting the $5D$ bulk field with $(P,\,P')=(\pm 1, \pm 1)$ by $\phi_{\pm\pm}$
one obtains the following Fourier expansions
\cite{Kawamura:2000ev}--\cite{Hebecker:2001wq}:
\begin{eqnarray}
\phi_{++}(x,y) =\sum_{n=0}^{\infty} \frac{1}{\sqrt{2^{\delta_{n,0}}}}
\sqrt{\frac{4}{\pi R}} \phi^{(2n)}_{++}(x)\cos\frac{2ny}{R}\,,
\label{24}
\end{eqnarray}
\begin{eqnarray}
\phi_{+-}(x,y) =\sum_{n=0}^{\infty} \sqrt{\frac{4}{\pi R}}
\phi^{(2n+1)}_{+-}(x)\cos\frac{(2n+1)y}{R}\,,
\label{25}
\end{eqnarray}
\begin{eqnarray}
\phi_{-+}(x,y) =\sum_{n=0}^{\infty} \sqrt{\frac{4}{\pi R}}
\phi^{(2n+1)}_{-+}(x)\sin\frac{(2n+1)y}{R}\,,
\label{26}
\end{eqnarray}
\begin{eqnarray}
\phi_{--}(x,y) =\sum_{n=0}^{\infty} \sqrt{\frac{4}{\pi R}}
\phi^{(2n+2)}_{--}(x)\sin\frac{(2n+2)y}{R}\,,
\label{27}
\end{eqnarray}
where $n$ is a non--negative integer. From the $4D$ perspective
the Fourier component fields $\phi^{(2n)}_{++}(x)$, $\phi^{(2n+1)}_{+-}(x)$,
$\phi^{(2n+1)}_{-+}(x)$ and $\phi^{(2n+2)}_{--}(x)$ acquire masses
$2n/R$, $(2n+1)/R$, $(2n+1)/R$ and $(2n+2)/R$ upon compactification.
Note that only $\phi_{++}(x,y)$ and $\phi_{+-}(x,y)$ can exist on
the $y=0$ brane. The fields $\phi_{++}(x,y)$ and $\phi_{-+}(x,y)$
are non--vanishing on the $y=\pi R/2$ brane, whereas the field
$\phi_{--}(x,y)$ vanishes on both branes.
Only $\phi_{++}(x,y)$ fields have zero--modes. Since full $SU(5)$
$5D$ multiplets $\Phi_i(x,y)$ can, in general, contain components
with even and odd parities, $P$ and $P'$, the matter content of the
massless sector can be smaller than that of the full $5D$ multiplet.
Unless all components of $\Phi(x,y)$ have common parities,
the gauge symmetry reduction occurs upon compactification.

As in the case of the simplest orbifold GUT scenarios
\cite{Kawamura:2000ev}--\cite{Hebecker:2001wq} we start
from the model with the minimal SUSY in $5D$ (with $8$ real supercharges,
corresponding to $N=2$ in 4D). We assume that the vector supermultiplets
associated with the $SU(5)$, $U(1)_{\chi}$ and $U(1)_{\psi}$ interactions
exist in the bulk $M^4\times S^1/(Z_2\times Z'_2)$. The $5D$ gauge
supermultiplets contain vector bosons $A_{M}$ ($M=0,1,2,3,5$) and
gauginos. The $5D$ gaugino is composed of two $4D$ Weyl fermions with
opposite $4D$ chirality, $\lambda$ and $\lambda'$. In addition
$5D$ vector supermultiplets have to involve real scalars $\sigma$
to ensure that the numbers of bosonic and fermionic degrees of freedom
are equal. Thus $5D$ gauge supermultiplets can be decomposed into
vector supermultiplets $V$ with physical components $(A_{\mu}, \lambda)$
and chiral multiplets $\Sigma$ with components
$\biggl((\sigma+i A_5)/\sqrt{2},\lambda'\biggr)$ under $N=1$
supersymmetry in $4D$. These two $N=1$ supermultiplets also form
$N=2$ vector supermultiplet in $4D$.

In addition to the $5D$ vector supermultiplets we assume the presence
of other $SU(5)$ representations as well as $SU(5)$ singlet superfields
that carry non--zero $U(1)_{\chi}$ and $U(1)_{\psi}$ charges in the $5D$
bulk. The corresponding representations also contain $5D$ fermions.
Since each $5D$ fermion state is composed of two $4D$ Weyl fermions,
$\psi$ and $\psi^c$, SUSY implies that each $5D$ supermultiplet includes
two complex scalars $\phi$ and $\phi^c$ as well. The states $\phi$, $\psi$,
$\phi^c$ and $\psi^c$ form one $4D$ $N=2$ hypermultiplet that consists
of two $4D$ $N=1$ chiral multiplets, $\hat{\Phi}\equiv (\phi,\,\psi)$
and $\hat{\Phi}^c \equiv (\phi^c,\,\psi^c)$, transforming as conjugate
representations with each other under the gauge group.

Taking into account that the derivative $\partial_5$ is odd under the
reflection $Z_2$ one can show that the 5D SUSY Lagrangian is invariant
under the following transformations \cite{Kawamura:2000ev}
\be
\ba{rcl}
A_{\mu}(x,y) & \to &A_{\mu}(x, -y) = P A_{\mu}(x,y) P^{-1}\,,\\[1mm]
A_{5}(x,y)   & \to &A_{5}(x, -y) = - P A_{5}(x,y) P^{-1}\,,\\[1mm]
\sigma(x,y)  & \to &\sigma(x, -y) = - P \sigma(x,y) P^{-1}\,,\\[1mm]
\lambda(x,y) & \to &\lambda(x, -y) = P \lambda(x,y) P^{-1}\,,\\[1mm]
\lambda'(x,y)& \to &\lambda'(x, -y) = - P \lambda'(x,y) P^{-1}\,,\\[1mm]
\phi_i(x,y)  & \to &\phi_i(x, -y) = P \phi_i(x, y) \,,\\[1mm]
\psi_i(x,y)  & \to &\psi_i(x, -y) = P \psi_i(x, y) \,,\\[1mm]
\phi_i^c(x,y)& \to &\phi_i^c(x, -y) = -P \phi_i^c(x, y) \,,\\[1mm]
\psi_i^c(x,y)& \to &\psi_i^c(x, -y) = -P \psi_i^c(x, y) \,,
\ea
\label{28}
\ee
where index $i$ represents different $SU(5)$ supermultiplets
that exist in the bulk $M^4\times S^1/(Z_2\times Z'_2)$. In the case
of $SU(5)$ the components of the corresponding $N=2$ vector
supermultiplet in Eq.~(\ref{28}) are given by $V(x, y)=V^{A}(x, y) T^{A}$
and $\Sigma(x, y)=\Sigma^A(x, y) T^A$, where $T^A$ is the set of the
$SU(5)$ generators ($A=1,2,...,24$). The transformations in Eq.~(\ref{28})
are associated with the $Z_2$ reflection symmetry. By replacing $y$ and
$P$ by $y'$ and $P'$ in Eq.~(\ref{28}) one obtains $Z'_2$ transformations.
Note that mass terms for $\phi_i$, $\psi_i$, $\phi^c_i$ and $\psi^c_i$
are allowed by $N=2$ SUSY but these terms are not compatible with the
$P$ and $P'$ parity assignments as follows from Eq.~(\ref{28}).
Therefore the zero--modes of these fields do not receive a bulk
mass contribution.

It is convenient to choose the matrix representation of the parity
assignment $P$, expressed in the fundamental representation of $SU(5)$,
to be $P=\mbox{diag}(+1, +1, +1, +1, +1)$ so that
$V^{A}(x, -y) T^{A}=V^{A}(x, y) T^{A}$. This boundary condition does
not break $SU(5)$ on the $O$ brane at $y=0$. However $4D$ $N=2$ supersymmetry
gets broken by this parity assignment to $4D$ $N=1$ SUSY. This can be seen
explicitly by examining the masses of the Kaluza--Klein (KK) towers of the fields.
Indeed, according to the parity assignment $P$ only $A_{\mu}$, $\lambda$,
$\phi$ and $\psi$ are allowed to have zero--modes whereas other components
of the $N=2$ vector supermultiplet ($\sigma, \lambda'$) and $N=2$ hypermultiplets
$(\phi^c_i,\,\psi^c_i)$ with odd parity P do not possess massless modes.
For the $SU(5)$ gauge symmetry to provide an understanding of the quark and
lepton quantum numbers, the three families of $27_i$ representations of $E_6$
should reside on the $O$ brane where the $SU(5)\times U(1)_{\chi}\times U(1)_{\psi}$
gauge symmetry and $N=1$ SUSY remains intact.
Then at low energies all ordinary quarks and leptons have to fill in complete
$SU(5)$ multiplets.

The $5D$ $SU(5)$ gauge symmetry is reduced to $4D$ $SU(3)_C\times SU(2)_W\times U(1)_Y$
gauge symmetry by choosing $P'=\mbox{diag}(-1, -1, -1, +1, +1)$ acting on
the fundamental representation of $SU(5)$. This boundary condition breaks not
only $SU(5)$ but also $4D$ $N=2$ SUSY to $4D$ $N=1$ SUSY on the $O'$ brane at $y=\ell$.
The parity assignment associated with the $Z'_2$ reflection symmetry leads to
the two types of the $SU(5)$ gauge generators $T^a$ and $T^{\hat{a}}$.
All generators of the SM gauge group satisfy the condition
\be
P'\,T^a\, P' = T^a\,.
\label{29}
\ee
Therefore the corresponding gauge fields $A^{a}_{\mu}(x,y)$ and gauginos $\lambda^a(x,y)$
are even under the reflections $Z_2$ and $Z'_2$ whereas $\sigma^a(x, y)$ and
$\lambda^{'a}(x,y)$ are odd. As a consequence the KK expansions of vector bosons
$A^{a}_{\mu}(x,y)$ and gauginos $\lambda^a(x,y)$ contain massless zero modes
$A^{a(0)}_{\mu}(x)$ and $\lambda^{a(0)}(x)$ corresponding to the unbroken
gauge symmetry of the SM. These zero modes form $4D$ $N=1$ vector supermultiplets.
The KK modes $A^{a(2n)}_{5}(x)$ are swallowed by $A^{a(2n)}_{\mu}(x)$ resulting in
the formation of vector boson state with mass $2n/R$. The KK gaugino modes
$\lambda^{a(2n)}(x)$ and $\lambda^{'a(2n)}(x)$ form $4D$ fermion state
with mass $2n/R$. The KK scalar mode $\sigma^{a(2n)}(x)$ also gains mass $2n/R$.

The other gauge generators $T^{\hat{a}}$ of $SU(5)$ obey the relationship
\be
P'\,T^{\hat{a}}\, P' = - T^{\hat{a}}\,,
\label{30}
\ee
which implies that $A^{\hat{a}}_{\mu}(x,y)$ and $\lambda^{\hat a}(x,y)$ are odd under
the $Z'_2$ symmetry while $\sigma^{\hat{a}}(x, y)$ and $\lambda^{'\hat{a}}(x,y)$ are even.
This means that all components of the $5D$ vector supermultiplet associated with
the broken $SU(5)$ generators $T^{\hat{a}}$ are odd either under the reflection $Z_2$
or $Z'_2$ so that their KK expansions does not possess massless modes.
The $Z_2$ and $Z'_2$ parity assignments for all components of the $5D$ bulk vector
supermultiplets are shown in Table~\ref{tab2}. The KK modes $A^{\hat{a}(2n+1)}_{\mu}(x)$,
$A^{\hat{a}(2n+1)}_{5}(x)$, $\sigma^{\hat{a}(2n+1)}(x)$, $\lambda^{\hat{a}(2n+1)}(x)$ and
$\lambda^{'\hat{a}(2n+1)}(x)$ form vector boson, scalar and fermion states
with masses $(2n+1)/R$.

\begin{table}[ht]
\centering
\begin{tabular}{|l|l|l|l|l|}
\hline
$5D$ fields &$SU(3)_C\times SU(2)_W$ &$Z_2\times Z'_2$& Mass\\
& quantum numbers & parity &\\
\hline
$A^a_{\mu},\,\lambda^a$                                   & $(8,1)+(1,3)+(1,1)$ & $(+,+)$ & $2n/R$\\
\hline
$A^{\hat{a}}_{\mu},\,\lambda^{\hat{a}}$                   & $(3,2)+(\bar{3},2)$ & $(+,-)$ & $(2n+1)/R$\\
\hline
$A^a_{5},\,\sigma^a,\,\lambda^{'a}$                       & $(8,1)+(1,3)+(1,1)$ & $(-,-)$ & $(2n+2)/R$\\
\hline
$A^{\hat{a}}_{5},\,\sigma^{\hat{a}},\,\lambda^{'\hat{a}}$ & $(3,2)+(\bar{3},2)$ & $(-,+)$ & $(2n+1)/R$\\
\hline
$A^{\chi}_{\mu},\,\lambda_{\chi}$                         & $(1,1)$             & $(+,+)$ & $2n/R$\\
\hline
$A^{\chi}_{5},\,\sigma_{\chi},\,\lambda'_{\chi}$          & $(1,1)$             & $(-,-)$ & $(2n+2)/R$\\
\hline
$A^{\psi}_{\mu},\,\lambda_{\psi}$                         & $(1,1)$             & $(+,+)$ & $2n/R$\\
\hline
$A^{\psi}_{5},\,\sigma_{\psi},\,\lambda'_{\psi}$          & $(1,1)$             & $(-,-)$ & $(2n+2)/R$\\
\hline
\end{tabular}
\caption{Parity assignments and KK masses of fields in the $5D$ bulk vector
supermultiplets associated with the $SU(5)$, $U(1)_{\psi}$ and $U(1)_{\chi}$
gauge interactions.}
\label{tab2}
\end{table}

At the fixed point $O'$ the gauge transformations generated by $T^{\hat{a}}$ as well as the
corresponding components of the $5D$ $SU(5)$ vector supermultiplet vanish. At the same time
at an arbitrary point in the bulk all generators of the $SU(5)$ gauge group are operative.
Thus orbifold procedure leads to a local explicit breaking of $SU(5)$ at the fixed point $O'$
due to the non--trivial orbifold quantum numbers of the gauge parameters.

The $Z_2$ and $Z'_2$ parity assignments for the components of the $U(1)_{\psi}$ and $U(1)_{\chi}$
bulk vector supermultiplets are such that the KK expansions of vector bosons $A^{\chi}_{\mu}(x,y)$
and $A^{\psi}_{\mu}(x,y)$ as well as the corresponding gaugino states $\lambda_{\chi}(x,y)$ and
$\lambda_{\psi}(x,y)$ contain massless zero modes $A^{\chi (0)}_{\mu}(x)$, $A^{\psi (0)}_{\mu}(x)$,
$\lambda^{(0)}_{\chi}(x)$ and $\lambda^{(0)}_{\psi}(x)$ associated with the unbroken $U(1)_{\psi}$
and $U(1)_{\chi}$ gauge symmetries (see Table~\ref{tab2}). Other KK modes form vector boson, scalar
and fermion states with masses $(2n+2)/R$ similar to the ones that appear in the case of unbroken
generators $T^{a}$ of $SU(5)$.

As in the simplest orbifold GUT scenarios
\cite{Kawamura:2000ev}--\cite{Altarelli:2001qj} we assume that all incomplete $SU(5)$
supermultiplets which are even under the custodial symmetry (the matter parity $Z_{2}^{M}$
in the case of the MSSM and the $\tilde{Z}^{H}_2$ symmetry in the case of the E$_6$SSM)
originate from the $5D$ bulk supermultiplets. In order to ensure that $H_u$ and
$\overline{H}_u$ as well as $H_d$ and $\overline{H}_d$ survive below the scale $M_X\sim 1/R$
we include two pairs of the $5D$ $SU(5)$ bulk supermultiplets $\Phi_{H_u}+\Phi_{\overline{H}_u}$
and $\Phi_{H_d}+\Phi_{\overline{H}_d}$ that decompose as follows
\be
\Phi_{H_u} = \Phi_{\overline{H}_u}= \ds\left(5,\,-\dfrac{2}{\sqrt{24}},\,\dfrac{2}{\sqrt{40}}\right),\qquad
\Phi_{H_d} = \Phi_{\overline{H}_d}= \ds\left(5,\,\dfrac{2}{\sqrt{24}},\,\dfrac{2}{\sqrt{40}}\right),
\label{31}
\ee
where first, second and third quantities in brackets are the $SU(5)$ representation,
extra $U(1)_{\psi}$ and $U(1)_{\chi}$ charges respectively. The multiplets $\Phi_{H_u}$ and
$\Phi_{\overline{H}_u}$ as well as $\Phi_{H_d}$ and $\Phi_{\overline{H}_d}$ transform
differently under $Z_2$ and $Z'_2$ (see Table~\ref{tab3}). Since $P'$ does not commute with
$SU(5)$ each $5D$ 5--plet is divided into four pieces associated with
different $N=1$ chiral supermultiplets:
\be
5=(3,1,-1/3)+(1,2,1/2)+(\bar{3},1,1/3)+(1,2,-1/2)\,.
\label{311}
\ee
In Eq.~(\ref{311}) first and second quantities in brackets are $SU(3)_C$ and $SU(2)_W$
quantum numbers whereas the third quantity is $U(1)_Y$ charge. As one can see from
Table~\ref{tab3} chiral supermultiplets in Eq.~(\ref{311}) have different $P$ and $P'$
parity assignments that result in different KK mode structures. These parity
assignments are such that the orbifold projection accomplishes doublet--triplet splitting,
in the sense that only one doublet superfield in Eq.~(\ref{311}) has zero mode
while the KK expansions of another doublet, triplet and antitriplet superfields do
not possess massless modes. Thus only $H_u$, $\overline{H}_u$, $H_d$ and $\overline{H}_d$
may survive to low--energies.

The $4D$ superfields $N^c_H$, $\overline{N}_H^c$, $S$ and $\overline{S}$ can stem from
the $5D$ SM singlet superfields that carry $U(1)_{\psi}$ and $U(1)_{\chi}$ charges
\be
\Phi_{S}= \Phi_{\overline{S}}= \ds\left(1,\,\dfrac{4}{\sqrt{24}},\,0\right),\qquad
\Phi_{N^c_H}= \Phi_{\overline{N}_H^c}= \ds\left(1,\,\dfrac{1}{\sqrt{24}},\,-\dfrac{5}{\sqrt{40}}\right)\,.
\label{32}
\ee
According to Eq.~(\ref{28}) only either $\phi_i$ and $\psi_i$ or $\phi^c_i$ and $\psi^c_i$
can have massless modes. Different parity assignments of $\Phi_{S}$ and $\Phi_{\overline{S}}$
as well as $\Phi_{N^c_H}$ and $\Phi_{\overline{N}_H^c}$ allow to project out different
components of these superfields so that only $4D$ superfields $N^c_H$, $\overline{N}_H^c$,
$S$ and $\overline{S}$ may be light (see Table~\ref{tab3}).

\begin{table}[ht]
\centering
\begin{tabular}{|l|l|l|l|l|}
\hline
$5D$ fields &$SU(3)_C\times SU(2)_W\times U(1)_Y\times U(1)_{\psi}\times U(1)_{\chi}$&$Z_2\times Z'_2$&
Mass\\
& quantum numbers & parity &\\
\hline
                                      & $(3,1,-1/3,-2,2)+(\bar{3},1,1/3,2,-2)$& $(+,-)$ & $(2n+1)/R$\\
$\Phi_{H_u}+\Phi_{\overline{H}_u}$    & $(1,2,1/2,-2,2)+(1,2,-1/2,2,-2)$      & $(+,+)$ & $2n/R$\\
                                      & $(\bar{3},1,1/3,2,-2)+(3,1,-1/3,-2,2)$& $(-,+)$ & $(2n+1)/R$\\
                                      & $(1,2,-1/2,2,-2)+(1,2,1/2,-2,2)$      & $(-,-)$ & $(2n+2)/R$\\
\hline
                                      & $(3,1,-1/3,2,2)+(\bar{3},1,1/3,-2,-2)$& $(-,+)$ & $(2n+1)/R$\\
$\Phi_{H_d}+\Phi_{\overline{H}_d}$    & $(1,2,1/2,2,2)+(1,2,-1/2,-2,-2)$      & $(-,-)$ & $(2n+2)/R$\\
                                      & $(\bar{3},1,1/3,-2,-2)+(3,1,-1/3,2,2)$& $(+,-)$ & $(2n+1)/R$\\
                                      & $(1,2,-1/2,-2,-2)+(1,2,1/2,2,2)$      & $(+,+)$ & $2n/R$\\
\hline
$\Phi_{S}+\Phi_{\overline{S}}$        & $(1,1,0,4,0)+(1,1,0,-4,0)$            & $(+,+)$ & $2n/R$\\
                                      & $(1,1,0,-4,0)+(1,1,0,4,0)$            & $(-,-)$ & $(2n+2)/R$\\
\hline
$\Phi_{N^c_H}+\Phi_{\overline{N}_H^c}$& $(1,1,0,1,-5)+(1,1,0,-1,5)$           & $(+,+)$ & $2n/R$\\
                                      & $(1,1,0,-1,5)+(1,1,0,1,-5)$           & $(-,-)$ & $(2n+2)/R$\\
\hline
                                      & $(3,1,-1/3,-1,-3)+(\bar{3},1,1/3,1,3)$& $(-,+)$ & $(2n+1)/R$\\
$\Phi_{L_4}+\Phi_{\overline{L}_4}$    & $(1,2,1/2,-1,-3)+(1,2,-1/2,1,3)$      & $(-,-)$ & $(2n+2)/R$\\
                                      & $(\bar{3},1,1/3,1,3)+(3,1,-1/3,-1,-3)$& $(+,-)$ & $(2n+1)/R$\\
                                      & $(1,2,-1/2,1,3)+(1,2,1/2,-1,-3)$      & $(+,+)$ & $2n/R$\\
\hline
                                      & $(3,1,-1/3,-1,-3)+(\bar{3},1,1/3,1,3)$& $(-,-)$ & $(2n+2)/R$\\
$\Phi_{d^c_4}+\Phi_{\overline{d^c}_4}$& $(1,2,1/2,-1,-3)+(1,2,-1/2,1,3)$      & $(-,+)$ & $(2n+1)/R$\\
                                      & $(\bar{3},1,1/3,1,3)+(3,1,-1/3,-1,-3)$& $(+,+)$ & $2n/R$\\
                                      & $(1,2,-1/2,1,3)+(1,2,1/2,-1,-3)$      & $(+,-)$ & $(2n+1)/R$\\
\hline
\end{tabular}
\caption{Parity assignments and KK masses of fields in the $4D$ chiral supermultiplets resulting from
the $5D$ bulk supermultiplets $\Phi_{H_u}$, $\Phi_{\overline{H}_u}$, $\Phi_{H_d}$, $\Phi_{\overline{H}_d}$
$\Phi_{S}$, $\Phi_{\overline{S}}$, $\Phi_{N^c_H}$, $\Phi_{\overline{N}_H^c}$, $\Phi_{L_4}$, $\Phi_{\overline{L}_4}$
$\Phi_{d^c_4}$ and $\Phi_{\overline{d^c}_4}$.}
\label{tab3}
\end{table}

Finally, the particle spectrum below the scale $M_X$ should be supplemented by either
$L_4$ and $\overline{L}_4$ or $d^c_4$ and $\overline{d^c}_4$ (but not both) to allow
the lightest exotic quarks to decay. These $4D$ $N=1$ chiral superfields can come from
either $\Phi_{L_4}$ and $\Phi_{\overline{L}_4}$ or $\Phi_{d^c_4}$ and $\Phi_{\overline{d^c}_4}$
which are $5D$ $SU(5)$ bulk supermultiplets with quantum numbers
\be
\Phi_{L_4}=\Phi_{\overline{L}_4}=\Phi_{d^c_4}=\Phi_{\overline{d^c}_4}=
\ds\left(5,\,-\dfrac{1}{\sqrt{24}},\,-\dfrac{3}{\sqrt{40}}\right)\,.
\label{33}
\ee
Again parity assignments guarantee that only two $4D$ doublet superfields $L_4$ and
$\overline{L}_4$ from $\Phi_{L_4}$ and $\Phi_{\overline{L}_4}$ can survive to low--energies
whereas the other $SU(2)_W$ doublet, color triplet and antitriplet partners do not have zero modes.
Using the freedom to flip the overall action of the $P'$ parity on the $SU(5)$ multiplets
by a sign relative to $\Phi_{L_4}+\Phi_{\overline{L}_4}$ one can get the KK spectrum
in which only triplet or antitriplet components of $SU(5)$ fundamental supermultiplets
possess massless modes. From Table~\ref{tab3} one can see that this freedom is used in the
case $\Phi_{d^c_4}$ and $\Phi_{\overline{d^c}_4}$ supermultiplets. Due to the different
structure of the KK spectrum only $4D$ triplet or antitriplet superfields, $\overline{d^c}_4$
and $d^c_4$, from $\Phi_{\overline{d^c}_4}$ and $\Phi_{d^c_4}$ are allowed to be light.


Since the three families of $27_i$ representations of $E_6$ are located on the $O$ brane,
where the $SU(5)\times U(1)_{\chi}\times U(1)_{\psi}$ gauge symmetry remains intact,
the Yukawa interactions of quarks and leptons are necessarily $SU(5)$ symmetric. In general
the $SU(5)$ invariance yields the prediction for the first and second generation fermion
mass ratios $m_s/m_d=m_{\mu}/m_{e}$, which is in conflict with the data. In $4D$ GUTs
acceptable mass relations can be obtained using higher dimensional operators and relatively
large representations which acquire VEVs breaking $SU(5)$ or $SO(10)$
\cite{Altarelli:2000fu}, \cite{Ellis:1979fg}. In the case of the simplest 5D orbifold GUTs
there are no $SU(5)$ breaking VEVs. Nevertheless in this case one can introduce two additional
$5D$ bulk supermultiplets with quantum numbers given by Eq.~(\ref{33}) that
transform under $Z_2$ and $Z'_2$ as either $\Phi_{L_4}$ and $\Phi_{\overline{L}_4}$
or $\Phi_{d^c_4}$ and $\Phi_{\overline{d^c}_4}$. Furthermore we assume that
these bulk supermultiplets are odd under $\tilde{Z}^{H}_2$ symmetry which is defined
on the $O$ brane. Hence the zero modes of these extra $5D$ supermultiplets,
which are either weak doublets ($L_5$ and $\overline{L}_5$) or $SU(3)_C$ triplet and
antitriplet ($\overline{d^c}_5$ and $d^c_5$), can mix with quark or lepton superfields
from $27_i$ spoiling the $SU(5)$ relations between the down type quark and charged
lepton masses. Indeed, suppose that zero modes are weak doublet superfields $L_5$ and
$\overline{L}_5$. Then $\overline{L}_5$ can get combined with the superposition of
lepton doublet superfields from $27_i$ so that the resulting vectorlike states gain
masses slightly below $M_X$. The remaining three families of lepton doublets,
that survive to low energies, are superpositions of the corresponding components
from $27_i$ and $L_5$ while three generations of down type quarks stem from
$27_i$ completely. As a consequence the $SU(5)$ relations between the down type quark
and charged lepton masses may get spoiled entirely if the Yukawa couplings of $L_5$
to Higgs doublet $H_d$ are relatively large ($\sim 0.01-0.1$).

\begin{table}[ht]
\centering
\begin{tabular}{|l|l|l|l|l|}
\hline
$5D$ fields &$SU(3)_C\times SU(2)_W\times U(1)_Y\times U(1)_{\psi}\times U(1)_{\chi}$&$Z_2\times Z'_2$&
Mass\\
& quantum numbers & parity &\\
\hline
                                      & $(\bar{3},1,-2/3,1,-1)+(3,1,2/3,-1,1)$& $(+,+)$ & $2n/R$\\
                                      & $(3,2,1/6,1,-1)+(\bar{3},2,-1/6,-1,1)$& $(+,-)$ & $(2n+1)/R$\\
$\Phi_{e^c}+\Phi_{\overline{e^c}}$    & $(1,1,1,1,-1)+(1,1,-1,-1,1)$          & $(+,+)$ & $2n/R$\\
                                      & $(3,1,2/3,-1,1)+(\bar{3},1,-2/3,1,-1)$& $(-,-)$ & $(2n+2)/R$\\
                                      & $(\bar{3},2,-1/6,-1,1)+(3,2,1/6,1,-1)$& $(-,+)$ & $(2n+1)/R$\\
                                      & $(1,1,-1,-1,1)+(1,1,1,1,-1)$          & $(-,-)$ & $(2n+2)/R$\\
\hline
\end{tabular}
\caption{The $(Z_2,Z'_2)$ transformation properties and KK masses of $4D$ chiral supermultiplets
that stem from $SU(5)$ bulk supermultiplets $\Phi_{e^c}$ and $\Phi_{\overline{e^c}}$.}
\label{tab4}
\end{table}

Although the discussed specific realization of the mechanism which allows to obtain
the realistic pattern of fermion masses is the simplest one it is worth to consider
another very attractive possibility. Instead of two additional $5D$ $SU(5)$ fundamental
supermultiplets one can include two larger representations of $SU(5)$ that decompose
under $SU(5)\times U(1)_{\chi}\times U(1)_{\psi}$ as follows:
\be
\Phi_{e^c}=\Phi_{\overline{e^c}}=\ds\left(10,\,\dfrac{1}{\sqrt{24}},\,-\dfrac{1}{\sqrt{40}}\right)\,.
\label{34}
\ee
As before we assume that $\Phi_{e^c}$ and $\Phi_{\overline{e^c}}$ supermultiplets
are odd under $\tilde{Z}^{H}_2$ symmetry. Due to $P$ and $P'$ parity assignments each
$SU(5)$ bulk decuplet is divided into six pieces associated with different $N=1$ chiral
supermultiplets:
\be
10=(\bar{3},1,-2/3)+(3,2,1/6)+(1,1,1)+(3,1,2/3)+(\bar{3},2,-1/6)+(1,1,-1)\,,
\label{35}
\ee
where quantities in brackets are $SU(3)_C$, $SU(2)_W$ and $U(1)_Y$ quantum numbers.
The $Z_2$ and $Z'_2$ parity assignments and mass spectrum for all components of
the $5D$ decuplets are given in Table~\ref{tab4}. These parity assignments
guarantee that only two $4D$ $SU(2)_W$ singlet superfields ($e^c_5$ and $\overline{e^c}_5$)
as well as $4D$ triplet and antitriplet supermultiplets ($u^c_5$ and $\overline{u^c}_5$)
from $\Phi_{e^c}$ and $\Phi_{\overline{e^c}}$ can survive below scale $M_X\sim 1/R$.
Again $\overline{e^c}_5$ and $\overline{u^c}_5$ can get combined with the superposition
of the appropriate components of $27_i$ forming vectorlike states which may have
masses slightly below $M_X$. At the same time $e^c_5$ can mix with the corresponding
components of $27_i$ spoiling the $SU(5)$ relations between the masses of the
down type quarks and charged leptons. It is worth noting that together bulk supermultiplets
(\ref{31}), (\ref{32}), (\ref{33}) and (\ref{34}) form two complete $27$ representations
of $E_6$. This simplifies the structure of bulk supermultiplets making the considered
$5D$ orbifold GUT model more elegant.

For the consistency of the considered model it is crucial that all anomalies get
cancelled. In $5D$ theories no bulk anomalies exist. Nevertheless orbifold
compactification may lead to anomalies at orbifold fixpoints
\cite{ArkaniHamed:2001is}--\cite{Asaka:2002my}.
At the fixed point brane anomaly reduces to the anomaly of the unbroken subgroup of
the original group, i.e. $SU(5)\times U(1)_{\chi}\times U(1)_{\psi}$ on the $O$ brane
and $SU(3)_C\times SU(2)_W\times U(1)_Y\times U(1)_{\chi}\times U(1)_{\psi}$
on the $O'$ brane. It was also shown that the sum of the contributions to the
$4D$ anomalies at the fixpoint equals to the sum of the contributions of
the zero modes localized at the corresponding brane
\cite{ArkaniHamed:2001is}--\cite{Asaka:2002my}.
In this context it is worth to emphasize that the contributions of three families
of $27_i$ representations of $E_6$, which reside on the $O$ brane, to the anomalies
associated with this fixpoint get cancelled automatically. Moreover from
Tables \ref{tab3} and \ref{tab4} one can see that the $P$ and $P'$ parity assignments
are chosen so that the zero modes of the bulk fields localized at the $O$ and
$O'$ branes always form pairs of $N=1$ supermultiplets with opposite quantum numbers.
Such choice of parity assignments guarantees that the contributions of zero modes
of the bulk superfields to the brane anomalies are cancelled as well.

Another important issue for any GUT model is proton stability which was
discussed in the context of $5D$ orbifold GUT models in \cite{Hall:2001pg},
\cite{5d-susy-ogut-proton}--\cite{5d-susy-ogut-proton-unif}. In orbifold GUT models
the dimension five operators, which are caused by an exchange of the color triplet
Higgsino multiplets and give rise to proton decay in ordinary GUTs, do not get induced.
Indeed, in the considered class of models colored Higgsinos acquire mass via the KK mode
expansion of operators $\psi_i \partial_{5} \psi_i^c$ that leads to the Dirac mass terms
of the form $\psi^{(2n+1)}_i \psi_i^{c(2n+1)}$. Since $\psi_i^{c(2n+1)}$ do not couple
directly to the quarks (squarks) and sleptons (leptons) the dimension five operators
are not generated. It turns out that the absence of tree-level amplitudes caused by
the colored Higgsino exchange which result in proton decay is deeply entangled with
the orbifold construction and continuous global $U(1)_R$ symmetry that $5D$ bulk
Lagrangian possesses \cite{Hall:2001pg}. Although the dimension five operators discussed
above do not get induced within orbifold GUT models one must also suppress the brane
interactions $[QQQL]_F$ and $[u^c u^c d^c e^c]_F$ that may be already present on the $O$
brane as non--renormalizable interactions. Such operators can give a substantial contribution
to the proton decay rate if the fundamental scale of gravity is close to the GUT scale.
In the $5D$ orbifold GUT model considered here these dangerous operators are forbidden
by $U(1)_{\chi}$ and $U(1)_{\psi}$ gauge symmetries. Nevertheless proton decay is
mediated by dimension six operators induced by the leptoquark gauge bosons
\cite{Ellis:1979hy}.

Finally, one should mention that in the $5D$ orbifold GUT models gauge couplings of
the $SU(3)_C$, $SU(2)_W$ and $U(1)_Y$ interactions do not exactly unify at the scale
$M_X\sim 1/R$ where $SU(5)$ gauge symmetry gets broken. The reason for this is that
the symmetry of the model on the GUT--breaking brane $O'$ remains limited to the
SM gauge group. In particular, on this brane there are brane--localized $4D$ kinetic
terms for the SM gauge fields with $SU(5)$--violating coefficients $1/g_{O'i}^2$.
The part of the $5D$ effective SUSY Lagrangian that contains kinetic terms for the
SM gauge fields can be written as follows
\be
\mathcal{L}_{eff}=\int d^2\theta
\biggl(\frac{1}{g_5^2}+\frac{1}{2g_{O}^2}\biggl\{\delta(y)+\delta(y-\pi R)\biggr\}\biggr)
\mbox{Tr}\,\mathcal{W}^{\alpha} \mathcal{W}_{\alpha}\qquad\qquad
\label{36}
\ee
$$
\qquad\qquad + \sum_i \int d^2\theta
\frac{1}{2g_{O'i}^2}\biggl\{\delta(y-\frac{\pi}{2}R)+\delta(y+\frac{\pi}{2}R)\biggr\}
\mbox{Tr}\,\mathcal{W}^{\alpha}_i \mathcal{W}^i_{\alpha}+\mbox{h.c.},
$$
where $\mathcal{W}^i_{\alpha}$ $(i=1,2,3)$ are the supersymmetric gauge field
strengths of the $U(1)_Y$, $SU(2)_W$ and $SU(3)_C$ gauge interactions on the $O'$ brane,
and $\mathcal{W}_{\alpha}$ is the $SU(5)$ gauge field strength on the $O$ brane and
in the bulk \footnote{Note the $O'$ brane contribution vanish for $\mathcal{W}_{\alpha}$
associated with the leptoquark gauge bosons which are odd under $Z'_2$.}. Integrating
over $y$ one obtains zero--mode $4D$ SM gauge couplings at the scale $M_X\sim 1/R$
\be
\frac{1}{g^2_i(M_X)}=\frac{2\pi R}{g_5^2}+\frac{1}{g_{O}^2}+\frac{1}{g_{O'i}^2}\,.
\label{37}
\ee
Since $SU(5)$--violating coefficients $1/g_{O'i}^2$ may differ from each other
substantially the $SU(3)_C$, $SU(2)_W$ and $U(1)_Y$ gauge couplings $g^2_i(M_X)$
are not identical. However if in the $5D$ model the bulk and brane gauge couplings
have almost equal strength then after integrating out $y$ the zero--mode gauge couplings
are dominated by the bulk contributions because of the spread of the wavefunction of the
zero--mode gauge bosons. In other words the $SU(5)$--violating brane kinetic terms
are dominated by the bulk contributions when the linear extent of the $5$th dimension
is sufficiently large. Because the bulk contributions to the gauge couplings
(\ref{37}) are necessarily $SU(5)$ symmetric, a $4D$ observer sees an approximate
unification of the SM gauge couplings. The gauge coupling unification within
$5D$ orbifold GUT models was discussed in
\cite{5d-susy-ogut-proton-unif}--\cite{5d-susy-ogut-unif}.

As one can see from Eqs.~(\ref{36})--(\ref{37}) the discrepancy between $g^2_i(M_X)$
is determined by the $SU(5)$--violating gauge kinetic terms on the $O'$ brane.
This discrepancy is small when $g^2_i(M_X)$ are relatively small whereas $g_{O'i}^2$
are large ($g_{O'i}^2 \sim 4\pi $). On the other hand one can expect that the
relative contribution of the $SU(5)$--violating brane corrections to $g^2_i(M_X)$
becomes more sizable in the case when the $SU(3)_C$, $SU(2)_W$ and $U(1)_Y$ gauge
couplings are large at the scale $M_X$.

\subsection{$E_6$ orbifold GUT model in six dimensions}

Having discussed in detail the simplest 5D orbifold GUT model, that may lead at low
energies to the gauge group and field content of the $E_6$ inspired SUSY model specified
in section 2, we next study $E_6$ gauge theory in $6D$ with $N=1$ supersymmetry.
We consider the compactification on a torus $T^2$ with two fixed radii $R_5$ and $R_6$
so that two extra dimensions $y (=x_5)$ and $z (=x_6)$ are compact, i.e.
$y\in (-\pi R_5, \pi R_5]$ and $z\in (-\pi R_6, \pi R_6]$. The physical region
associated with the compactification on the orbifold $T^2/Z_2$ is a pillow with the
four fixed points of the $Z_2$ transformations ($y\to -y$, $z\to -z$) as corners.
The orbifold $T^2/Z_2$ has the following fixpoints $(0,0)$, $(\pi R_5,0)$, $(0,\pi R_6)$
and $(\pi R_5,\pi R_6)$.

Here we discuss $E_6$ gauge theory in $6D$ compactified on the orbifold
$T^2/(Z_2 \times Z^{I}_2 \times Z^{II}_2)$. The $Z_2$, $Z^{I}_2$ and $Z^{II}_2$
symmetries are reflections. The $Z_2$ transformations are defined as before,
i.e. $y\to -y$, $z\to -z$. The $Z^{I}_2$ reflection symmetry transformations
act as $y'\to -y'$, $z\to -z$ with $y' = y - \pi R_5/2$. The reflection $Z^{II}_2$
corresponds to $y\to -y$, $z'\to -z'$ where $z' = z - \pi R_6/2$. The
$Z^{I}_2$ and $Z^{II}_2$ reflection symmetries introduce additional fixed
points. As in the case of $5D$ orbifold GUT models extra reflection symmetries
lead to the reduction of the physical region which is again limited by the
appropriate fixed points. The $Z_2$, $Z^{I}_2$ and $Z^{II}_2$ reflection symmetries
allow to work with the theory obtained by truncating to the physically irreducible
space in which $y\in [0, \pi R_5/2]$ and $z\in [0, \pi R_6/2]$ with the four
$4D$ walls (branes) located at its corners.

Again, we assume that the considered orbifold GUT model contains a set of $E_6$
bulk supermultiplets and another set of $N=1$ superfields which are confined on
one of the branes. The set of superfields that propagate in the bulk
$M^4\times T^2/(Z_2 \times Z^{I}_2 \times Z^{II}_2)$ includes $E_6$ gauge
supermultiplet and a few 27--plets. As before all quark and lepton superfields
are expected to be confined on one brane.

The $E_6$ gauge supermultiplet that exist in the bulk must involve vector bosons
$A_{M}$ ($M=0,1,2,3,5,6$) and $6D$ Weyl fermions (gauginos) which are composed of
two $4D$ Weyl fermions, $\lambda$ and $\lambda'$. These fields can be conveniently
grouped into vector and chiral multiplets of the $N=1$ supersymmetry in $4D$, i.e.
\be
V=(A_{\mu}, \lambda)\,,\qquad\qquad \Sigma=\biggl((A_5+i A_6)/\sqrt{2},\lambda'\biggr)\,,
\label{38}
\ee
where $V$, $A_{M}$, $\lambda$ and $\lambda'$ are matrices in the adjoint
representation of $E_6$. Two $N=1$ supermultiplets (\ref{38}) form $N=2$ vector
supermultiplet in $4D$. The bulk $27'$ supermultiplets also include $6D$ Weyl
fermion states (that involve two $4D$ Weyl fermions, $\psi_i$ and $\psi^c_i$)
together with two complex scalars $\phi_i$ and $\phi^c_i$. The fields
$\psi_i, \psi^c_i, \phi_i$ and $\phi^c_i$ compose $4D$ $N=2$ hypermultiplet
containing two $4D$ $N=1$ chiral superfields: $\hat{\Phi}_i=(\phi_i,\,\psi_i)$
and its conjugate $\hat{\Phi}^c_i = (\phi^c_i,\,\psi^c_i)$ with opposite
quantum numbers. Thus each bulk $27'$ supermultiplet involves two $4D$ $N=1$
supermultiplets $27'$ and $\overline{27'}$.

To ensure the consistency of the construction the Lagrangian of the considered
orbifold GUT model has to be invariant under $Z_2$, $Z^{I}_2$ and $Z^{II}_2$
symmetries.
As in the case of 5D orbifold GUT models each reflection symmetry, $Z_2$,
$Z^{I}_2$ and $Z^{II}_2$, has its own orbifold parity, $P$, $P_{I}$ and $P_{II}$.
The components $\hat{\Phi}$ and $\hat{\Phi}^c$ of the bulk $27'$ supermultiplet
$\Phi$ transform under $Z_2$, $Z^{I}_2$ and $Z^{II}_2$ as follows
\be
\ba{ll}
\hat{\Phi}(x, -y, -z) = P \hat{\Phi}(x, y, z)\,,&\qquad
\hat{\Phi}^c(x, -y, -z) = -P \hat{\Phi}^c (x, y, z)\,,\\
\hat{\Phi}(x, -y', -z) = P_{I} \hat{\Phi}(x, y', z)\,,&\qquad
\hat{\Phi}^c(x, -y', -z) = -P_{I} \hat{\Phi}^c (x, y', z)\,,\\
\hat{\Phi}(x, -y, -z') = P_{II} \hat{\Phi}(x, y, z')\,,&\qquad
\hat{\Phi}^c(x, -y, -z') = -P_{II} \hat{\Phi}^c (x, y, z')\,,
\ea
\label{39}
\ee
where $P$, $P_{I}$ and $P_{II}$ are diagonal matrices with eigenvalues $\pm 1$
that act on each component of the fundamental representation of $E_6$.

It is convenient to specify the matrix representation of the orbifold parity
assignments in terms of the $E_6$ weights $\alpha_{j}$ and gauge shifts,
$\Delta$, $\Delta_{I}$ and $\Delta_{II}$, associated with $Z_2$, $Z^{I}_2$ and
$Z^{II}_2$. The diagonal elements of the matrices $P$, $P_{I}$ and $P_{II}$
can be presented in the following form \cite{Braam:2010sy}
\be
\ba{c}
(P)_{jj}=\sigma\exp\{2\pi i \Delta \alpha_j\}\,,\qquad\qquad
(P_{I})_{jj}=\sigma_{I}\exp\{2\pi i \Delta_{I} \alpha_j\}\,,\\
(P_{II})_{jj}=\sigma_{II}\exp\{2\pi i \Delta_{II} \alpha_j\}\,,
\ea
\label{40}
\ee
where $\sigma$, $\sigma_{I}$ and $\sigma_{II}$ are parities of the
bulk $27'$ supermultiplet, i.e. $\sigma, \sigma_{I}, \sigma_{II} \in \{+,-\}$.
The particle assignments of the weights in the fundamental representation
of $E_6$ are well known (see, for example \cite{Braam:2010sy}). Here we choose
the following gauge shifts
\be
\ba{c}
\Delta=\biggl(\dfrac{1}{2},\,\dfrac{1}{2},\,0,\,\dfrac{1}{2},\,\dfrac{1}{2},\,0\biggr)\,,\qquad
\Delta_{I}=\biggl(\dfrac{1}{2},\,\dfrac{1}{2},\,\dfrac{1}{2},\,\dfrac{1}{2},\,\dfrac{1}{2},\,0\biggr)\,,\\
\Delta_{II}=\biggl(\dfrac{1}{2},\,\dfrac{1}{2},\,0,\,0,\,\dfrac{1}{2},\,0\biggr)\,,
\ea
\label{41}
\ee
that correspond to the orbifold parity assignments shown in Table~\ref{tab5}.

\begin{table}[ht]
\centering
\begin{tabular}{|c|c|c|c|c|c|c|c|c|c|c|c|}
\hline
          & $Q$ & $u^c$ & $e^c$ & $L$ & $d^c$ & $N^c$ & $S$ & $H^u$ & $D$ & $H^d$ & $\overline{D}$ \\
\hline
$Z_2$     & $-$ & $-$   & $-$   & $-$ & $-$   & $-$   & $+$ & $+$   & $+$ & $+$   & $+$ \\
\hline
$Z_2^{I}$ & $-$ & $+$   & $+$   & $-$ & $+$   & $+$   & $+$ & $-$   & $+$ & $-$   & $+$ \\
\hline
$Z_2^{II}$& $-$ & $-$   & $-$   & $+$ & $+$   & $+$   & $-$ & $+$   & $+$ & $-$   & $-$ \\
\hline
\end{tabular}
\caption{Orbifold parity assignments in the bulk $27'$ supermultiplet
with $\sigma=\sigma_{I}=\sigma_{II}=+1$.}
  \label{tab5}
\end{table}

The components $V$ and $\Sigma$ of the $E_6$ gauge supermultiplet
transform under $Z_2$, $Z^{I}_2$ and $Z^{II}_2$ as follows
\be
\ba{ll}
V(x, -y, -z) = P V(x, y, z) P^{-1}\,,&\qquad
\Sigma(x, -y, -z) = -P \Sigma (x, y, z) P^{-1}\,,\\
V(x, -y', -z) = P_{I} V(x, y', z) P^{-1}_{I}\,,&\qquad
\Sigma(x, -y', -z) = -P_{I} \Sigma (x, y', z) P^{-1}_{I}\,,\\
V(x, -y, -z') = P_{II} V(x, y, z') P^{-1}_{II}\,,&\qquad
\Sigma(x, -y, -z') = -P_{II} \Sigma (x, y, z') P^{-1}_{II}\,,
\ea
\label{42}
\ee
where $V(x, y, z)=V^{A}(x, y, z) T^{A}$ and $\Sigma(x, y, z)=\Sigma^A(x, y, z) T^A$
while $T^A$ is the set of generators of the $E_6$ group. The boundary conditions
given by Eqs.~(\ref{39}) and (\ref{42}) break $4D$ $N=2$ supersymmetry because
different components of the $N=2$ supermultiplets transform differently
under $Z_2$, $Z^{I}_2$ and $Z^{II}_2$ reflection symmetries. Moreover since
$P$, $P_{I}$ and $P_{II}$ are not unit matrices the $E_6$ gauge symmetry also
gets broken by these parity assignments.

The $P$ parity assignment indicates that on the $O$ brane at $y=z=0$ associated
with the $Z_2$ reflection symmetry the $E_6$ gauge group is broken down to
$SO(10)\times U(1)_{\psi}$ subgroup. Indeed, according to Table~\ref{tab5} the
$SO(10)$ representations that compose bulk $27'$ supermultiplet ($27\to 16+10+1$)
transform differently under $Z_2$ symmetry, i.e. $16\to -16$, $10\to 10$ and
$1\to 1$. Since the considered symmetry breaking mechanism preserves the rank
of the group the unbroken subgroup at the fixed point $O$ should be
$SO(10)\times U(1)_{\psi}$.

On the brane $O_{I}$ located at the fixed point $y=\pi R_5/2$, $z=0$
and associated with the $Z_2^{I}$ symmetry the $E_6$ gauge symmetry is broken
to $SU(6)\times SU(2)_W$. Again this follows from the $P_{I}$ parity assignment
in the bulk $27'$ supermultiplet. The fundamental representation of $E_6$
decomposes under the $SU(6)\times SU(2)_W$ as follows:
$$
27\to (\overline{15},\, 1) + (6,\,2)\,,
$$
where the first and second quantities in brackets are the $SU(6)$ and
$SU(2)_W$ representations respectively. The multiplet $(6,\,2)$ is formed
by all $SU(2)_W$ doublets which are contained in $27$--plet. From Table~\ref{tab5}
one can see that all $SU(2)_W$ doublet components of the $27'$ supermultiplet
transform differently under the $Z^{I}_2$ reflection symmetry as compared with
other components of this supermultiplet which form $(\overline{15},\, 1)$.

The $E_6$ gauge symmetry is also broken on the brane $O_{II}$ placed at the
fixed point $y=0$, $z=\pi R_6/2$ of the $Z_2^{II}$ symmetry transformations.
The $P_{II}$ parity assignment is such that $16$ components of the $27'$
are odd whereas $10+1$ components are even or viceversa. This implies that
$E_6$ group gets broken down to its $SO(10)'\times U(1)'$ subgroup.
It is worth to emphasize here that $SO(10)$ and $SO(10)'$ are not
the same $SO(10)$ subgroups of $E_6$. In particular, from Table~\ref{tab5}
one can see that the 16-plets of $SO(10)$ and $SO(10)'$ are formed by
different components of the fundamental representation of $E_6$. The
$U(1)_{\psi}$ and $U(1)'$ charge assignments should be also different.

In addition to the three branes mentioned above there is a fourth brane
located at the corner $O_{III}=(\pi R_5/2, \pi R_6/2)$ of the physically
irreducible space. The $Z^{III}_2$ reflection symmetry associated with
this brane is obtained by combining the three symmetries $Z_2$, $Z^{I}_2$
and $Z^{II}_2$ defined above. As a consequence the corresponding parity
assignment $P_{III}=P\, P_{I}\, P_{II}$. Combining three parity assignments
$P$, $P_{I}$ and $P_{II}$ it is easy to see that on the brane $O_{III}$
the unbroken subgroup is $SO(10)''\times \tilde{U}(1)$.

The unbroken gauge group of the effective $4D$ theory is given by the
intersection of the $E_6$ subgroups at the fixed points. Since $P$ and
$P_{II}$ commute with $SU(5)$ the intersection of the $E_6$ subgroups
$SO(10)\times U(1)_{\psi}$ and $SO(10)'\times U(1)'$ is
$SU(5)\times U(1)_{\chi}\times U(1)_{\psi}$. The intersection of
$SU(6)\times SU(2)_W$ and $SU(5)\times U(1)_{\chi}\times U(1)_{\psi}$
gives the SM gauge group with two additional $U(1)$ factors, $U(1)_{\psi}$
and $U(1)_{\chi}$.

The mode expansion for the $6D$ bulk fields $\phi(x,y,z)$ with any
combinations of parities reads \cite{Asaka:2001eh}:
\begin{eqnarray}
\phi_{+++}(x,y,z) = \sum_{n,m}^{\infty} \frac{1}{2^{\delta_{n,0}\delta_{m,0}}\pi \sqrt{R_5 R_6}}
\phi^{(2n,2m)}_{+++}(x)\cos\biggl(\frac{2ny}{R_5}+\frac{2mz}{R_6}\biggr)\,,
\label{43}
\end{eqnarray}
\begin{eqnarray}
\phi_{+-+}(x,y,z) = \sum_{n,m}^{\infty} \frac{1}{\pi \sqrt{R_5 R_6}}
\phi^{(2n+1,2m)}_{+-+}(x)\cos\biggl(\frac{(2n+1)y}{R_5}+\frac{2mz}{R_6}\biggr)\,,
\label{44}
\end{eqnarray}
\begin{eqnarray}
\phi_{++-}(x,y,z) = \sum_{n,m}^{\infty} \frac{1}{\pi \sqrt{R_5 R_6}}
\phi^{(2n,2m+1)}_{++-}(x)\cos\biggl(\frac{2ny}{R_5}+\frac{(2m+1)z}{R_6}\biggr)\,,
\label{45}
\end{eqnarray}
\begin{eqnarray}
\phi_{+--}(x,y,z) = \sum_{n,m}^{\infty} \frac{1}{\pi \sqrt{R_5 R_6}}
\phi^{(2n+1,2m+1)}_{+--}(x)\cos\biggl(\frac{(2n+1)y}{R_5}+\frac{(2m+1)z}{R_6}\biggr)\,,
\label{46}
\end{eqnarray}
\begin{eqnarray}
\phi_{-++}(x,y,z) = \sum_{n,m}^{\infty} \frac{1}{\pi \sqrt{R_5 R_6}}
\phi^{(2n+1,2m+1)}_{-++}(x)\sin\biggl(\frac{(2n+1)y}{R_5}+\frac{(2m+1)z}{R_6}\biggr)\,,
\label{47}
\end{eqnarray}
\begin{eqnarray}
\phi_{--+}(x,y,z) = \sum_{n,m}^{\infty} \frac{1}{\pi \sqrt{R_5 R_6}}
\phi^{(2n,2m+1)}_{--+}(x)\sin\biggl(\frac{2ny}{R_5}+\frac{(2m+1)z}{R_6}\biggr)\,,
\label{48}
\end{eqnarray}
\begin{eqnarray}
\phi_{-+-}(x,y,z) = \sum_{n,m}^{\infty} \frac{1}{\pi \sqrt{R_5 R_6}}
\phi^{(2n+1,2m)}_{--+}(x)\sin\biggl(\frac{(2n+1)y}{R_5}+\frac{2mz}{R_6}\biggr)\,,
\label{49}
\end{eqnarray}
\begin{eqnarray}
\phi_{---}(x,y,z) = \sum_{n,m}^{\infty} \frac{1}{\pi \sqrt{R_5 R_6}}
\phi^{(2n,2m)}_{---}(x)\sin\biggl(\frac{2ny}{R_5}+\frac{2mz}{R_6}\biggr)\,,
\label{50}
\end{eqnarray}
where $n$ and $m$ are non--negative integers. As follows from Eqs.~(\ref{43})--(\ref{50})
each bosonic and fermionic KK mode $\phi^{(k,\ell)}(x)$ is characterized by two integer
numbers and from the $4D$ perspective acquires mass
$\sqrt{\left(\dfrac{k}{R_5}\right)^2+\left(\dfrac{\ell}{R_5}\right)^2}$
upon compactification. Only fields for which all parities are positive
have zero modes, i.e. modes with $k=0$ and $\ell=0$. Such modes form
$4D$ $N=1$ massless vector multiplet of the unbroken
$SU(3)_C\times SU(2)_W\times U(1)_Y\times U(1)_{\psi}\times U(1)_{\chi}$ subgroup
of $E_6$. The corresponding $6D$ bulk fields are non--vanishing on all branes.
All other KK modes of the bulk gauge fields combine to massive states.
In particular, one linear combination of $A^{a(k,\ell)}_{5}(x)$ and
$A^{a(k,\ell)}_{6}(x)$ play the role of the Nambu--Goldstone boson, i.e.
it is swallowed by $A^{a(k,\ell)}_{\mu}(x)$ leading to the formation of
the $4D$ vector boson state with mass
$\sqrt{\left(\dfrac{k}{R_5}\right)^2+\left(\dfrac{\ell}{R_5}\right)^2}$.
Thus the mass generation of the vector boson states is analogous to the
Higgs mechanism. The orthogonal superposition of $A^{a(k,\ell)}_{5}(x)$ and
$A^{a(k,\ell)}_{6}(x)$ compose a scalar state with the same mass.
The KK gaugino modes $\lambda^{a(k,\ell)}(x)$ and $\lambda^{'a(k,\ell)}(x)$
form $4D$ fermion state which is degenerate with the corresponding
vector and scalar states.

As before we assume that all incomplete $E_6$ supermultiplets in the E$_6$SSM,
which are even under the $\tilde{Z}^{H}_2$ symmetry, stem from the $6D$ bulk
superfields. Hereafter we also require that the three complete families of
$27_i$ representations of $E_6$ are located on the $O$ brane where $E_6$ gauge
group is broken down to $SO(10)\times U(1)_{\psi}$. The $4D$ superfields $H_u$
and $\overline{H}_u$ can originate from the bulk $27'$--plets $\Phi^{\prime}_{H_u}$
and $\Phi^{\prime}_{\overline{H}_u}$ that decompose as follows
\be
\Phi^{\prime}_{H_u} = \ds\left(27,\,+,\,-,\,+\right),\qquad
\Phi^{\prime}_{\overline{H}_u}= \ds\left(27,\,-,\,+,\,-\right)\,,
\label{51}
\ee
where first, second, third and fourth quantities in brackets are the $E_6$
representation as well as $\sigma$, $\sigma_{I}$ and $\sigma_{II}$
associated with this representation respectively. The parities of these bulk
$27'$--plets are chosen so that $H_u$ and $\overline{H}_u$ components of the
$N=1$ chiral superfields $\hat{\Phi}^{\prime}_{H_u}$ and
$\hat{\Phi}^{\prime c}_{\overline{H}_u}$ have positive parities with respect
to $Z_2$, $Z^{I}_2$ and $Z^{II}_2$ reflection symmetries (see Table~\ref{tab5}).
In this context it is essential to keep in mind that the invariance of the
$6D$ action requires that the parities of the $4D$ chiral supermultiplets
$\hat{\Phi}^{\prime}_{\overline{H}_u}$ and $\hat{\Phi}^{\prime c}_{\overline{H}_u}$
are opposite. Since the parities of $H_u$ and $\overline{H}_u$ are positive
the KK expansions of the bulk $27'$--plets $\Phi^{\prime}_{H_u}$ and
$\Phi^{\prime}_{\overline{H}_u}$ contain zero modes that form $N=1$ chiral
superfields with quantum numbers of $H_u$ and $\overline{H}_u$.

The $SU(2)_W$ doublet chiral superfields $H_u$ and $\overline{H}_u$ are not
the only supermultiplets from $\Phi^{\prime}_{H_u}$ and $\Phi^{\prime}_{\overline{H}_u}$
that may survive below the scale $M_X\sim 1/R$. Indeed, the parity assignments
in Eq.~(\ref{51}) indicate that the $\overline{u}^{c}$ and $\overline{e}^{c}$
components of the $\hat{\Phi}^{\prime c}_{H_u}$ as well as $u^c$ and $e^c$
components of the $\hat{\Phi}^{\prime}_{\overline{H}_u}$ also have positive
parities with respect to $Z_2$, $Z^{I}_2$ and $Z^{II}_2$ symmetries.
It means that the KK mode structures of the bulk supermultiplets
$\Phi^{\prime}_{H_u}$ and $\Phi^{\prime}_{\overline{H}_u}$ involve zero
modes that correspond to $N=1$ chiral superfields $u^c$, $e^c$, $\overline{u}^{c}$
and $\overline{e}^{c}$. Because the $E_6$ gauge symmetry is broken down to
the $SO(10)\times U(1)_{\psi}$ subgroup on the $O$ brane the zero modes
that come from the same bulk $27'$--plet but belong to different $SO(10)$
representations are not required to have the same transformation properties
under the custodial $\tilde{Z}^{H}_2$ symmetry. This permits us to assume that
$4D$ chiral superfields $u^c$, $e^c$, $\overline{u}^{c}$ and $\overline{e}^{c}$
are odd under the $\tilde{Z}^{H}_2$ symmetry. Then these supermultiplets
are expected to mix with the appropriate components from other $27$--plets
forming vectorlike states with masses slightly below $M_X$ and spoiling the
$SO(10)$ relations between the Yukawa couplings of quarks and leptons to $H_u$
and $H_d$ as it is discussed in the previous subsection.

The $4D$ superfields $H_d$ and $\overline{H}_d$ can originate from another
pair of bulk $27'$--plets
\be
\Phi^{\prime}_{H_d} = \ds\left(27,\,+,\,-,\,-\right),\qquad
\Phi^{\prime}_{\overline{H}_d}= \ds\left(27,\,-,\,+,\,+\right)\,.
\label{52}
\ee
Using the orbifold parity assignments presented in Table~\ref{tab5} it is
easy to check that all parities of $H_d$ and $\overline{H}_d$ components of
the $N=1$ superfields $\hat{\Phi}^{\prime}_{H_d}$ and
$\hat{\Phi}^{\prime c}_{\overline{H}_d}$ are positive so that the
KK expansions of $6D$ superfields  $\Phi^{\prime}_{H_u}$ and
$\Phi^{\prime}_{\overline{H}_u}$ contain the appropriate zero modes.
On the other hand one can also find that $\overline{d}^{c}$ and $\overline{N}^{c}$
components of the $\hat{\Phi}^{\prime c}_{H_d}$ as well as $d^c$ and $N^c$
components of the $\hat{\Phi}^{\prime}_{\overline{H}_d}$ also have positive
parities with respect to $Z_2$, $Z^{I}_2$ and $Z^{II}_2$ reflection
symmetries. Therefore the particle content below the scale $M_X$ includes
bosonic and fermionic states from $N=1$ chiral supermultiplets $d^c$, $N^c$,
$\overline{d}^{c}$ and $\overline{N}^{c}$ as well. The scalar components
of the $4D$ superfields $N^c$ and $\overline{N}^{c}$ can be used to break
$U(1)_{\psi}$ and $U(1)_{\chi}$ down to $U(1)_N\times Z_2^M$.
Because of this the supermultiplets $d^c$, $N^c$, $\overline{d}^{c}$ and
$\overline{N}^{c}$ are expected to be even under the $\tilde{Z}^{H}_2$
symmetry and therefore can not mix with the components of $27_i$ localised
on the $O$ brane. The large VEVs of $N^c$ and $\overline{N}^{c}$ ($\lesssim M_X$)
can give rise to the masses of the bosonic and fermionic components of
$N^c$ and $\overline{N}^{c}$ as well as $d^c$ and $\overline{d}^{c}$ which
are just slightly below $M_X$. 

In order to achieve the appropriate breakdown of the $SU(2)_W\times U(1)_Y\times U(1)_{N}$
gauge symmetry at low energies the particle spectrum below the scale $M_X$ should be
supplemented by the $4D$ chiral superfields $S$ and $\overline{S}$ which are even under
the $\tilde{Z}^{H}_2$ symmetry. The corresponding zero modes can come from the pair of
bulk $27'$--plets
\be
\Phi^{\prime}_{S} = \ds\left(27,\,+,\,+,\,-\right),\qquad
\Phi^{\prime}_{\overline{S}}= \ds\left(27,\,-,\,-,\,+\right)\,.
\label{53}
\ee
The $S$ and $\overline{S}$ components of the $N=1$ superfields $\hat{\Phi}^{\prime}_{S}$
and $\hat{\Phi}^{\prime c}_{\overline{S}}$ have positive orbifold parities.
The $\overline{D}$ component of $\hat{\Phi}^{\prime}_{S}$ and the companion component
from the $\hat{\Phi}^{\prime c}_{\overline{S}}$ superfield have also positive parities
with respect to $Z_2$, $Z^{I}_2$ and $Z^{II}_2$ symmetries. It is convenient to assume
that the states associated with these exotic quark supermultiplets are odd under
the $\tilde{Z}^{H}_2$ symmetry so that the corresponding zero modes can mix with the
appropriate components of the $27$--plets localised on the $O$ brane leading to the
formation of the vectorlike states with masses slightly below $M_X$ and spoiling the
$SO(10)$ relations between the Yukawa couplings of $S$ to the inert Higgs and exotic
quark states. In addition to the components of $\hat{\Phi}^{\prime}_{S}$ and
$\hat{\Phi}^{\prime c}_{\overline{S}}$ mentioned above the orbifold parities of
$\overline{L}$ and $L$ components of $\hat{\Phi}^{\prime c}_{S}$ and
$\hat{\Phi}^{\prime}_{\overline{S}}$ are positive. If the zero modes
associated with these components survive to low energies and the corresponding
$N=1$ supermultiplets are even under the $\tilde{Z}^{H}_2$ symmetry then
the Yukawa couplings of these superfields to $Q_i$ and $\overline{D}_k$
allow the lightest exotic quarks to decay like in the case of Scenario A.

The discussion above indicate that the simplest $6D$ orbifold GUT model
based on the $E_6$ gauge group, which may lead at low energies to the gauge
group and field content of the Scenario A 
specified in section 2, include six bulk $27'$--plets. The consistency of 
this orbifold GUT model requires the absence of anomalies. In the 6D orbifold 
models there are two types of anomalies: $4D$ anomalies \cite{Adler:1969gk} 
intrinsic to the fixed points and bulk anomalies \cite{Asaka:2002my},
\cite{vonGersdorff:2006nt}--\cite{E6-anomaly-2}
which are induced by box diagrams with four gauge currents. For the $6D$
orbifold GUT model to be consistent it is necessary that both the fixed point
and the bulk anomalies must cancel. The contributions of the
anomalous box diagrams with four gauge currents to the $6D$ bulk anomalies
are determined by the trace of four generators of gauge group. This
trace contains nonfactorizable part and part which can be reduced to
the product of traces of two generators. The nonfactorizable part is
associated with the irreducible gauge anomaly while the factorized
contribution corresponds to what is known as reducible anomaly. The reducible
anomalies can be canceled by the Green--Schwarz mechanism \cite{Green:1984sg}.
For the consistency the chiral field content of the $6D$ orbifold model must
lead to the cancellation of the irreducible anomalies which is normally highly
restrictive requirement \cite{Hebecker:2001jb}. However $6D$ orbifold GUT
models based on the $E_6$ gauge group do not have irreducible bulk anomaly
\cite{vonGersdorff:2006nt}--\cite{E6-anomaly-2}. Moreover using the results
obtained in \cite{E6-anomaly-2} one can show that the reducible gauge
anomaly gets cancelled if the field content of the $6D$ orbifold model
involves six bulk $27'$--plets. The $4D$ anomalies at the fixpoints get
also cancelled within the $6D$ orbifold GUT model discussed above. Indeed,
the contributions of $27_i$ supermultiplets, that reside on the $O$ brane,
to the anomalies vanish. Since the orbifold parity assignments are such
that the KK modes of the bulk $27'$ superfields localized at the fixpoints
always form pairs of $N=1$ supermultiplets with opposite quantum numbers
the contributions of the bulk $27'$--plets to the $4D$ fixed point anomalies
are cancelled automatically as well.

Phenomenological viability of the $5D$ and $6D$ orbifold GUT models considered
in this section requires the adequate suppression of the baryon and lepton
number violating operators which can be induced at the scale $M_X$ giving rise
to proton decay. As it was mentioned before the dimension five operators,
that lead to the proton decay, are forbidden by the gauge symmetry in these
models. However baryon and lepton number violating operators, which are mediated
by the exchange of the leptoquark gauge bosons, are enhanced compared to the
usual $4D$ case due to the presence of KK towers of such states. The proton
decay rate in the $6D$ orbifold GUT models based on the $SO(10)$ gauge group
was studied in \cite{Buchmuller:2004eg} where it was
shown that in order to satisfy the experimental
lower limit on the proton lifetime the scale $M_X$ should be larger than
$9\cdot 10^{15}\,\mbox{GeV}$. This restriction on the scale $M_X$ can be
used in the case of the $E_6$ inspired SUSY models as well. However the analysis
of the RG flow of the gauge couplings, which we are going to consider next,
indicates that the value of $g^2_i(M_X)$ in these models are 3-5 times larger
than in the MSSM. This implies that the lower bound on the scale $M_X$
in the considered $E_6$ inspired models is expected to be
$1.5-2\cdot 10^{16}\,\mbox{GeV}$. It is worth noting here again that
the simplest $5D$ and $6D$ orbifold GUT models discussed in this section
do not lead to the exact gauge coupling unification at the scale $M_X$
due to the brane contributions to the gauge couplings. The relative
contribution of these brane corrections is expected to become more
sizable with increasing $g_i^2(M_X)$ as it was discussed before.
The gauge coupling unification in the $6D$ orbifold GUT models was
considered in \cite{6d-susy-ogut-unif}.

\section{RG flow of gauge couplings in the E$_6$SSM}
In this section we discuss the RG flow of the SM gauge couplings $g_i(t)$ above
the EW scale. The running of these couplings between $M_{X}$ and $M_Z$ is described
by a system of renormalisation group equations (RGEs). To simplify our analysis
we assume that $U(1)_{\psi}\times U(1)_{\chi}$ gauge symmetry is broken
down to $U(1)_{N}\times Z_{2}^{M}$ near the scale $M_X$. This permits us to
restrict our consideration to the analysis of the RG flow of four diagonal gauge
couplings $g_3(t)$, $g_2(t)$, $g_1(t)$ and $g'_1(t)$ which correspond to $SU(3)_C$,
$SU(2)_W$, $U(1)_Y$ and $U(1)_N$ gauge interactions respectively. Besides the
evolution of these gauge couplings is affected by a kinetic term mixing. The mixing
effect can be concealed in the interaction between the $U(1)_{N}$ gauge field and
matter fields that can be parametrized in terms of off--diagonal gauge coupling
$g_{11}$ (see \cite{Langacker:1998tc}, \cite{King:2005jy}, \cite{9}). In this
framework the RG equations can be written as follows:
\be
\ds\frac{d G}{d t}=G\times B\,,\qquad\qquad
\frac{d g_2}{dt}=\ds\frac{\beta_2 g_2^3}{(4\pi)^2}\,,\qquad\qquad
\frac{d g_3}{dt}=\frac{\beta_3 g_3^3}{(4\pi)^2}\,,
\label{54}
\ee
where $t=\ln\left(q/M_Z\right)$, $q$ is a renormalisation scale while $B$ and $G$
are $2\times 2$ matrices
\be
G=\left(
\ba{cc}
g_1 & g_{11}\\[2mm]
0   & g'_1
\ea
\right)\,,\qquad
B=\ds\frac{1}{(4\pi)^2}
\left(
\ba{cc}
\beta_1 g_1^2 & 2g_1g'_1\beta_{11}+2g_1g_{11}\beta_1\\[2mm]
0 & g^{'2}_1\beta'_1+2g'_1 g_{11}\beta_{11}+g_{11}^2\beta_1
\ea
\right)\,.
\label{55}
\ee
In Eqs.~(\ref{54})--(\ref{55}) $\beta_i$ and $\beta_{11}$ are beta functions.

Here we examine the RG flow of gauge couplings in the two--loop approximation.
In general the two--loop diagonal $\beta_i$ and off--diagonal $\beta_{11}$ beta
functions may be presented as a sum of one--loop and two--loop contributions.
However the previous analysis performed in \cite{King:2007uj} revealed that an
off--diagonal gauge coupling $g_{11}$ being set to zero at the scale $M_X$ remains
very small at any other scale below $M_X$. Since it seems to be rather natural
to assume that just after the breakdown of the $E_6$ symmetry there is no mixing
in the gauge kinetic part of the Lagrangian between the field strengths associated
with the $U(1)_Y$ and $U(1)_{N}$ gauge interactions $g_{11}$ tends to be substantially
smaller than the diagonal gauge couplings. Because of this we can neglect two--loop
corrections to the off--diagonal beta function $\beta_{11}$. In the case of scenario A
the one--loop off--diagonal beta function is given by $\beta_{11}=-\ds\frac{\sqrt{6}}{5}$
while in the scenario B $\beta_{11}=\ds\frac{3\sqrt{6}}{10}$.

In the scenario A the two--loop diagonal beta functions $\beta_i$ are given by:
\be
\ba{rcl}
\beta_3&=&-9+3N_g+\ds\frac{1}{16\pi^2}\Biggl[g_3^2(-54+34 N_g)+3 N_g\,g_2^2+ N_g\, g_1^2\\[3mm]
&&+N_g\,g_1^{'2}-4h_t^2-4h_b^2-2\Sigma_{\kappa}\Biggr]\,,\\[3mm]
\beta_2&=&-5+3N_g+\ds\frac{1}{16\pi^2}\Biggl[8N_g g_3^2+(-17+21 N_g)g_2^2+ \left(\ds\frac{3}{5}+N_g\right) g_1^2\\[3mm]
&&+\left(\ds\frac{2}{5}+N_g\right) g_1^{'2}
-6 h_t^2-6 h_b^2-2h_{\tau}^2-2\Sigma_{\lambda}\Biggr]\,,\\[3mm]
\beta_1&=&\ds\frac{3}{5}+3N_g+\ds\frac{1}{16\pi^2}\Biggl[8N_g g_3^2+\left(\ds\frac{9}{5}+3N_g\right)g_2^2+
\left(\ds\frac{9}{25}+3 N_g\right) g_1^2\\[3mm]
&&+\left(\ds\frac{6}{25}+N_g\right) g_1^{'2}-\ds\frac{26}{5} h_t^2-\ds\frac{14}{5}h_b^2-
\ds\frac{18}{5}h_{\tau}^2-\ds\frac{6}{5}\Sigma_{\lambda}-\ds\frac{4}{5}\Sigma_{\kappa}\Biggr]\,,\\[3mm]
\beta'_1&=&\ds\frac{2}{5}+3N_g+\ds\frac{5}{4}n+
\ds\frac{1}{16\pi^2}\Biggl[8N_g g_3^2+\left(\ds\frac{6}{5}+3N_g\right)g_2^2+
\left(\ds\frac{6}{25}+ N_g\right) g_1^2\\[3mm]
&&+\left(\ds\frac{4}{25}+3N_g+\ds\frac{25}{8}n \right) g_1^{'2}-
\ds\frac{9}{5} h_t^2-\ds\frac{21}{5}h_b^2-\ds\frac{7}{5}h_{\tau}^2-
\ds\frac{19}{5}\Sigma_{\lambda}-\ds\frac{57}{10}\Sigma_{\kappa}\Biggr]\,,\\[3mm]
\Sigma_{\lambda}&=&\lambda_1^2+\lambda_2^2+\lambda^2\,,\qquad\qquad\qquad
\Sigma_{\kappa}=\kappa_1^2+\kappa_2^2+\kappa_3^2\,,
\ea
\label{56}
\ee
where $N_g$ is a number of generations forming complete $E_6$ fundamental representations that
the considered model involves at low energies, i.e. $N_g=3$, whereas $n$ is a number of $S$ and
$\overline{S}$ supermultiplets from $27'_S$ and $\overline{27'}_S$ that survive to low energies
(i.e. $n=0$ or $1$). Here we assume that the structure of the Yukawa interactions appearing in the
superpotential (\ref{13}) is relatively simple, i.e. $\lambda_{\alpha\beta}=\lambda_{\alpha}\delta_{\alpha\beta}$,
and $\kappa_{ij}=\kappa_i\delta_{ij}$ while $\tilde{f}_{\alpha\beta}$, $f_{\alpha\beta}$, $g^D_{ij}$
and $h^E_{i\alpha}$ are small and can therefore be ignored ($i,\,j=1,\,2,\,3$ and $\alpha,\,\beta=1,\,2$).
We have also neglected all Yukawa couplings that may be associated with the presence of extra $S$
and $\overline{S}$ supermultiplets at low energies. In Eqs.~(\ref{56}) $h_t$, $h_b$ and $h_{\tau}$
are top quark, $b$-quark and $\tau$--lepton Yukawa couplings respectively. In the limit of $n=0$
the RG equations (\ref{56}) coincide with the ones presented in \cite{King:2007uj}.

In the scenario B the two--loop diagonal beta functions $\beta_i$ can be written
in the following form:
$$
\ba{rcl}
\beta_3&=&-8+3N_g+\ds\frac{1}{16\pi^2}\Biggl[g_3^2\left(-\ds\frac{128}{3}+34 N_g\right)+3 N_g\,g_2^2
+ \left(N_g+\ds\frac{4}{15}\right)\,g_1^2\\[3mm]
&& + \left(N_g+\ds\frac{2}{5}\right)\,g_1^{'2}-4h_t^2-4h_b^2-2\Sigma_{\kappa}\Biggr]\,,
\ea
$$
\be
\ba{rcl}
\beta_2&=&-4+3N_g+\ds\frac{1}{16\pi^2}\Biggl[8N_g g_3^2+(-10+21 N_g)g_2^2+ \left(\ds\frac{6}{5}+N_g\right) g_1^2\\[3mm]
&&+\left(\ds\frac{13}{10}+N_g\right) g_1^{'2}
-6 h_t^2-6 h_b^2-2h_{\tau}^2-2\tilde{\Sigma}_{\lambda}\Biggr]\,,\\[3mm]
\beta_1&=&\ds\frac{8}{5}+3N_g+\ds\frac{1}{16\pi^2}\Biggl[\left(8N_g + \ds\frac{32}{15}\right)g_3^2+
\left(\ds\frac{18}{5}+3N_g\right)g_2^2+
\left(\ds\frac{62}{75}+3 N_g\right) g_1^2\\[3mm]
&&+\left(\ds\frac{47}{50}+N_g\right) g_1^{'2}-\ds\frac{26}{5} h_t^2-\ds\frac{14}{5}h_b^2-
\ds\frac{18}{5}h_{\tau}^2-\ds\frac{6}{5}\tilde{\Sigma}_{\lambda}-\ds\frac{4}{5}\Sigma_{\kappa}\Biggr]\,,\\[3mm]
\beta'_1&=&\ds\frac{19}{10}+3N_g+\ds\frac{5}{4}n+
\ds\frac{1}{16\pi^2}\Biggl[\left(8N_g+\ds\frac{16}{5}\right)g_3^2+\left(\ds\frac{39}{10}+3N_g\right)g_2^2\\[4mm]
&&+\left(\ds\frac{47}{50}+ N_g\right) g_1^2
+\left(\ds\frac{121}{100}+3N_g+\ds\frac{25}{8}n \right) g_1^{'2}\\[3mm]
&&-\ds\frac{9}{5} h_t^2-\ds\frac{21}{5}h_b^2-\ds\frac{7}{5}h_{\tau}^2-
\ds\frac{19}{5}\tilde{\Sigma}_{\lambda}-\ds\frac{57}{10}\Sigma_{\kappa}\Biggr]\,,
\ea
\label{57}
\ee
where $\tilde{\Sigma}_{\lambda}=\lambda_1^2+\lambda_2^2+\lambda_3^2+\lambda^2$.
As before we assume relatively simple structure of the Yukawa interactions in the
superpotential (\ref{18}), i.e. $\lambda_{ij}=\lambda_{i}\delta_{ij}$,
$\kappa_{ij}=\kappa_i\delta_{ij}$, and ignore $\tilde{f}_{\alpha i}$, $f_{\alpha i}$,
$g^{q}_{ij}$, $h^D_{ij}$ as well as all Yukawa couplings of extra $S$ and
$\overline{S}$ supermultiplets.

As one can see from Eqs.~(\ref{56})--(\ref{57}) $N_g=3$ is the critical value for
the one--loop beta function of the strong interactions in the case of scenario A.
Indeed, in the one--loop approximation the $SU(3)_C$ gauge coupling is equal to zero
in this case. In the scenario B the one--loop contribution to $\beta_3$ remains rather
small ($b_3=1$). Because of this any reliable analysis of the RG flow of
gauge couplings requires the inclusion of two--loop corrections to the diagonal
beta functions.

One can obtain an approximate solution of the two--loop RGEs presented above
(see \cite{Chankowski:1995dm}). At high energies this solution for the SM gauge
couplings can be written as
\be
\ds\frac{1}{\alpha_i(t)}=\frac{1}{\alpha_i(M_Z)}-\ds\frac{b_i}{2\pi} t-\frac{C_i}{12\pi}-\Theta_i(t)
+\ds\frac{b_i-b_i^{SM}}{2\pi}\ln\frac{T_i}{M_Z}\,,
\label{58}
\ee
where $\alpha_i(t)=\ds\frac{g_i^2(t)}{(4\pi)}$, $b_i$ and $b_i^{SM}$ are the coefficients
of the one--loop beta functions in the E$_6$SSM and SM respectively, the third term in the
right--hand side of Eq.~(\ref{58}) is the $\overline{MS}\to\overline{DR}$ conversion factor
with $C_1=0$, $C_2=2$, $C_3=3$ \cite{MS-DR}, while
\be
\Theta_i(t)=\ds\frac{1}{2\pi}\int_0^t (\beta_i-b_i)d\tau\,,\qquad\qquad
T_i=\prod_{k=1}^N\biggl(m_k\biggr)^{\ds\frac{\Delta b^k_i}{b_i-b_i^{SM}}}\,.
\label{59}
\ee
In Eq.~(\ref{59}) $m_k$ and $\Delta b_i^k $ are masses and one--loop contributions to
the beta functions due to new particles appearing in the E$_6$SSM. For the calculation of
$\Theta_i(t)$ the solutions of the one--loop RGEs are normally used.
In Eqs.~(\ref{58})--(\ref{59}) only leading one--loop threshold effects are
taken into account.

Using the approximate solution of the two--loop RGEs in Eqs.~(\ref{58})--(\ref{59})
one can establish the relationships between the values of the gauge couplings at
low energies and GUT scale. Then by using the expressions describing the RG flow
of $\alpha_1(t)$ and $\alpha_2(t)$ it is rather easy to find the scale $M_X$ where
$\alpha_1(M_X)=\alpha_2(M_X)=\alpha_0$ and the value of the overall gauge coupling
$\alpha_0$ at this scale. Substituting $M_X$ and $\alpha_0$ into the solution of
the RGE for the strong gauge coupling one finds the value of $\alpha_3(M_Z)$ for
which exact gauge coupling unification occurs (see \cite{Carena:1993ag}):
\be
\ba{c}
\ds\frac{1}{\alpha_3(M_Z)}=\frac{1}{b_1-b_2}\biggl[\ds\frac{b_1-b_3}{\alpha_2(M_Z)}-
\ds\frac{b_2-b_3}{\alpha_1(M_Z)}\biggr]-\frac{1}{28\pi}+\Theta_s+\frac{19}{28\pi}\ln\frac{T_{S}}{M_Z}\,,\\[4mm]
\Theta_s=\biggl(\ds\frac{b_2-b_3}{b_1-b_2}\Theta_1-\frac{b_1-b_3}{b_1-b_2}\Theta_2+\Theta_3\biggr)\,,
\qquad \Theta_i=\Theta_i(M_X)\,.
\ea
\label{60}
\ee
The combined threshold scale $T_{S}$, that appears in Eq.~(\ref{60}), can be expressed
in terms of the effective threshold scales $T_1$, $T_2$ and $T_3$.
The expression for $T_{S}$ is model--dependent. In the scenario A $T_{S}$ is given by
$$
\begin{array}{rcl}
T_{S}&=&\ds\frac{T_2^{172/19}}{T_1^{55/19} T_3^{98/19}}\,,\\[0mm]
T_1&=&\tilde{M}_1^{5/11} \mu_{L}^{4/55} m_{L}^{2/55}
\Biggl(\prod_{i=1,2,3}m_{\tilde{D}_i}^{4/165}\mu_{D_i}^{8/165}\Biggr)
\Biggl(\prod_{\alpha=1,2}m_{H_{\alpha}}^{2/55}\mu_{\tilde{H}_{\alpha}}^{4/55}\Biggr)\,,\\[0mm]
\end{array}
$$
\vspace{-2mm}
\begin{eqnarray}
T_2&=&\tilde{M}_2^{25/43} \mu_{L}^{4/43} m_{L}^{2/43}
\Biggl(\prod_{\alpha=1,2} m_{H_{\alpha}}^{2/43}\mu_{\tilde{H}_{\alpha}}^{4/43}\Biggr)\,,\qquad\qquad\qquad\qquad\qquad\quad\nonumber\\[0mm]
T_3&=&\tilde{M}_3^{4/7}\Biggl(\prod_{i=1,2,3}m_{\tilde{D}_i}^{1/21}\mu_{D_i}^{2/21}\Biggr)\,,
\label{61}
\end{eqnarray}
where $\mu_{D_i}$ and $m_{\tilde{D}_i}$ are the masses of exotic quarks and their superpartners,
$m_{H_{\alpha}}$ and $\mu_{\tilde{H}_{\alpha}}$ are the masses of Inert Higgs and Inert Higgsino
fields, $m_{L}$ and $\mu_{L}$ are the masses of the scalar and fermion components of $L_4$ and
$\overline{L}_4$ while $\tilde{M}_1$, $\tilde{M}_2$ and $\tilde{M}_3$ are the effective threshold
scales in the MSSM
\begin{eqnarray}
\tilde{M}_1&=& \mu^{4/25} m_{A}^{1/25}
\Biggl(\prod_{i=1,2,3} m_{\tilde{Q}_i}^{1/75} m_{\tilde{d}_i}^{2/75} m_{\tilde{u}_i}^{8/75}
m_{\tilde{L}_i}^{1/25} m_{\tilde{e}_i}^{2/25}\Biggr)\,,\nonumber \\[0mm]
\tilde{M}_2&=& M_{\tilde{W}}^{8/25} \mu^{4/25} m_A^{1/25}
\Biggl(\prod_{i=1,2,3} m_{\tilde{Q}_i}^{3/25} m_{\tilde{L}_i}^{1/25}\Biggr)\,,\nonumber\\[0mm]
\tilde{M}_3&=& M_{\tilde{g}}^{1/2}
\Biggl(\prod_{i=1,2,3} m_{\tilde{Q}_i}^{1/12} m_{\tilde{u}_i}^{1/24} m_{\tilde{d}_i}^{1/24}\Biggr)\,.
\label{62}
\end{eqnarray}
In Eqs.~(\ref{62}) $M_{\tilde{g}}$ and $M_{\tilde{W}}$ are masses of gluinos and winos
(superpartners of $SU(2)_W$ gauge bosons), $\mu$ and $m_A$ are effective $\mu$--term and
masses of heavy Higgs states respectively; $m_{\tilde{u}_i}$, $m_{\tilde{d}_i}$ and
$m_{\tilde{Q}_i}$ are the masses of the right--handed and left--handed squarks and
$m_{\tilde{L}_i}$ and $m_{\tilde{e}_i}$ are the masses of the left--handed and right--handed
sleptons.

In the case of scenario B we find
\begin{eqnarray}
\tilde{T}_{S}&=&\ds\frac{\tilde{T}_2^{196/19}}{\tilde{T}_1^{65/19} \tilde{T}_3^{112/19}}\,,\nonumber \\[0mm]
\tilde{T}_1&=&\tilde{M}_1^{5/13} \mu_{d_4}^{8/195} m_{d_4}^{4/195}
\mu_{H_u}^{4/65} m_{H_u}^{2/65} \mu_{H_d}^{4/65} m_{H_d}^{2/65}
\Biggl(\prod_{i=1,2,3}m_{\tilde{D}_i}^{4/195}\mu_{D_i}^{8/195}\Biggr)
\Biggl(\prod_{\alpha=1,2}m_{H_{\alpha}}^{2/65}\mu_{\tilde{H}_{\alpha}}^{4/65}\Biggr)\,,\nonumber\\[0mm]
\tilde{T}_2&=&\tilde{M}_2^{25/49} \mu_{H_u}^{4/49} m_{H_u}^{2/49} \mu_{H_d}^{4/49} m_{H_d}^{2/49}
\Biggl(\prod_{\alpha=1,2} m_{H_{\alpha}}^{2/49}\mu_{\tilde{H}_{\alpha}}^{4/49}\Biggr)\,,\nonumber\\[0mm]
\tilde{T}_3&=&\tilde{M}_3^{1/2} \mu_{d_4}^{1/12} m_{d_4}^{1/24}
\Biggl(\prod_{i=1,2,3} m_{\tilde{D}_i}^{1/24} \mu_{D_i}^{1/12}\Biggr)\,,
\label{63}
\end{eqnarray}
where $\mu_{d_4}$, $\mu_{H_u}$ and $\mu_{H_d}$ are the masses of the fermionic components
of $d^c_4$ and $\overline{d^c}_4$, $H^u_{i}$ and $\overline{H}_u$ as well as $H^d_{i}$ and $\overline{H}_d$,
that form vector-like states at low energies, whereas $m_{d_4}$, $m_{H_u}$ and $m_{H_d}$ are the
masses of the scalar components of the corresponding supermultiplets.

\begin{figure}
\bc
\hspace*{-11cm}{$\alpha_i(t)$}\\[1mm]
\includegraphics[height=70mm,keepaspectratio=true]{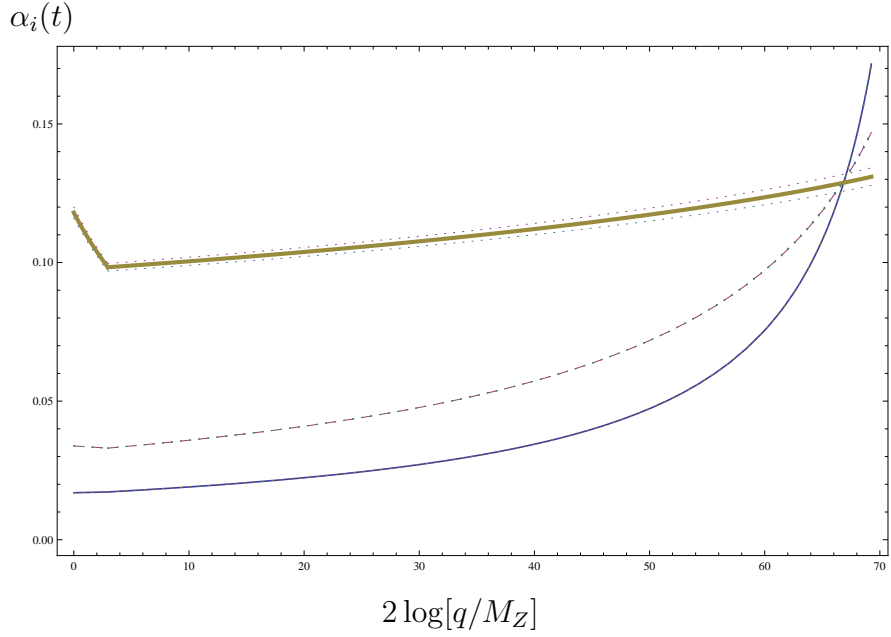}\\
\hspace*{0cm}{$2\log[q/M_Z]$}\\[1mm]
\hspace*{0cm}{\bf (a)}\\[3mm]
\hspace*{-11cm}{$\alpha_i(t)$}\\[1mm]
\includegraphics[height=70mm,keepaspectratio=true]{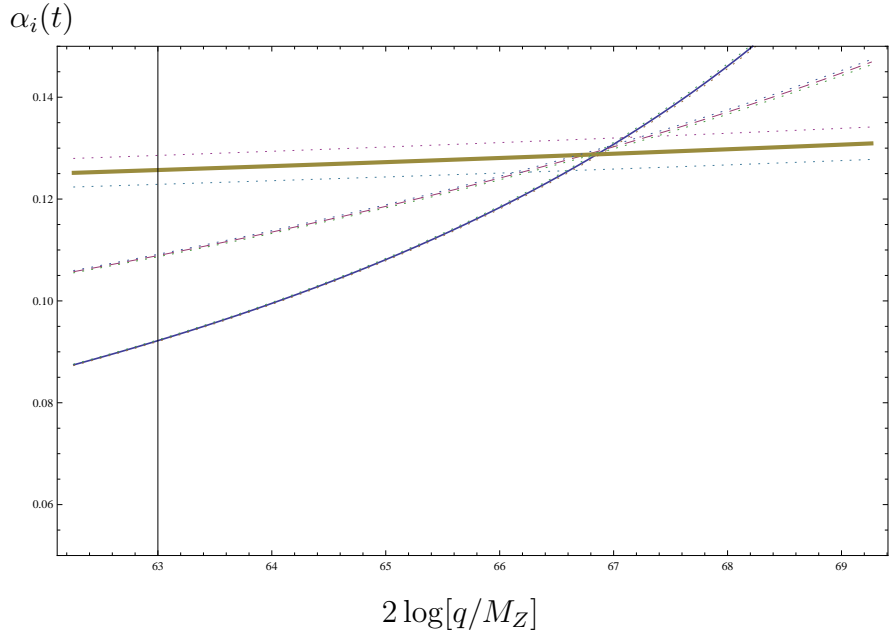}\\
\hspace*{0cm}{$2\log[q/M_Z]$}\\[1mm]
\hspace*{0cm}{\bf (b) }\\
\vspace{-3mm}
\caption{Two--loop RG flow of gauge couplings in the Scenario A:
{\it (a)} RG flow of $SU(3)_C$, $SU(2)_W$ and $U(1)_Y$ couplings
from $M_Z$ to $M_X$ for $T_S=400\,\mbox{GeV}$ and $n_S=1$;
{\it (b)} running of SM gauge couplings in the vicinity of $M_X$
for $T_S=400\,\mbox{GeV}$ and $n_S=1$. Thick, dashed and solid lines
correspond to the running of $SU(3)_C$, $SU(2)_W$ and $U(1)_Y$ couplings
respectively. We used  $\tan\beta=10$, $\alpha_s(M_Z)=0.118$,
$\alpha(M_Z)=1/127.9$, $\sin^2\theta_W=0.231$ and $\kappa_1(T_S)=\kappa_2(T_S)
=\kappa_3(T_S)=\lambda_1(T_S)=\lambda_2(T_S)=\lambda_3(T_S)=g^{'}_1(T_S)$.
The dotted lines represent the uncertainty in $\alpha_i(t)$ caused by
the variation of the strong gauge coupling from 0.116 to 0.120 at the
EW scale.}
\ec
\label{essmfig1}
\end{figure}

\begin{figure}
\bc
\hspace*{-11cm}{$\alpha_i(t)$}\\[1mm]
\includegraphics[height=70mm,keepaspectratio=true]{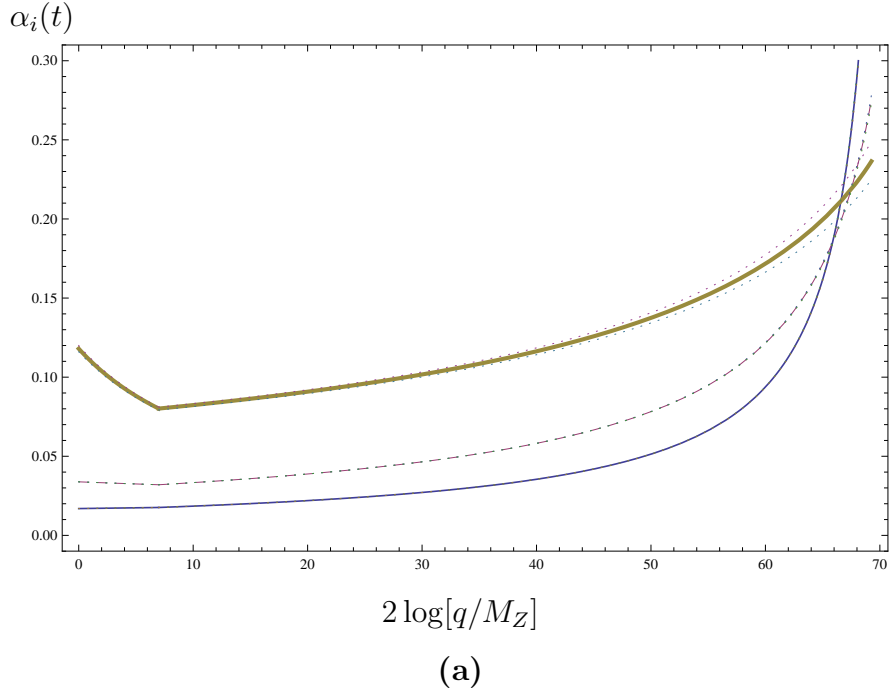}\\
\hspace*{0cm}{$2\log[q/M_Z]$}\\[1mm]
\hspace*{0cm}{\bf (a)}\\[3mm]
\hspace*{-11cm}{$\alpha_i(t)$}\\[1mm]
\includegraphics[height=70mm,keepaspectratio=true]{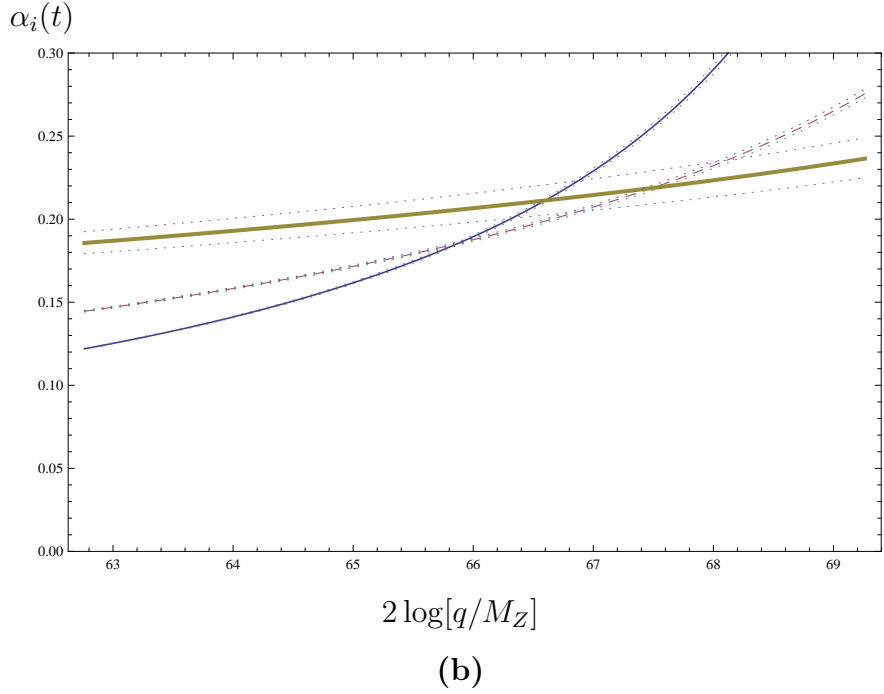}\\
\hspace*{0cm}{$2\log[q/M_Z]$}\\[1mm]
\hspace*{0cm}{\bf (b) }\\
\vspace{-3mm}
\caption{Two--loop RG flow of gauge couplings in the Scenario B:
{\it (a)} evolution of $SU(3)_C$, $SU(2)_W$ and $U(1)_Y$ couplings
from the EW scale to the GUT scale for $\tilde{T}_S=3\,\mbox{TeV}$ and $n_S=0$;
{\it (b)} running of SM gauge couplings near the scale $M_X$ for
$\tilde{T}_S=3\,\mbox{TeV}$ and $n_S=0$. The parameters and notations
are the same as in Fig.~\ref{essmfig1}.
}
\ec
\label{essmfig2}
\end{figure}

In general the effective threshold scales derived above can be quite different.
Since our purpose is to establish the range of the values of $T_S$ and $\tilde{T}_{S}$
that leads to the unification of gauge couplings we shall set these effective
threshold scales equal to each other. Then from Eqs.~(\ref{61}) and (\ref{63}) it
follows that $T_1=T_2=T_3=T_S$ and $\tilde{T}_1=\tilde{T}_2=\tilde{T}_3=\tilde{T}_S$.
The results of our numerical studies of the two--loop RG flow of gauge couplings in the
case of scenarios A and B are summarized in Figs.~\ref{essmfig1} and \ref{essmfig2}
respectively. We use the two--loop SM beta functions to describe the running of gauge
couplings between $M_Z$ and $T_1=T_2=T_3=T_S$ (or $\tilde{T}_1=\tilde{T}_2=\tilde{T}_3=\tilde{T}_S$),
then we apply the two--loop RGEs of the E$_6$SSM to compute the flow of $g_i(t)$ from
$T_S$ (or $\tilde{T}_S$) to $M_X$ which is equal to $3\cdot 10^{16}\,\mbox{GeV}$
in the case of the E$_6$SSM. The low energy values of $g'_1$ and $g_{11}$ are chosen
so that all four diagonal gauge couplings are approximately equal near the GUT scale
and $g_{11}=0$ at this scale. For the calculation of the evolution of Yukawa couplings
a set of one--loop RGEs is used. The corresponding one--loop RG equations are specified
in \cite{King:2005jy}.

In Fig.~\ref{essmfig1} we fix the effective threshold scale to be equal to $400\,\mbox{GeV}$.
In Fig.~1a we plot the running of the gauge couplings from $M_Z$ to $M_X$ assuming
that the low energy matter content involves three 27-plets of $E_6$ as well as $L_4$,
$\overline{L}_4$, $S$ and $\overline{S}$ supermultiplets. Fig.~1b shows a blow--up
of the crucial region in the vicinity of the GUT scale. Dotted lines show the interval
of variations of gauge couplings caused by $1\,\sigma$ deviations of $\alpha_3(M_Z)$
around its average value, i.e.  $\alpha_3(M_Z)\simeq 0.118\pm 0.002$. The results
of the numerical analysis presented in Fig.~\ref{essmfig1} demonstrate that in the
scenario A almost exact unification of the SM gauge couplings can be achieved for
$\alpha_3(M_Z)=0.118$ and $\tilde{T}_S=400\,\mbox{GeV}$. With increasing (decreasing)
the effective threshold scale the value of $\alpha_3(M_Z)$, at which exact gauge coupling
unification takes place, becomes lower (greater). Thus in this case the gauge coupling
unification can be achieved for any phenomenologically reasonable value of $\alpha_3(M_Z)$,
consistent with the central measured low energy value, unlike in the MSSM
where it is rather problematic to get the exact unification of gauge couplings
\cite{Chankowski:1995dm}, \cite{gc-unif-mssm-2}--\cite{gc-unif-mssm-1}.
Indeed, it is well known that in order to achieve gauge coupling
unification in the MSSM with $\alpha_s(M_Z)\simeq 0.118$, the combined threshold
scale, which is given by \cite{Chankowski:1995dm}, \cite{Carena:1993ag},
\cite{gc-unif-mssm-1}--\cite{Langacker:1992rq}
\be
\tilde{M}_{S}=\ds\frac{\tilde{M}_2^{100/19}}{\tilde{M}_1^{25/19} \tilde{M}_3^{56/19}}
\simeq \mu/6\,,
\label{64}
\ee
must be around $\tilde{M}_S\approx 1\,\mbox{TeV}$. However the correct pattern of EW symmetry
breaking requires $\mu$ to lie within the $1-2\,\mbox{TeV}$ range which implies
$\tilde{M}_S<200-300\,\mbox{GeV}$, so that, ignoring the effects of high energy threshold
corrections, the exact gauge coupling unification in the MSSM requires significantly
higher values of $\alpha_3(M_Z)$, well above the experimentally measured central value
\cite{Chankowski:1995dm}, \cite{Carena:1993ag}, \cite{gc-unif-mssm-1}--\cite{gc-unif-mssm-3}.
It was argued that it is possible to get the unification of gauge couplings
in the minimal SUSY model for $\alpha_3(M_Z)\simeq 0.123$ \cite{gc-unif-mssm-4}.

On the other hand in the case of scenario A the combined threshold scale $T_{S}$
can be substantially larger than in the MSSM. This can be seen directly from the
explicit expression for $T_S$. Combining Eqs.~(\ref{61}) we find
\begin{eqnarray}
T_{S}&=&\tilde{M}_{S}\cdot
\Biggl(\ds\frac{\mu_{L}^{12/19} m_{L}^{6/19}}{\mu_{D_3}^{12/19} m_{\tilde{D}_3}^{6/19}}\Biggr)
\Biggl(\prod_{\alpha=1,2}
\ds\frac{m_{H_{\alpha}}^{6/19}\mu_{\tilde{H}_{\alpha}}^{12/19}}{m_{\tilde{D}_{\alpha}}^{6/19}\mu_{D_{\alpha}}^{12/19}}
\Biggr)\,.
\label{65}
\end{eqnarray}
From Eq.~(\ref{65}) it is obvious that $T_{S}$ is determined by the masses of the scalar
and fermion components of $L_4$ and $\overline{L}_4$. The term $\mu_L L_4\overline{L}_4$
in the superpotential (\ref{13}) is not involved in the process of EW symmetry breaking.
As a consequence the parameter $\mu_L$ remains arbitrary\footnote{When $\mu_L$ is considerably
larger than the SUSY breaking scale $m_L\simeq \mu_L$.}. In particular, since the corresponding
mass term is not suppressed by the $E_6$ symmetry the components of the doublet superfields
$L_4$ and $\overline{L}_4$ may be much heavier than the masses of all exotic states resulting in
the large combined threshold scale $T_{S}$ that lies in a few hundred GeV range even
when scale $\tilde{M}_S$ is relatively low. The large range of variation of $T_{S}$ allows
to achieve the exact unification of gauge couplings in the scenario A for any value of
$\alpha_3(M_Z)$ which is in agreement with current data.

It is worth noting here that, in principle, one could naively expect that large two--loop
corrections to the diagonal beta functions would spoil the unification of the SM gauge couplings
entirely in the considered case. Indeed, in the scenario A these corrections affect the RG
flow of gauge couplings much more strongly than in the case of the MSSM because at any intermediate
scale the values of the gauge couplings in the E$_6$SSM are substantially larger as compared to
the ones in the MSSM. Nevertheless the results of our analysis discussed above are not as surprising
as they may first appear. The analysis of the RG flow of the SM gauge couplings performed in
\cite{King:2007uj} revealed that the two--loop corrections to $\alpha_i(M_X)$ are a few times
bigger in the E$_6$SSM than in the MSSM. At the same time due to the remarkable cancellation
of different two--loop corrections the absolute value of $\Theta_s$ is more than three times
smaller in the E$_6$SSM as compared with the MSSM. This cancellation is caused by the structure
of the two--loop corrections to the diagonal beta functions in the considered model.
As a result, the prediction for the value of $\alpha_3(M_Z)$ at which exact gauge coupling
unification takes place is considerably lower in the E$_6$SSM than in the MSSM.

The only difference between the E$_6$SSM scenario, which was studied in \cite{King:2007uj},
and scenario A discussed above is in the possible presence of extra $S$ and $\overline{S}$
supermultiplets at low energies. From Eqs.~(\ref{56}) it follows that these supermultiplets
do not contribute to the diagonal beta functions of the SM gauge couplings. Our analysis of
the RG flow of $g_i(t)$ reveals that the evolution of the SM gauge couplings does not change
much when the low energy particle spectrum is supplemented by the bosonic and fermionic
components that originate from the extra $S$ and $\overline{S}$ chiral superfields. This
explains why our results are so similar to those previously obtained in \cite{King:2007uj}.

It is also worthwhile to point out that at high energies the uncertainty in $\alpha_3(t)$
caused by the variations of $\alpha_3(M_Z)$ is much bigger in the E$_6$SSM than in the MSSM.
This is because in the E$_6$SSM the strong gauge coupling grows slightly with increasing
renormalisation scale whereas in the MSSM it decreases at high energies. This implies that
the uncertainty in the high energy value of $\alpha_3(t)$ in the E$_6$SSM is approximately equal
to the low energy uncertainty in $\alpha_3(t)$ while in the MSSM the interval of variations of
$\alpha_3(t)$ near the scale $M_X$ shrinks drastically. The relatively large uncertainty in
$\alpha_3(M_X)$ in the E$_6$SSM, compared to the MSSM, allows one to achieve exact unification
of gauge couplings for values of $\alpha_3(M_Z)$ which are within one standard deviation of its
measured central value.

The RG flow of the SM gauge couplings changes substantially in the case of scenario B
as can be seen from Figs.~\ref{essmfig2}. As before we assume that the effective threshold
scales are equal, i.e. $\tilde{T}_1=\tilde{T}_2=\tilde{T}_3=\tilde{T}_S$. Our numerical
analysis reveals that the evolution of $\alpha_i(t)$ depends very strongly on $\tilde{T}_S$.
When $\tilde{T}_S\lesssim 1\,\mbox{TeV}$ the gauge couplings become rather large
near the GUT scale, i.e. $\alpha_i(M_X) \sim 1$, where as before we set
$M_X\simeq 3\cdot 10^{16}\,\mbox{GeV}$. For so large values of $\alpha_i(t)$
the perturbation theory method becomes inapplicable.
Therefore in our analysis we consider the range
of scales $\tilde{T}_S$ which are much higher than $1\,\mbox{TeV}$. In Figs.~\ref{essmfig2}
we set the threshold scale $\tilde{T}_S$ to be equal to $3\,\mbox{TeV}$. As one
can see from these figures for $\tilde{T}_S=3\,\mbox{TeV}$ the values of $\alpha_i(M_X)$
are about $0.2$ that still allows us to use the perturbation theory up to
the scale $M_X$.

The effective threshold scale that we consider in our analysis $\tilde{T}_S$
is in the multi TeV range. At first glance, it is not clear if so large values of
$\tilde{T}_i$ and $\tilde{T}_S$ can be obtained for a reasonable set of parameters.
In particular, to satisfy naturalness requirements the third generation sfermions
as well as neutralino and chargino states which are superparners of the SM gauge
bosons and Higgs fields are expected to have masses below $1\,\mbox{TeV}$. Because
of this in the MSSM naturalness arguments constrain the combined threshold scale
$\tilde{M}_{S}$ to be lower than $200-300\,\mbox{GeV}$ as it was mentioned above.
In the case of scenario B the analytical expression for the threshold scale
$\tilde{T}_{S}$ can be obtained by combining Eqs.~(\ref{63}) that gives
\begin{eqnarray}
\tilde{T}_{S}&=&\tilde{M}_{S}\cdot
\Biggl(\ds\frac{\mu_{H_u}^{12/19} m_{H_u}^{6/19} \mu_{H_d}^{12/19} m_{H_d}^{6/19}}
{\mu_{d_4}^{12/19} m_{d_4}^{6/19} \mu_{D_3}^{12/19} m_{\tilde{D}_3}^{6/19}}\Biggr)
\Biggl(\prod_{\alpha=1,2}
\ds\frac{m_{H_{\alpha}}^{6/19}\mu_{\tilde{H}_{\alpha}}^{12/19}}{m_{\tilde{D}_{\alpha}}^{6/19}\mu_{D_{\alpha}}^{12/19}}
\Biggr)\,.
\label{66}
\end{eqnarray}
Eq.~(\ref{66}) indicates that the combined threshold scale $\tilde{T}_{S}$ tends to be
very large if, for example, $\mu_{H_u}\simeq m_{H_u}\simeq \mu_{H_d}\simeq m_{H_d}$
are considerably larger than the masses of the scalar and fermion components of $d^c_4$ and
$\overline{d^c}_4$ as well as the masses of all exotic states. In this case $\tilde{T}_{S}$
can be as large as $10\,\mbox{TeV}$ even when $\tilde{M}_{S}$ lies in a few hundred GeV range
and $\mu_{H_u}\simeq m_{H_u}\simeq \mu_{H_d}\simeq m_{H_d}\lesssim 10\,\mbox{TeV}$.
This can be achieved if the components of $d^c_4$ and $\overline{d^c}_4$ and
some of the exotic quark and squark states have masses below $1\,\mbox{TeV}$.
The effective threshold scales $\tilde{T}_1$, $\tilde{T}_2$ and $\tilde{T}_3$ can be
also as large as a few $\mbox{TeV}$ if the scalar superpartners of the first and second
generation fermions and some of the exotic states have masses above $10\,\mbox{TeV}$.
Naturalness does not require these states to be light and, in fact, allowing them to be
heavy ameliorates SUSY flavor and CP problems. As a consequence the several TeV threshold
scales $\tilde{T}_1$, $\tilde{T}_2$, $\tilde{T}_3$ and $\tilde{T}_S$ can naturally emerge
in the scenario B.

In Fig.~2a we show the running of the SM gauge couplings from the EW scale to high
energies. We assume that in this case the low energy matter content includes three
27-plets of $E_6$ as well as $d^c_4$, $\overline{d^c}_4$, $H_u$, $\overline{H}_u$
$H_d$ and $\overline{H}_d$ supermultiplets. Fig.~2b shows the same RG flow of the
SM gauge couplings but just around the scale where the values of $\alpha_i(t)$ become
rather close. Again dotted lines in Figs.~2a and 2b represent the changes of the
evolution of the SM gauge couplings induced by the variations of $\alpha_3(M_Z)$ within
$1\,\sigma$ around its average value.

From Figs.~2a and 2b one can see that the interval of variations of $\alpha_3(t)$
enlarges with increasing renormalisation scale. The growth of the uncertainty in
the high energy value of $\alpha_3(t)$ is caused by the raise of this coupling itself.
As follows from Figs.~\ref{essmfig1} and \ref{essmfig2} in the scenario B the SM gauge
couplings grow faster with increasing renormalisation scale than in the case of
scenario A. This happens because the one--loop beta functions of these couplings are
larger in the scenario B as compared to the ones in the scenario A. As a consequence
the interval of variations of $\alpha_3(t)$ at high energies is also a bit bigger in
the former than in the latter. However as one can see from Figs.~2a and 2b this does
not facilitate the gauge coupling unification in scenario B. In fact, these figures
demonstrate that large two--loop corrections spoil the unification of gauge couplings
in this case. Indeed, in the one--loop approximation Eq.~(\ref{60}) leads to the same
prediction for $\alpha_3(M_Z)$ in the scenarios A and B because extra matter in these
scenarios form complete $SU(5)$ representations which contribute equally to the
one--loop beta functions of the $SU(3)_C$, $SU(2)_W$ and $U(1)_Y$ interactions so
that the differences of the coefficients of the one--loop beta functions $b_i-b_j$
remain intact. At the same time the contributions of two--loop corrections to
$\alpha_i(M_X)$ ($\Theta_i$) and $\alpha_3(M_Z)$ ($\Theta_s$) are different in these
cases. Our numerical analysis reveals that for $\tilde{T}_S\simeq 3\,\mbox{TeV}$ the
exact gauge coupling unification can be achieved in the scenario B only if the value
of $\alpha_3(M_Z)$ is around $0.112$. For higher scale $T_S$ the exact unification of
$\alpha_i(t)$ requires even smaller values of $\alpha_3(M_Z)$ which are disfavoured
by the recent fit to experimental data. The lower scales $T_S\lesssim 3\,\mbox{TeV}$
lead to the larger values of $\alpha_i(M_X)$ making questionable the validity of our
calculations.

As before extra $S$ and $\overline{S}$ superfields, that may survive to
low energies, do not contribute to the diagonal beta functions of the SM gauge couplings
and, therefore, do not change much the RG flow of $\alpha_i(t)$. As a result the value
of $\alpha_3(M_Z)$ at which exact gauge coupling unification takes place does not
change much as well after the inclusion of the bosonic and fermionic components of
these supermultiplets. Thus it seems to be rather difficult to reconcile the
unification of gauge couplings with present data in the Scenario B. Nevertheless
the values of $\alpha_i(M_X)$ are not so much different from each other. From Fig.~2b
it follows that the relative discrepancy of $\alpha_i(M_X)$ is about 10\% . This brings
us back to the orbifold GUT framework which was discussed in the previous section.
As it has been already mentioned orbifold GUTs do not imply the exact gauge
coupling unification near the scale $M_X$, which is associated with the size of
compact extra dimensions, due to the brane contributions to the gauge couplings
(see Eq.~(\ref{37})). Since one can expect that these brane corrections
become more sizable when $\alpha_i(M_X)$ are large, the relative discrepancy
of 10\% between $\alpha_i(M_X)$ should not be probably considered as a big problem
in the case of scenario B.

\section{Phenomenological implications}

We now consider cosmological implications and collider signatures of the $E_6$
inspired SUSY models discussed above. The phenomenological implications of these
models are determined by the structure of the particle spectrum that can vary
substantially depending on the choice of the parameters. For example, the masses
of the $Z'$ boson, exotic quarks, Inert Higgsinos and Inert singlinos are set by
the VEVs of the Higgs fields. In this section we primarily focus on the simplest
case when only $H_u$, $H_d$ and $S$ acquire non--zero VEVs breaking
$SU(2)_W\times U(1)_Y\times U(1)_{N}$ symmetry to $U(1)_{em}$ associated with
electromagnetism. Assuming that $f_{\alpha\beta}$ and $\tilde{f}_{\alpha\beta}$
are sufficiently small the masses of the exotic quarks, Inert Higgsino states and
$Z'$ boson are given by
\be
\mu_{D_i}=\dfrac{\kappa_i}{\sqrt{2}}\,s\,, \qquad\qquad
\mu_{H_{\alpha}}=\dfrac{\lambda_{\alpha}}{\sqrt{2}}\,s\,, \qquad\qquad
M_{Z'}\simeq g^{'}_1 \tilde{Q}_S s\,,
\label{67}
\ee
where $s$ is a VEV of the field $S$, i.e. $\langle S \rangle=s/\sqrt{2}$.
Here without loss of generality we set $\kappa_{ij}=\kappa_i\delta_{ij}$
and $\lambda_{\alpha\beta}=\lambda_{\alpha}\delta_{\alpha\beta}$.
Since $\mu_{D_i}$, $\mu_{H_{\alpha}}$ and $M_{Z'}$ are determined by $s$,
that remains a free parameter, the $Z'$ boson mass and the masses of exotic
quarks and Inert Higgsinos cannot be predicted. Because recent measurements
from the LHC experiments exclude $E_6$ inspired $Z'$ with masses lower than
$2-2.15\,\mbox{TeV}$ \cite{Chatrchyan:2012it} the singlet field $S$
must acquire a large VEV ($s\gtrsim 5.5-6\,\mbox{TeV}$) to induce sufficiently
large $M_{Z'}$. The couplings $\kappa_i$ should be also large enough to ensure
that the exotic fermions are sufficiently heavy to avoiding conflict with direct
particle searches at present and former accelerators. However the exotic
fermions (quarks and Inert Higgsinos) can be relatively light in the E$_6$SSM.
This happens, for example, when the Yukawa couplings of the exotic particles
have hierarchical structure similar to the one observed in the ordinary
quark and lepton sectors. Then $Z'$ mass lie beyond $10\,\mbox{TeV}$ and
the only manifestation of the considered models may be the presence of
light exotic quark and/or Inert Higgsino states in the particle spectrum.

Since the qualitative pattern of the particle spectrum and associated collider
signatures are so sensitive to the parameter choice it is worth to discuss
first the robust predictions that the considered models have.
It is well known that SUSY models predict that the mass of the lightest Higgs
particle is limited from above. The E$_6$SSM is not an exception.
In the simplest case when only $H_u$, $H_d$ and $S$ develop the VEVs, so that
$\langle H_d\rangle =\ds\frac{v_1}{\sqrt{2}}$,\,$\langle H_u\rangle
=\ds\frac{v_2}{\sqrt{2}}$ and $\langle S\rangle =\ds\frac{s}{\sqrt{2}}$,
the Higgs sector involves ten degrees of freedom. However four of them are
massless Goldstone modes which are swallowed by the $W^{\pm}$, $Z$ and $Z'$
gauge bosons that gain non-zero masses. If CP--invariance is preserved
the other degrees of freedom form two charged, one CP--odd and three CP-even
Higgs states. When the SUSY breaking scale is considerably larger than
the EW scale, the mass matrix of the CP-even Higgs sector has a hierarchical
structure and can be diagonalised using the perturbation
theory \cite{Nevzorov:2001um}-\cite{Nevzorov:2004ge}. In this case the mass
of one CP--even Higgs particle is always very close to the $Z'$ boson mass
$M_{Z'}$. The masses of another CP--even, the CP--odd and the charged Higgs
states are almost degenerate. When $\lambda\gtrsim g'_1$, the qualitative
pattern of the Higgs spectrum is rather similar to the one which arises
in the PQ symmetric NMSSM \cite{Nevzorov:2004ge}-\cite{Miller:2005qua}.
In the considered limit the heaviest CP--even, CP--odd and charged states
are almost degenerate and lie beyond the $\mbox{TeV}$ range \cite{King:2005jy}.
Finally, like in the MSSM and NMSSM, one of the CP--even Higgs bosons is
always light irrespective of the SUSY breaking scale. However, in contrast
with the MSSM, the lightest Higgs boson in the E$_6$SSM can be heavier
than $110-120\,\mbox{GeV}$ even at tree level. In the two--loop approximation
the lightest Higgs boson mass does not exceed $150-155\,\mbox{GeV}$ \cite{King:2005jy}.

\subsection{Dark matter}

The structure of the Yukawa interactions in the E$_6$SSM leads to another
important prediction. Using the method proposed in \cite{Hesselbach:2007te}
one can argue that there are theoretical upper bounds on the masses of the
lightest and second lightest inert neutralino states \cite{Hall:2010ix}.
To simplify the analysis we assume that the fermion components of
the supermultiplets $\overline{S}$, $\overline{H}_u$ and $\overline{H}_d$,
which may survive below the scale $M_X$, get combined with the corresponding
superpositions of the fermion components of the superfields $S_i$, $H^u_i$
and $H^d_i$ resulting in a set of heavy vectorlike states. Furthermore
we also assume that these vectorlike states completely decouple so that
the particle spectrum below the TeV scale contains only two generations of
inert Higgsinos ($\tilde{H}^u_{\alpha}$ and $\tilde{H}^d_{\alpha}$) and
two generations of inert singlinos $\tilde{S}_{\alpha}$. The Yukawa
interactions of these superfields are described by the superpotential
\begin{eqnarray}
W_{IH}=\lambda_{\alpha\beta} S (H^d_{\alpha} H^u_{\beta})+
f_{\alpha\beta} S_{\alpha} (H_d H^u_{\beta})+
\tilde{f}_{\alpha\beta} S_{\alpha} (H^d_{\beta} H_u)\,,
\label{essm2}
\end{eqnarray}
where $\alpha,\beta=1,2$\,.

Thus below the TeV scale the inert neutralino states are linear superposition
of the inert singlino states ($\tilde{S}_1$, $\tilde{S}_2$) and neutral components
of inert Higgsinos ($\tilde{H}^{d0}_1$, $\tilde{H}^{d0}_2$, $\tilde{H}^{u0}_1$,
$\tilde{H}^{u0}_2$). The charged components of the inert Higgsinos
$(\tilde{H}^{u+}_2,\,\tilde{H}^{u+}_1,\,\tilde{H}^{d-}_2,\,\tilde{H}^{d-}_1)$,
form inert chargino sector. In order to avoid the LEP lower limit on the masses
of inert charginos the couplings $\lambda_{\alpha\beta}$ and $s$ must be chosen
so that all inert chargino states are heavier than $100\,\mbox{GeV}$.
In addition, the requirement of the validity of perturbation theory up to
the GUT scale constrains the allowed range of Yukawa couplings $\lambda_{\alpha\beta}$,
$f_{\alpha\beta}$ and $\tilde{f}_{\alpha\beta}$. The restrictions specified
above set very stringent limits on the masses of two lightest inert neutralinos.
The analysis performed in \cite{Hall:2010ix} indicates that the lightest and
second lightest inert neutralinos ($\tilde{H}^0_{1}$ and $\tilde{H}^0_{2}$)
are typically lighter than $60-65\,\mbox{GeV}$. These neutralinos are predominantly
inert singlinos so that they can have rather small couplings to the $Z$--boson.
Therefore any possible signal which these neutralinos could give rise to at LEP
would be extremely suppressed. On the other hand the couplings of $\chi^0_{1}$ and
$\chi^0_{2}$ to the lightest CP--even Higgs boson $h_1$ are proportional to the
$\mbox{mass}/\sqrt{v_1^2+v_2^2}$ in the leading approximation \cite{Hall:2010ix}.
As a consequence the couplings of two lightest inert neutralino to the lightest Higgs
state are always large if the corresponding states have appreciable masses.

The discussion above indicates that the lightest and second lightest inert
neutralinos tend to be the lightest states which are odd under the $Z_{2}^{E}$ symmetry.
It is worth to remind here that in the considered $E_6$ inspired SUSY models
$U(1)_{\psi}\times U(1)_{\chi}$ gauge symmetry is broken down to $U(1)_{N}\times Z_{2}^{M}$
where $Z_{2}^{M}=(-1)^{3(B-L)}$ is the so--called matter parity which is a discrete
subgroup of $U(1)_{\psi}$ and $U(1)_{\chi}$. Since the low--energy effective
Lagrangian is invariant under both $Z_{2}^{M}$ and $\tilde{Z}^{H}_2$ symmetries and
$\tilde{Z}^{H}_2 = Z_{2}^{M}\times Z_{2}^{E}$ (see Table~\ref{tab1}), the $Z_{2}^{E}$
symmetry is also conserved. This means that the lightest exotic state, which is odd
under the $Z_{2}^{E}$ symmetry, is absolutely stable and contributes to the relic
density of dark matter.

Because the lightest inert neutralino is also the lightest
$R$--parity odd state either the lightest $R$--parity even exotic state or
the lightest $R$--parity odd state with $Z_{2}^{E}=+1$ must be absolutely
stable. When $f_{\alpha\beta}$ and $\tilde{f}_{\alpha\beta}$ are large enough
($f_{\alpha\beta}\sim \tilde{f}_{\alpha\beta}\sim 0.5$) the large mixing in
the inert Higgs sector may lead to the lightest CP--even (or CP--odd) inert Higgs
state with mass of the order of the EW scale. The corresponding exotic state
is $R$--parity even neutral particle. If it is substantially lighter than the
lightest ordinary neutralino state $\chi_1^0$ and the decay of $\chi_1^0$ into
the lightest inert neutralino and the lightest inert Higgs scalar (pseudoscalar)
is kinematically allowed then this lightest inert Higgs scalar (pseudoscalar) is
absolutely stable and may result in considerable contribution to the relic dark
matter density.

Although the possibility mentioned above looks very attractive a substantial
fine-tuning is normally required to make the lightest inert Higgs scalar (pseudoscalar)
lighter than $\chi_1^0$. Most commonly $\chi_1^0$ is considerably lighter than
the lightest inert Higgs scalar (pseudoscalar) so that the lightest CP--even
(CP--odd) inert Higgs state can decay into $\chi_1^0$ and the lightest inert
neutralino state. In other words, in the considered $E_6$ inspired SUSY models
the lightest $R$--parity odd state with
$Z_{2}^{E}=+1$, i.e. $\chi_1^0$, tend to be substantially lighter than
the $R$--parity even exotic states. As a result the lightest neutralino
state $\chi_1^0$ is a natural candidate for a cold component of dark matter
in these models.

In the neutralino sector of the E$_6$SSM there are two extra neutralinos besides
the four MSSM ones. One of them is an extra gaugino $\tilde{B}'$ coming from the
$Z'$ vector supermultiplet. The other one is an additional singlino $\tilde{S}$
which is a fermion component of the SM singlet superfield $S$. Extra neutralinos form
two eigenstates $(\tilde{B}'\pm\tilde{S})/\sqrt{2}$ with masses around $M_{Z'}$\,
\cite{King:2005jy}. Since LHC experiments set very stringent lower bound on the mass
of the $Z'$ boson extra neutralino eigenstates tend to be the heaviest ones
and decouple. The mixing between these heavy neutralino states and other gauginos and
Higgsinos is very small. Therefore the lightest neutralino states in the E$_6$SSM,
that determine the composition of $\chi_1^0$ and as a consequence its contribution
to the relic dark matter density, become almost indistinguishable from the ones
in the MSSM. This means that in the E$_6$SSM, like in the MSSM, the lightest neutralino
$\chi_1^0$ can give a substantial contribution to the relic density which is in
agreement with the measured abundance of cold dark matter
$\Omega_{\mathrm{CDM}}h^2 = 0.1099 \pm 0.0062$ \cite{cdm}.

In the E$_6$SSM the lightest inert neutralino can account for all or some of the
observed cold dark matter relic density if $\chi_1^0$ has mass close to half the
$Z$ mass. In this case the lightest inert neutralino states annihilate mainly through
an $s$--channel $Z$--boson, via its Inert Higgsino doublet components which couple
to the $Z$--boson \cite{Hall:2010ix}, \cite{Hall:2009aj}. When $|m_{\tilde{H}^0_{1}}|\ll M_Z$
the lightest inert neutralino states are almost inert singlinos and the couplings of
$\tilde{H}^0_1$ to gauge bosons, Higgs states, quarks (squarks) and leptons (sleptons)
are quite small leading to a relatively small annihilation cross section for
$\tilde{H}^0_1\tilde{H}^0_1\to \mbox{SM particles}$. Since the dark matter
number density is inversely proportional to the annihilation cross section
at the freeze-out temperature the lightest inert neutralino state with mass
$|m_{\tilde{H}^0_{1,2}}|\ll M_Z$ gives rise to a relic density which is typically
much larger than its measured value\footnote{When $f_{\alpha\beta},\,
\tilde{f}_{\alpha\beta}\to 0$ the masses of $\tilde{H}^0_1$ and $\tilde{H}^0_2$
tend to zero and inert singlino states essentially decouple from the rest of the
spectrum. In this limit the lightest non-decoupled Inert neutralino may be rather
stable and can play the role of dark matter \cite{Hall:2011zq}. The presence of very
light neutral fermions in the particle spectrum might have interesting implications
for the neutrino physics (see, for example \cite{Frere:1996gb}).}.

Because the scenarios with $|m_{\tilde{H}^0_{1,2}}|\sim M_Z/2$ imply that the couplings
of $\tilde{H}^0_1$ and $\tilde{H}^0_2$ to the lightest Higgs boson are much larger than
the $b$--quark Yukawa coupling the lightest Higgs state decays more than 95\% of the
time into $\tilde{H}^0_1$ and $\tilde{H}^0_2$ in these cases while the total branching
ratio into SM particles varies from 2\% to 4\% \cite{Hall:2010ix}. At the same time
the LHC production cross section of the lightest Higgs state in the considered $E_6$
inspired SUSY models is almost the same as in the MSSM. Therefore the evidence for the
Higgs boson recently presented by ATLAS \cite{:2012gk} and CMS \cite{:2012gu} indicates
that the corresponding scenarios are basically ruled out.

In this context one should point out another class of scenarios that might have interesting
cosmological implications. Let us consider a limit when
$f_{\alpha\beta}\sim \tilde{f}_{\alpha\beta}\sim 10^{-5}$. So small values of the
Yukawa couplings $f_{\alpha\beta}$ and $\tilde{f}_{\alpha\beta}$ result in extremely light
inert neutralino states $\tilde{H}^0_1$ and $\tilde{H}^0_2$ which are basically inert singlinos.
These states have masses about $1\,\mbox{eV}$.
Since $\tilde{H}^0_1$ and $\tilde{H}^0_2$ are so light and absolutely stable they form hot
dark matter in the Universe\footnote{In the context of $E_6$ inspired SUSY models warm dark
matter was recently discussed in \cite{King:2012wg}.}.
These inert neutralinos have
negligible couplings to $Z$ boson and would not have been observed at earlier collider
experiments. These states also do not change the branching ratios of the $Z$ boson and
Higgs decays. Moreover if $Z'$ boson is sufficiently heavy the presence of such light
Inert neutralinos does not affect Big Bang Nucleosynthesis \cite{Hall:2011zq}.
When the masses of $\tilde{H}^0_1$ and $\tilde{H}^0_2$ are about $1\,\mbox{eV}$
these states give only a very minor contribution to the dark matter density while
the lightest neutralino may account for all or some of the observed dark matter density.
In this case one can expect that the lifetime of the next-to-lightest exotic state
(for example, inert chargino) is given by
\begin{equation}
\tau_{NLES}\sim \frac{8\pi^2}{f^2 M_{NLES}}\,,
\label{73}
\end{equation}
where $f_{\alpha\beta}\sim \tilde{f}_{\alpha\beta}\sim f$ and
$M_{NLES}$ is the mass of the next-to-lightest exotic state.
Assuming that $M_{NLES}\sim 1\,\mbox{TeV}$ we get $\tau_{NLES}\sim 10^{-15}\,s$.
With increasing $f_{\alpha\beta}$ and $\tilde{f}_{\alpha\beta}$ the masses of the
lightest inert neutralino states grow and their contribution to the relic density of
dark matter becomes larger. This may lead to some interesting cosmological implications.
The detailed study of these implications is beyond the scope of this paper and will
be considered elsewhere.

\subsection{LHC signatures}

We can now turn to the possible collider signatures of the $E_6$ inspired SUSY models
with exact custodial $\tilde{Z}^{H}_2$ symmetry. The presence of $Z'$ boson and exotic
multiplets of matter in the particle spectrum is a very peculiar feature that may permit
to distinguish the considered $E_6$ inspired SUSY models from the MSSM or NMSSM.
Although the masses of the $Z'$ boson and exotic states cannot be predicted there are
serious reasons to believe that the corresponding particles should be relatively light.
Indeed, in the simplest scenario the VEVs of $H_u$, $H_d$ and $S$ are determined by the
corresponding soft scalar masses. Since naturalness arguments favor SUSY models with
$O(1\,\mbox{TeV})$ soft SUSY breaking terms the VEV $s$ is expected to be of the order
of $1-10\,\mbox{TeV}$. On the other hand the requirement of the validity of perturbation
theory up to the GUT scale sets stringent upper bounds on the low--energy values of the
Yukawa couplings $\kappa_i$ and $\lambda_{\alpha}$ whereas the gauge coupling unification
implies that $g^{'}_1(q)\simeq g_1(q)$. As a consequence the $Z'$ boson and exotic states
are expected to have masses below $10\,\mbox{TeV}$.

Collider experiments and precision EW tests set stringent limits on the mass of the $Z'$
boson and $Z-Z'$ mixing. The direct searches at the Fermilab Tevatron
$(p\overline{p}\to Z'\to l^{+}l^{-})$ exclude $Z'$, which is associated with $U(1)_N$,
with mass below $892\,\mbox{GeV}$ \cite{Accomando:2010fz}
\footnote{Slightly weaker lower bound on the mass of the $Z_N'$ boson was obtained
in \cite{Erler:2011ud}.}. Recently ATLAS and CMS experiments ruled out $E_6$ inspired $Z'$
with masses lower than $2-2.15\,\mbox{TeV}$ \cite{Chatrchyan:2012it}.
The analysis performed in \cite{ZprimeE6} revealed that $Z'$ boson in the $E_6$ inspired
models can be discovered at the LHC if its mass is less than $4-4.5\,\mbox{TeV}$.
The determination of its couplings should be possible if $M_{Z'}\lesssim 2-2.5\,\mbox{TeV}$
\cite{Dittmar:2003ir}. The precision EW tests bound the $Z-Z'$ mixing angle to be around
$[-1.5,\,0.7]\times 10^{-3}$ \cite{Erler:2009jh}. Possible $Z'$ decay channels in $E_6$
inspired supersymmetric models were studied in \cite{Accomando:2010fz}, \cite{Gherghetta:1996yr}.
The potential influence of gauge kinetic mixing on $Z'$ production at the 7 TEV LHC was
considered in \cite{Rizzo:2012rf}.

The production of a TeV scale exotic states will also provide spectacular LHC signals.
Several experiments at LEP, HERA, Tevatron and LHC have searched for colored objects
that decay into either a pair of quarks or quark and lepton. But most searches focus on
exotic color states, i.e leptoquarks or diquarks, have integer--spin. So they are either
scalars or vectors. These colored objects can be coupled directly to either
a pair of quarks or to quark and lepton. Moreover it is usually assumed that leptoquarks and
diquarks have appreciable couplings to the quarks and leptons of the first generation.
The most stringent constraints on the the masses of leptoquarks come from the nonobservation
of these exotic color states at the ATLAS and CMS experiments. Recently ATLAS collaboration ruled
out first and second generation scalar leptoquarks (i.e. leptoquarks that couple to the first and
second generation fermions respectively) with masses below $600-700\,\mbox{GeV}$ \cite{Aad:2011ch}.
The CMS collaboration excluded first and second generation scalar leptoquarks which are lighter
than $640-840\,\mbox{GeV}$ \cite{:2012dn}. The experimental lower bounds on the masses of dijet
resonances (in particular, diquarks) tend to be considerably higher (see, for example, \cite{90}).

However the LHC lower bounds on the masses of exotic quarks mentioned above are not directly
applicable in the case of the $E_6$ inspired SUSY models considered here.
Since $Z_{2}^{E}$ symmetry is conserved every interaction vertex contains an even number of
exotic states. As a consequence each exotic particle must eventually decay into a final state
that contains at least one lightest Inert neutralino (or an odd number of the lightest Inert
neutralinos). Since stable lightest Inert neutralinos cannot be detected directly each exotic
state should result in the missing energy and transverse momentum in the final state.
The $Z_{2}^{E}$ symmetry conservation also implies that in collider experiments exotic
particles can only be created in pairs.

In this context let us consider the production and sequential decays of the lightest exotic
quarks at the LHC first. Because $D$ and $\overline{D}$ states are odd under the $Z_{2}^{E}$
symmetry they can be only pair produced via strong interactions. In the scenario A the lifetime
and decay modes of the lightest exotic quarks are determined by the operators
$g^D_{ij} (Q_i L_4) \overline{D}_j$ and $h^E_{i\alpha} e^c_{i} (H^d_{\alpha} L_4)$ in the
superpotential (\ref{13}). These operators ensure that the lightest exotic quarks decay into
$$
D\to u_i (d_i) + \ell (\nu) + E^{\rm miss}_{T} + X\,,
$$
where $\ell$ is either electron or muon. Here $X$ may contain extra
charged leptons that can originate from the decays of intermediate
states (like Inert chargino or Inert neutralino). Since lightest exotic
quarks are pair produced these states may lead to a substantial enhancement
of the cross section $pp\to jj\ell^{+}\ell^{-}+E^{\rm miss}_{T}+X$ if they
are relatively light. In the scenario B the decays of the lightest exotic quarks
are induced by the operators $g^{q}_{ij}\overline{D}_i d^c_4 u^c_j$ and $h^D_{ij} d^c_4 (H^d_{i} Q_j)$.
As a consequence the lightest diquarks decay into
$$
D\to u^c_i + d^c_j + E^{\rm miss}_{T}+X\,,
$$
where $X$ again can contain charged leptons that may come from the decays
of intermediate states. In this case the presence of light $D$-fermions in the particle spectrum
could result in an appreciable enhancement of the cross section $pp\to jjjj+E^{\rm miss}_{T}+X$.

In general exotic squarks are expected to be substantially heavier than the exotic quarks
because their masses are determined by the soft SUSY breaking terms. Nevertheless the exotic
squark associated with the heavy exotic quark maybe relatively light. Indeed, as in the case
of the superpartners of the top quark in the MSSM, the large mass of the heaviest exotic quark
in the E$_6$SSM gives rise to the large mixing in the corresponding exotic squark sector
that may result in the large mass splitting between the appropriate mass eigenstates. As a
consequence the lightest exotic squark can have mass in TeV range. Moreover, in principle,
the lightest exotic squark can be even lighter than the lightest exotic quark. If this is
a case then the decays of the lightest exotic squark are induced by the same operators
which give rise to the decays of the lightest exotic quarks when all exotic squarks are heavy.
Therefore the decay patterns of the lightest exotic color states are rather similar in both
cases. In other words when exotic squark is the lightest exotic color state in the particle
spectrum it decays into either
$$
\tilde{D}\to
u_i (d_i) + \ell (\nu) + E^{\rm miss}_{T} + X\,,
$$
if exotic squark is a scalar leptoquark or
$$
\tilde{D}\to u^c_i + d^c_j + E^{\rm miss}_{T} + X\,,
$$
if it is a scalar diquark. Due to the $Z_{2}^{E}$ symmetry conservation $E^{\rm miss}_{T}$
should always contain contribution associated with the lightest exotic particle. However
since the lightest exotic squark is $R$--parity even state whereas the lightest Inert neutralino
is $R$--parity odd particle the final state in the decay of $\tilde{D}$ should also involve
the lightest neutralino to ensure that $R$--parity is conserved. Again, $X$ may contain
charged leptons that can stem from the decays of intermediate states. Because the $Z_{2}^{E}$
symmetry conservation implies that the lightest exotic squarks can be only pair produced in
the considered case the presence of light $\tilde{D}$ is expected to lead to an appreciable
enhancement of the cross section of either $pp\to jj\ell^{+}\ell^{-}+E^{\rm miss}_{T}+X$ if
$\tilde{D}$ is scalar leptoquark or $pp\to jjjj+E^{\rm miss}_{T}+X$ if $\tilde{D}$ is scalar
diquark.

Thus one can see that in both scenarios when the lightest exotic color state is either
$D$-fermion or $\tilde{D}$-scalar the collider signatures associated with these new states
are rather similar. Moreover since the decays of the lightest exotic color particles lead
to the missing energy and transverse momentum in the final state it might be rather
problematic to distinguish the corresponding signatures from the ones which are associated
with the MSSM. For example, the pair production of gluinos at the LHC should also result
in the enhancement of the cross section of $pp\to jjjj+E^{\rm miss}_{T}+X$. In this context
the presence of additional charged leptons in $X$ can play an important role leading to
characteristic signatures such as $\ell^{+}\ell^{-}$ pairs together with large missing energy
in the final state. The situation also becomes a bit more promising if one assumes
that the Yukawa couplings of the exotic particles have hierarchical structure similar
to the one observed in the ordinary quark and lepton sectors. In this case all states
which are odd under the $Z^{E}_2$ symmetry couple to the third generation fermions and
sfermions mainly\footnote{This possibility was discussed at length in
\cite{King:2005jy}--\cite{Accomando:2006ga}, \cite{8}.}. As a consequence the
presence of the relatively light exotic color states should give rise to the enhancement
of the cross section of either $pp\to t\bar{t}\ell^{+}\ell^{-}+E^{\rm miss}_{T}+X$ or
$pp\to t\bar{t}b\bar{b}+E^{\rm miss}_{T}+X$.

Here it is worthwhile to point out that the collider signatues associated with the light
scalar leptoquarks or diquarks in the considered $E_6$ inspired SUSY models are very
different from the commonly established ones which have been thoroughly studied.
For instance, it is expected that scalar diquarks may be produced singly at the LHC
and decay into quark--quark without missing energy in the final state. The scalar
leptoquarks can be only pair produced at the LHC but it is commonly assumed that
these states decay into quark--lepton without missing energy as well. On the other
hand in the $E_6$ inspired SUSY models considered here the $Z_{2}^{E}$ symmetry
conservation necessarily leads to the missing energy and transverse momentum
in the corresponding final state.

The presence of relatively light exotic quark and squark can substantially modify
the collider signatures associated with the production and decay of gluinos
\footnote{Novel gluino decays in the $E_6$ inspired models were recently considered
in \cite{Belyaev:2012si}}.
Indeed, if all squarks except the lightest exotic squark are rather heavy and
the decay of the gluino into exotic quark and squark are kinematically allowed
then the gluino pair production at the LHC results in $D\bar{D}\tilde{D}\tilde{\bar{D}}$
in the corresponding final state. The sequential decays of exotic quarks and squarks
give rise to the enhancement of either $pp\to 4\,\ell+4\, j + E^{\rm miss}_{T}+X$
if exotic color states are leptoquarks or $pp\to 8\,j + E^{\rm miss}_{T}+X$
if exotic color states are diquarks, modulo of course effects of QCD
radiation and jet merging. The modification of the gluino collider
signatures discussed above might be possible only if there are non--zero
flavor-off-diagonal couplings $\theta^g_{ij}$ of gluino to $D_i$ and $\tilde{D}_j$
($i\ne j$). This is a necessary condition because the lightest exotic squark is
normally associated with the heaviest exotic quark. Rough estimates indicate
that the corresponding modification of the gluino collider signatures can occur
even when the gluino flavour-off-diagonal couplings $\theta^g_{ij}$ are relatively
small, i.e. $\theta^g_{ij}\gtrsim 0.01$.

If gluino is heavier than the lightest exotic color state, but is substantially lighter
than the second lightest exotic color state than the branching ratios of the nonstandard
gluino decays mentioned above are suppressed. In this case the second lightest exotic
color state can decay mostly into the lightest exotic color state and gluino if the
corresponding decay channel is kinematically allowed. This happens when the lightest exotic
color state is exotic $D$-fermion while the second lightest exotic color state is
$\tilde{D}$-scalar or vice versa.

Other possible manifestations of the $E_6$ inspired SUSY models considered
here are related to the presence of vectorlike states $d^c_4$ and $\overline{d^c}_4$
as well as $L_4$ and $\overline{L}_4$. In the case of scenario B the fermionic components
of the supermultiplets $d^c_4$ and $\overline{d^c}_4$ can have mass below the TeV scale.
One of the superpartners of this vectorlike quark state may be also relatively light
due to the mixing in the corresponding squark sector. If these quark and/or squark
states are light they can be pair produced at the LHC via strong interactions. Since
the superfields $d^c_4$ and $\overline{d^c}_4$ are odd under the $Z^{E}_2$
symmetry the decays of the corresponding quarks ($d_4$) and squarks ($\tilde{d}_4$) must
always lead to the missing energy in the final state. In the limit when the lightest
exotic color states include $d_4$ and/or $\tilde{d}_4$ whereas all other exotic states
and sparticles are much heavier, the operators $h^D_{ij} d^c_4 (H^d_{i} Q_j)$ give
rise to the following decay modes of $d_4$ and $\tilde{d}_4$
$$
d_4 \to  q_i + E^{\rm miss}_{T} + X\,,\qquad\qquad\qquad
\tilde{d}_4 \to  d_i + E^{\rm miss}_{T} + X\,,
$$
where $q_i$ can be either up-type or down-type quark while $X$ may contain charged leptons
which can appear as a result of the decays of intermediate states. As in the case of
exotic squark the final state in the decay of $d_4$ should contain the lightest neutralino
and the lightest Inert neutralino to ensure the conservation of $R$--parity and $Z^{E}_2$
symmetry. Again due to the $Z_{2}^{E}$ symmetry conservation $d_4$ and $\tilde{d}_4$ can be only
pair produced at the LHC resulting in an enhancement of $pp\to jj+E^{\rm miss}_{T}+X$.
If $d_4$ and $\tilde{d}_4$ couple predominantly to the third generation fermions and sfermions
then the pair production of these quarks/squarks should lead to the presence of two heavy
quarks in the final state. As before these collider signatures do not permit to distinguish
easily the considered $E_6$ inspired SUSY models from other supersymmetric models.
For example, squark pair production at the LHC can also lead to two jets and missing energy
in the final state. Again, the presence of additional charged leptons in $X$ can lead to
the signatures that may help to distinguish the considered $E_6$ inspired SUSY models from
the simplest SUSY extensions of the SM.

In the case of scenario A the fermionic components of the supermultiplets $L_4$ and
$\overline{L}_4$ as well as one of the superpartners of this vectorlike state may have masses
below the TeV scale. If all other exotic states and sparticles are rather heavy the
corresponding bosonic ($\tilde{L}_4$) and fermionic ($L_4$) states can be produced at the
LHC via weak interactions only. Because of this their production cross section is relatively
small. In the considered limit the decays of $L_4$ and/or $\tilde{L}_4$ are induced by
the operators $h^E_{i\alpha} e^c_{i} (H^d_{\alpha} L_4)$. As a consequence the decays
of $L_4$ and/or $\tilde{L}_4$ always lead to either $\tau$--lepton or electron/muon
as well as missing energy in the final state. In the case of $\tilde{L}_4$ decays the
missing energy in the final state can be associated with only one lightest Inert neutralino
whereas the final state of the $L_4$ decays must contain at least one lightest Inert neutralino
and one lightest ordinary neutralino to ensure the conservation of $R$--parity and $Z^{E}_2$
symmetry. More efficiently $L_4$ and/or $\tilde{L}_4$ can be produced through the decays of the
lightest exotic color states (i.e. $D$ and/or $\tilde{D}$) if these states are relatively light
and the corresponding decay channels are kinematically allowed.

The Inert Higgs bosons and/or Inert neutralino and chargino states, which are predominantly Inert
Higgsinos, can be also light or heavy depending on their free parameters. Indeed, as follows from
Eq.~(\ref{67}) the lightest Inert Higgsinos may be light if the corresponding Yukawa coupling
$\lambda_{\alpha}$ is rather small. On the other hand if at least one coupling $\lambda_{\alpha}$
is large it can induce a large mixing in the Inert Higgs sector that may lead to relatively
light Inert Higgs boson states. Since Inert Higgs and Higgsino states do not couple to quarks
directly at the LHC the corresponding states can be produced in pairs via off--shell $W$ and
$Z$--bosons. Therefore their production cross section remains relatively small even when these
states have masses below the TeV scale. The lightest Inert Higgs and Higgsino states are
expected to decay via virtual lightest Higgs, $Z$ and $W$ exchange. The conservation of $R$--parity
and $Z^{E}_2$ symmetry implies that the final state in the decay of Inert Higgsino involves at
least one lightest Inert neutralino while the final state in the decay of Inert Higgs state
should contain at least one lightest ordinary neutralino and one lightest Inert neutralino.

As it was mentioned in the beginning of this subsection in the simplest scenario, when
only $H_u$, $H_d$ and $S$ acquire VEVs at low energies, there are serious reasons to believe
that the $Z'$ boson and all exotic states from three complete $27_i$ representations of $E_6$
have masses below $10\,\mbox{TeV}$. However the situation may change dramatically when
$\tilde{Z}^H_2$ even superfield $\overline{S}$ survive to low energies. In order to demonstrate
this, let us consider a simple toy model, where $U(1)_N$ gauge symmetry is broken by VEVs of
a pair of SM singlet superfields $S$ and $\overline{S}$. Assuming that the superpotential of the
considered model involves bilinear term $\mu_S\, S \overline{S}$ the part of the tree--level scalar
potential, which depends on the scalar components of the superfields $S$ and $\overline{S}$ only,
can be written as
\begin{equation}
V_S = (m^2_S+\mu_S^2) |S|^2 + (m^2_{\overline{S}}+\mu_S^2) |\overline{S}|^2
+ (B_S \mu_S S \overline{S} + h.c.)
+\ds\frac{Q_S^2 g^{'2}_1}{2}\left(|S|^2-|\overline{S}|^2\right)^2\,,
\label{74}
\end{equation}
where $m_S^2$, $m^2_{\overline{S}}$ and $B_S$ are soft SUSY breaking parameters and $Q_S$ is
a $U(1)_N$ charge of the SM singlet superfields $S$. The last term in Eq.~(\ref{74}), which
is the $U(1)_N$ D--term contribution to the scalar potential, forces the minimum of the
corresponding potential to be along the $D$--flat direction
$\langle S \rangle = \langle \overline{S} \rangle$. Indeed, in the limit
$\langle S \rangle = \langle \overline{S} \rangle$ the quartic terms in the potential (\ref{74})
vanish. In the considered case the scalar potential (\ref{74}) remains positive definite only if
$(m^2_S + m^2_{\overline{S}}+ 2 \mu_S^2 - 2|B_S \mu_S|)>0$. Otherwise physical vacuum becomes
unstable, i.e. $\langle S \rangle = \langle \overline{S} \rangle \to \infty$.

The scalar potential can be easily stabilized the if bilinear term $\mu_S\, S \overline{S}$
in the superpotential is replaced by
\begin{equation}
W_S = \lambda_0 \tilde{\phi} S \overline{S} + f(\tilde{\phi})\,,
\label{75}
\end{equation}
where $\tilde{\phi}$ is $\tilde{Z}^H_2$ even superfield that does not participate in the
$SU(3)_C\times SU(2)_W\times U(1)_Y\times U(1)_{\psi}\times U(1)_{\chi}$ gauge interactions.
When $\lambda_0$ is small (i.e. $\lambda_0\ll 0.1$) the $U(1)_N$ D--term contribution to
the scalar potential still forces the minimum of the scalar potential to be along the
nearly $D$--flat direction if $m^2_S + m^2_{\overline{S}}<0$. This condition can be satisfied
because sufficiently large values of $\kappa_i$ affect the evolution of $m_S^2$ rather
strongly resulting in negative values of $m_S^2$ at low energies \cite{8}.
If $m^2_S + m^2_{\overline{S}}<0$ and $\lambda_0$ is small then the scalar components of
the superfields $\tilde{\phi}$, $S$ and $\overline{S}$ acquire very large VEVs, i.e.
\begin{equation}
\langle \tilde{\phi} \rangle \sim \langle S \rangle \simeq \langle \overline{S} \rangle \sim M_{SUSY}/\lambda_0\,,
\label{76}
\end{equation}
where $M_{SUSY}$ is a supersymmetry breaking scale.
If $\lambda_0\simeq 10^{-3}-10^{-4}$ the VEVs of the SM singlet superfields $S$ and $\overline{S}$
are of the order of $10^{3}-10^4\,\mbox{TeV}$ even when $M_{SUSY}\sim 1\,\mbox{TeV}$.
So large VEV of the superfield $S$ may give rise to the extremely heavy spectrum of exotic
particles and $Z'$. This can lead to the MSSM type of particle spectrum at the $\mbox{TeV}$
scale.

Nevertheless even in this case the broken $U(1)_N$ symmetry leaves its imprint on the
MSSM sfermion mass spectrum. Since $m^2_S\ne m^2_{\overline{S}}$ the VEVs of the SM singlet
superfields $S$ and $\overline{S}$ deviates from the $D$--flat direction
\begin{equation}
Q_S^2 g^{'2}_1 \left(\langle S \rangle^2 - \langle \overline{S} \rangle^2\right)\simeq m^2_{\overline{S}}-m^2_S\,.
\label{77}
\end{equation}
As a consequence all sfermions receive an additional contribution to the mass that come
from the $U(1)_N$ $D$--term quartic interactions in the scalar potential \cite{Kolda:1995iw}.
This contribution $\Delta_i$ is proportional to the $U(1)_N$ charge of the corresponding sfermion
$Q_i$, i.e.
\begin{equation}
\Delta_{i}=\dfrac{g^{'2}_1}{2}\biggl(Q_1 v_1^2 + Q_2 v_2^2 +
2 Q_S \left(\langle S \rangle^2 - \langle \overline{S} \rangle^2\right)\biggr)
Q_{i}=M_0^2\, \sqrt{40}\, Q_{i} \,,
\label{78}
\end{equation}
where $Q_1$ and $Q_2$ are the $U(1)_N$ charges of $H_d$ and $H_u$.
Thus for the superpartners of the first and second generation quarks and leptons one finds
$$
\begin{array}{rcl}
m_{\tilde{d}_{L\,i}}^2&\simeq &m_{Q_i}^2+\left(-\dfrac{1}{2}+\dfrac{1}{3}\sin^2\theta_W\right)M_Z^2\cos 2\beta + M_0^2\,,\\
m_{\tilde{u}_{L\,i}}^2&\simeq &m_{Q_i}^2+\left(\dfrac{1}{2}-\dfrac{2}{3}\sin^2\theta_W\right)M_Z^2\cos 2\beta + M_0^2\,,\\
m_{\tilde{u}_{R\,i}}^2&\simeq &m_{u^c_i}^2+\dfrac{2}{3} M_Z^2 \sin^2\theta_W \cos 2\beta + M_0^2\,,\\
m_{\tilde{d}_{R\,i}}^2&\simeq &m_{d^c_i}^2-\dfrac{1}{3} M_Z^2 \sin^2\theta_W \cos 2\beta + 2 M_0^2\,,\\[2mm]
m_{\tilde{e}_{L\,i}}^2&\simeq &m_{L_i}^2+\left(-\dfrac{1}{2}+\sin^2\theta_W\right)M_Z^2\cos 2\beta + 2 M_0^2\,,\\
m_{\tilde{\nu}_{i}}^2&\simeq &m_{L_i}^2+\dfrac{1}{2} M_Z^2\cos 2\beta + 2 M_0^2\,,\\
m_{\tilde{e}_{R\,i}}^2&\simeq &m_{e^c_i}^2- M_Z^2 \sin^2\theta_W \cos 2\beta + M_0^2\,.
\end{array}
$$

\section{Conclusions}

In this paper we have considered the $E_6$ inspired SUSY models in which a single discrete
$\tilde{Z}^{H}_2$ symmetry forbids the tree--level flavor--changing transitions and baryon
number violating operators. We assumed that the breakdown of $E_6$ symmetry or its subgroup
lead to the rank--6 SUSY models below the GUT scale $M_X$. These models are based on the
Standard Model (SM) gauge group together with extra $U(1)_{\psi}$ and $U(1)_{\chi}$ gauge
symmetries. We also allow three copies of $27_i$ representations of $E_6$ to survive below
the scale $M_X$ so that anomalies get canceled generation by generation. If extra exotic
states from $27_i$--plets survive to low energies they give rise to tree--level non--diagonal
flavor transitions and rapid proton decay. In order to suppress baryon number violating
operators one can impose $\tilde{Z}^{H}_2$ discrete symmetry. We assumed that all matter
superfields, that fill in complete $27_i$ representations of $E_6$, are odd under this
discrete symmetry. Thus $\tilde{Z}^{H}_2$ symmetry is defined analogously to the matter
parity $Z_{2}^{M}$ in the simplest $SU(5)$ SUSY GUTs, that lead to the low--energy spectrum
of the MSSM.

In addition to three complete fundamental representations of $E_6$ we further assumed the
presence of of $M_{l}$ and $\overline{M}_l$ supermultiplets from the incomplete $27'_l$ and
$\overline{27'}_l$ representation just below the GUT scale. Because multiplets $M_{l}$ and
$\overline{M}_l$ have opposite $U(1)_{Y}$, $U(1)_{\psi}$ and $U(1)_{\chi}$ charges their
contributions to the anomalies get cancelled identically.
As in the MSSM we allowed the set of multiplets $M_{l}$ to be used for the breakdown of
gauge symmetry and therefore assumed that all multiplets $M_{l}$ are even under $\tilde{Z}^{H}_2$
symmetry. In order to ensure that the $SU(2)_W\times U(1)_Y\times U(1)_{\psi}\times U(1)_{\chi}$
symmetry is broken down to $U(1)_{em}$ associated with the electromagnetism the set of multiplets
$M_{l}$ should involve $H_u$, $H_d$, $S$ and $N^c_H$.

We argued that $U(1)_{\psi}\times U(1)_{\chi}$ gauge symmetry can be broken by the VEVs
of $N^c_H$ and $\overline{N}_H^c$ down to $U(1)_{N}\times Z_{2}^{M}$ because matter parity
is a discrete subgroup of $U(1)_{\psi}$ and $U(1)_{\chi}$. Such breakdown of $U(1)_{\psi}$ and
$U(1)_{\chi}$ gauge symmetries guarantees that the exotic states which originate from $27_i$
representations of $E_6$ as well as ordinary quark and lepton states survive to low energies.
On the other hand the large VEVs of $N^c_H$ and $\overline{N}_H^c$ can induce the large Majorana
masses for right-handed neutrinos allowing them to be used for the see--saw mechanism.
For this reason we assumed that the $U(1)_{\psi}\times U(1)_{\chi}$ symmetry is broken down
to $U(1)_{N}\times Z_{2}^{M}$ just below the GUT scale.

The $\tilde{Z}^{H}_2$ symmetry allows the Yukawa interactions in the superpotential that originate
from $27'_l \times 27'_m \times 27'_n$ and $27'_l \times 27_i \times 27_k$. Since the set of
multiplets $M_{l}$ contains only one pair of doublets $H_d$ and $H_u$ the $\tilde{Z}^{H}_2$ symmetry
defined above forbids not only the most dangerous baryon and lepton number violating operators
but also unwanted FCNC processes at the tree level. Nevertheless if the set of $\tilde{Z}^{H}_2$
even supermultiplets $M_{l}$ involve only $H_u$, $H_d$, $S$ and $N^c_H$ then the lightest exotic
quarks are extremely long--lived particles because $\tilde{Z}^{H}_2$ symmetry forbids all Yukawa
interactions in the superpotential that allow the lightest exotic quarks to decay. Since models
with stable charged exotic particles are ruled out by different terrestrial experiments the set
of supermultiplets $M_{l}$ in the phenomenologically viable $E_6$ inspired SUSY models should be
supplemented by some components of $27$-plet that carry $SU(3)_C$ colour or lepton number.

In this work we required that extra matter beyond the MSSM fill in complete $SU(5)$ representations
because in this case the gauge coupling unification remains almost exact in the one--loop
approximation. As a consequence we restricted our consideration to two scenarios that result in
different collider signatures associated with the exotic quarks. In the scenario A the set of
$\tilde{Z}^{H}_2$ even supermultiplets $M_{l}$ involves lepton superfields $L_4$. To ensure the
unification of gauge couplings we assumed that $\overline{H}_u$ and $\overline{H}_d$ are odd under
the $\tilde{Z}^{H}_2$ symmetry whereas supermultiplet $\overline{L}_4$ is even. Then $\overline{H}_u$
and $\overline{H}_d$ from the $\overline{27'}_l$ get combined with the superposition of the
corresponding components from $27_i$ so that the resulting vectorlike states gain masses of order
of $M_X$. In contrast, $L_4$ and $\overline{L}_4$ should form vectorlike states at low energies
facilitating the decays of exotic quarks. The superfield $\overline{S}$ can be either odd or even
under the $\tilde{Z}^{H}_2$ symmetry. The bosonic and fermionic components of $\overline{S}$ may or may
not survive to low energies. In the scenario A the exotic quarks are leptoquarks.

Another scenario, that permits the lightest exotic quarks to decay within a reasonable time, implies
that the set of multiplets $M_{l}$ together with $H_u$, $H_d$, $S$ and $N^c_H$ contains extra $d^c_4$
supermultiplet. Because in this scenario B the $\tilde{Z}^{H}_2$ even supermultiplets $d^c_4$ and
$\overline{d^c}_4$ give rise to the decays of the lightest exotic color states they are expected
to form vectorlike states with the TeV scale masses. Then to ensure that the extra matter
beyond the MSSM fill in complete $SU(5)$ representations $\overline{H}_u$ and $\overline{H}_d$ should
survive to the TeV scale as well. Again we assumed that $\overline{H}_u$ and $\overline{H}_d$ are
odd under the $\tilde{Z}^{H}_2$ symmetry so that they can get combined with the superposition of the
corresponding components from $27_i$ forming vectorlike states at low energies. As in the case of
scenario A the superfield $\overline{S}$ can be either even or odd under the $\tilde{Z}^{H}_2$
symmetry and may or may not survive to the TeV scale. In the scenario B the exotic quarks manifest
themselves in the Yukawa interactions as superfields with baryon number $\left(\pm\dfrac{2}{3}\right)$.

The gauge group and field content of the $E_6$ inspired SUSY model discussed here can originate from the
5D and 6D orbifold GUT models in which the splitting of GUT multiplets can be naturally achieved.
In particular, we studied $SU(5)\times U(1)_{\chi}\times U(1)_{\psi}$ SUSY GUT model in 5D compactified on
the orbifold $S^1/(Z_2\times Z'_2)$. At low energies this model may lead to the scenarios A and B. We also
considered $E_6$ gauge theory in $6D$ compactified on the orbifold $T^2/(Z_2 \times Z^{I}_2 \times Z^{II}_2)$
that can lead to the scenario A at low energies. In these orbifold GUT models all anomalies get cancelled
and GUT relations between Yukawa couplings get spoiled. The adequate suppression of the operators, that
give rise to proton decay, can be also achieved if the GUT scale $M_X\sim 1/R$ is larger than
$1.5-2\cdot 10^{16}\,\mbox{GeV}$.

We examined the RG flow of gauge couplings from $M_Z$ to $M_X$ in the case of scenarios A and B using both
analytical and numerical techniques. We derived the corresponding two--loop RG equations and studied the
running of the gauge couplings with and without extra $S$ and $\overline{S}$ superfields at the TeV scale.
In the scenario A the gauge coupling unification can be achieved for any phenomenologically reasonable value
of $\alpha_3(M_Z)$ consistent with the central measured low energy value. This was already established in
the case of the SUSY model with extra $U(1)_N$ gauge symmetry and low energy matter content that involves
three 27-plets of $E_6$ as well as $L_4$ and $\overline{L}_4$ \cite{King:2007uj}. Our analysis here revealed
that the evolution of the SM gauge couplings does not change much when the low energy particle spectrum is
supplemented by the $S$ and $\overline{S}$ chiral superfields. Thus this is not so surprising that the
unification of the SM gauge couplings can be so easily achieved even in this case. In the scenario B
large two--loop corrections spoil the unification of gauge couplings. Indeed, in this case the exact gauge
coupling unification can be achieved only if $\alpha_3(M_Z)\lesssim 0.112$. As before the inclusion of
extra $S$ and $\overline{S}$ superfields does not change much the RG flow of $\alpha_i(t)$ and therefore
does not improve gauge coupling unification. However the relative discrepancy of $\alpha_i(M_X)$ is about 10\% .
At the same time orbifold GUT framework does not imply the exact gauge coupling unification near the scale
$M_X\sim 1/R$ because of the brane contributions to the gauge couplings. Therefore relative discrepancy of
10\% between $\alpha_i(M_X)$ should not be probably considered as a big problem.

Finally we also discussed the cosmological implications and collider signatures of the $E_6$ inspired
SUSY models discussed above. As it was mentioned the low--energy effective Lagrangian of these models is invariant
under both $Z_{2}^{M}$ and $\tilde{Z}^{H}_2$ symmetries. Since $\tilde{Z}^{H}_2 = Z_{2}^{M}\times Z_{2}^{E}$ the
$Z_{2}^{E}$ symmetry associated with exotic states is also conserved. As a result the lightest exotic
state, which is odd under the $Z_{2}^{E}$ symmetry, must be stable. In the scenarios A and B the lightest and
second lightest inert neutralinos tend to be the lightest exotic states in the particle spectrum. On the other
hand the $Z_{2}^{M}$ symmetry conservation implies that $R$--parity is conserved. Because the lightest inert
neutralino $\tilde{H}^0_1$ is also the lightest $R$--parity odd state either the lightest $R$--parity even exotic
state or the lightest $R$--parity odd state with $Z_{2}^{E}=+1$ must be absolutely stable. Most commonly the second
stable state is the lightest ordinary neutralino $\chi_1^0$ ($Z_{2}^{E}=+1$). Both stable states are natural
dark matter candidates in the considered $E_6$ inspired SUSY models.

When $|m_{\tilde{H}^0_{1}}|\ll M_Z$ the lightest inert neutralino is predominantly inert singlino and its couplings
to the gauge bosons, Higgs states, quarks and leptons are very small resulting in too small annihilation cross
section for $\tilde{H}^0_1\tilde{H}^0_1\to \mbox{SM particles}$. As a consequence the cold dark matter density is much
larger than its measured value. In principle, $\tilde{H}^0_1$ could account for all or some of the observed cold
dark matter density if it had mass close to half the $Z$ mass. In this case the lightest inert neutralino states
annihilate mainly through an $s$--channel $Z$--boson. However the usual SM-like Higgs boson decays more than 95\%
of the time into either $\tilde{H}^0_1$ or $\tilde{H}^0_2$ in these cases while the total branching ratio into
SM particles is suppressed. Because of this the corresponding scenarios are basically ruled out nowadays.
The simplest phenomenologically viable scenarios imply that the lightest and second lightest inert neutralinos are
extremely light. For example, these states can have masses about $1\,\mbox{eV}$. The lightest and second lightest
inert neutralinos with masses about $1\,\mbox{eV}$ form hot dark matter in the Universe but give only a very minor
contribution to the dark matter density while the lightest ordinary neutralino may account for all or some of
the observed dark matter density.

The presence of two types of dark matter is a very peculiar feature that affect the collider
signatures of the considered $E_6$ inspired SUSY models. The most spectacular LHC signals
associated with these models may come from the TeV scale exotic color states and $Z'$.
The production of the $Z'$ boson, that corresponds to the $U(1)_N$ gauge
symmetry, should lead to unmistakable signal $pp\to Z'\to l^{+}l^{-}$ at the LHC.
The $Z_{2}^{E}$ symmetry conservation implies that in collider experiments exotic particles
can only be created in pairs. Moreover each exotic particle has to decay into a final state
that contains at least one lightest inert neutralino resulting in the missing energy.
Because of this the lightest exotic color state, that can be either $D$-fermion or $\tilde{D}$-scalar,
decay into either $u_i (d_i) + \ell (\nu) + E^{\rm miss}_{T} + X$ if exotic
quark (squark) is leptoquark or $u^c_i + d^c_j + E^{\rm miss}_{T} + X$ if exotic quark
(squark) is diquark. The $Z_{2}^{E}$ symmetry conservation requires that $E^{\rm miss}_{T}$
should always contain contribution associated with the lightest inert neutralino. Since the
lightest exotic squark is $R$--parity even state while the lightest inert neutralino is $R$--parity
odd particle the final state in the decay of $\tilde{D}$ should also involve the lightest
ordinary neutralino to ensure $R$--parity conservation. Thus the pair production of the lightest
exotic color state is expected to lead to a substantial enhancement of the cross section of
either $pp\to jj\ell^{+}\ell^{-}+E^{\rm miss}_{T}+X$ or
$pp\to jjjj+E^{\rm miss}_{T}+X$. If the Yukawa couplings of the exotic particles
have hierarchical structure similar to the one observed in the ordinary quark and lepton
sectors then all states which are odd under the $Z^{E}_2$ symmetry couple to the third
generation fermions and sfermions mainly. As a result the TeV scale exotic color
states should give rise to the enhancement of the cross section of either
$pp\to t\bar{t}\ell^{+}\ell^{-}+E^{\rm miss}_{T}+X$ or $pp\to t\bar{t}b\bar{b}+E^{\rm miss}_{T}+X$.

Our consideration indicates that $\tilde{D}$-scalars in the considered $E_6$ inspired
SUSY models lead to rather unusual collider signatures. Indeed, it is commonly expected
that scalar diquarks decay into quark--quark without missing energy in the final state
while the scalar leptoquarks decay into quark--lepton without missing energy as well.
In the models considered here the $Z_{2}^{E}$ symmetry conservation necessarily
leads to the missing energy in the corresponding final states. In addition relatively
light exotic quark and squark can modify the collider signatures associated with gluinos
if the decay of the gluino into exotic quark and squark is kinematically allowed.
In this case gluino pair production at the LHC may result in $D\bar{D}\tilde{D}\tilde{\bar{D}}$
in the final state. The sequential decays of $D$-fermions and $\tilde{D}$-scalars give rise to the
enhancement of either $pp\to 4\,\ell+4\, j + E^{\rm miss}_{T}+X$ or $pp\to 8\,j + E^{\rm miss}_{T}+X$.

In the scenario B the fermionic components of the supermultiplets $d^c_4$ and $\overline{d^c}_4$
that form vectorlike quark state as well as their superpartner may have TeV scale masses.
Then these quark and/or squark states can be pair produced at the LHC via strong interactions
and decay into $q_i + E^{\rm miss}_{T} + X$ where $q_i$ can be either up-type or down-type quark.
This may lead to an enhancement of $pp\to jj+E^{\rm miss}_{T}+X$.

The discovery of $Z'$ and new exotic particles predicted by the $E_6$ inspired SUSY models considered
here will open a new era in elementary particle physics. This would not only represent a revolution
in particle physics, but would also point towards an underlying $E_6$ gauge structure at high energies.

\section*{Acknowledgements}
\vspace{-2mm}
R.N. thanks X.~Tata for sharing his valuable ideas in connection with this work. 
R.N. acknowledges fruitful discussions with S.~F.~King, J.~Kumar, S.~Moretti, S.~Pakvasa 
and T.~Rizzo. R.N. is also grateful to P.~Athron, J.~Bjorken, K.~R.~Dienes, J.~Hewett, 
S.~Kraml, D.~J.~Miller, M.~M\"{u}hlleitner, M.~Sher, M.~A.~Shifman, L.~B.~Okun, B.~D.~Thomas, 
D.~G.~Sutherland, A.~I.~Vainshtein, M.~I.~Vysotsky for valuable comments and remarks. 
The work of R.N. was supported by the U.S. Department of Energy under Contract DE-FG02-04ER41291.

\end{document}